\documentclass[notitlepage,aps,prd,showpacs,longbibliography,preprintnumbers,nofootinbib]{revtex4-1}

\usepackage{graphicx}
\usepackage{bm,latexsym,amsmath,amssymb,amsfonts}

\usepackage{hyperref}
\usepackage[utf8]{inputenc}

\usepackage[format=plain,small,bf,justification=centerlast]{caption}
\usepackage[subrefformat=parens]{subcaption}

\newcommand{\calG}{\mathcal{G}}

\allowdisplaybreaks

\makeatletter
\def\l@subsubsection#1#2{}
\makeatother

\begin{document}

\title{Wave propagation and shock formation in the most general scalar-tensor theories}

\author{Norihiro~Tanahashi}
\email[Email: ]{tanahashi``at''het.phys.sci.osaka-u.ac.jp}
\affiliation{Department of Physics, Osaka University, Toyonaka, Osaka 560-0043, Japan}

\author{Seiju~Ohashi}
\email[Email: ]{seiju.ohashi``at''informatix.co.jp}
\affiliation{Informatix Inc., Kawasaki, Kanagawa 212-0014, Japan}

\begin{abstract}
This work studies wave propagation in the most general scalar-tensor theories, particularly focusing on the causal structure realized in these theories and also the shock formation process induced by nonlinear effects.
For these studies we use the Horndeski theory and its generalization to the two scalar field case. We show that propagation speeds of gravitational wave and scalar field wave in these theories may differ from the light speed depending on background field configuration, and find that a Killing horizon becomes a boundary of causal domain if the scalar fields share the symmetry of the background spacetime. About the shock formation, we focus on transport of discontinuity in second derivatives of the metric and scalar field in the shift-symmetric Horndeski theory. 
We find that amplitude of the discontinuity generically diverges within finite time, 
which corresponds to shock formation.
It turns out that the canonical scalar field and the scalar DBI model, among other theories described by the Horndeski theory, are free from such shock formation even when the background geometry and scalar field configuration are nontrivial.
We also observe that gravitational wave is protected against shock formation when the background has some symmetries at least. This fact may indicate that the gravitational wave in this theory is more well-behaved compared to the scalar field, which typically suffers from shock formation.
\end{abstract}

\preprint{OU-HET-929}

\maketitle

\tableofcontents

\section{Introduction}
\label{Sec:intro}

Theories of gravitation modified by scalar degrees of freedom have a long history of study, and nowadays such theories are utilized in various research fields including gravitational physics and cosmology.
Just to list some of the applications, inflation describing our universe in the earliest era~\cite{Starobinsky:1980te,*Sato:1980yn,*Guth:1980zm} is typically realized using a scalar field called inflaton that is coupled to gravity. Also, many attempts have been made to explain the accelerated expansion of our universe at late time~\cite{Schmidt:1998ys,*Riess:1998cb,*Perlmutter:1998np} by modifying gravitation at cosmologically large scale rather than by attributing it to the cosmological constant, whose origin it yet to be known.
For these modifications to be viable, they must be consistent with various tests of gravity at scales ranging from sub-millimeter to astrophysical ones. To conduct such experimental tests and also to examine their theoretical consistencies, behaviors of gravity and scalar fields in those modified theories have been studied from various viewpoints as reviewed in, e.g., Ref.~\cite{Clifton:2011jh}.

Quite a number of such gravitation theories with modifications has been proposed by now, and then it is desirable to have a theory that could be used as a framework to treat them in a unified manner.
Such a theory must be free from pathological behaviors including ghost instability, and also it should encompass a wide variety of other theories as its subclass.
Within this context,
the Galileon theory was proposed as a theory that is free from the Ostrogradsky ghost instability although its Lagrangian contains higher derivative terms \cite{Nicolis:2008in}. Later on, this theory was covariantized to include gravity and also was generalized to incorporate more parameters into the theory by \cite{Deffayet:2009wt,Deffayet:2011gz}. The resultant theory was dubbed the generalized Galileon theory, which was shown by \cite{Kobayashi:2011nu} to be equivalent to the Horndeski theory~\cite{Horndeski:1974wa} constructed in 1970's as a generalization of Lovelock theories of gravity~\cite{Lovelock:1971yv}.
This theory has been studied in wide variety of contexts in cosmology and gravitational physics, 
e.g.\ by \cite{Kobayashi:2011nu} which pursued inflation scenario based on it.

The Horndeski theory is the most general scalar-tensor theory with a single scalar field whose Euler-Lagrange equation has derivatives of the metric and the scalar field only up to second order.
There are a several ways to extend this theory further.
One of the simplest extensions would be to incorporate multiple scalar fields into the theory, which was realized in the generalized multi-Galileon theory \cite{Padilla:2010de,Padilla:2010tj,Hinterbichler:2010xn,Padilla:2010ir,Padilla:2012dx}.
Later on, it was realized that the multi-DBI inflation models \cite{RenauxPetel:2011dv,RenauxPetel:2011uk,Koyama:2013wma,Langlois:2008wt,Langlois:2008qf}
are not included in this theory~\cite{Kobayashi:2013ina}, which motivated to construct the two scalar field version of the Horndeski theory based on the derivation in the single scalar field case.
Such a construction was attempted in \cite{Ohashi:2015fma}. As a result, the most general equations of motion of such a theory was successfully constructed, while the Lagrangian corresponding to those equations has not been found so far. We call this theory the bi-Horndeski theory in this work. 
There are also some other theories with multiple scalar fields proposed based on different methods (see e.g.\ \cite{Padilla:2013jza,Saridakis:2016ahq,Saridakis:2016mjd,Allys:2016hfl,Akama:2017jsa}) as well as generalized theories for vector fields~\cite{doi:10.1063/1.522837,Deffayet:2013tca,Tasinato:2014eka,Heisenberg:2014rta,Allys:2015sht,Jimenez:2016isa,Jimenez:2016upj}.

Another type of extension is to construct scalar-tensor theories that encompass the Horndeski theory as their subclass. Such a theory was first introduced in \cite{Zumalacarregui:2013pma,Gleyzes:2014dya,Gleyzes:2014qga}, where a subclass of the extended theory was related to the Horndeski theory by disformal transformation.
This theory was dubbed the beyond-Horndeski theory, and later on it was extended further to incorporate higher order terms in Lagrangian without re-introducing the ghost instability in \cite{Gao:2014soa,Deffayet:2015qwa,Domenech:2015tca,Langlois:2015cwa,Langlois:2015skt,Crisostomi:2016tcp,Crisostomi:2016czh,Achour:2016rkg,Ezquiaga:2016nqo,Motohashi:2016ftl,deRham:2016wji,BenAchour:2016fzp,Crisostomi:2017aim,Langlois:2017mxy}.
There are similar extended theories involving vector degrees of freedom and also derivatives of spacetime curvature \cite{Heisenberg:2016eld,Naruko:2015zze},
which have been attracting wide interest recently.

In this work, 
among the extended theories mentioned above, 
we focus on the Horndeski theory and also the bi-Horndeski theory.
The latter is related to various other theories such as the Horndeski theory and the multi-Galileon theory as shown in appendix~\ref{App:toOtherTheories}, hence the results obtained for this theory will be applicable to those theories as well.
Also, the mathematical structure of these theories are relatively simpler and then easier to deal with
compared to 
the other extended theories mentioned above.
Adding to that,
the Horndeski theory is known to be related to Gauss-Bonnet and Lovelock theories of gravity in higher dimensions via dimensional reduction (see e.g.\ \cite{Charmousis:2012dw,Charmousis:2014mia}).
This fact suggests that some properties of these higher-dimensional theories, such as those studied in \cite{Izumi:2014loa,Reall:2014pwa,Reall:2014sla}, may persist even in the bi-Horndeski theory and its descendants.

One of the most basic properties of a theory is the wave propagation speed, 
since 
it governs the wave dynamics 
and also the causal structure 
of that theory.
In the general relativity (GR) with a minimally-coupled canonical scalar field, the situation is relatively simple because any wave simply propagates at the speed of light. However, in the Horndeski theory and its generalizations, propagation speeds of the gravitational wave and scalar field wave may differ from the speed of light, and also they may depend on the environment, that is, the background field configuration on which the wave propagates.
Phenomena associated with wave propagation in extended gravity theories have been studied in various contexts (see e.g.\ \cite{Adams:2006sv,Babichev:2006vx,Babichev:2007dw,Appleby:2011aa,Neveu:2013mfa,deFromont:2013iwa,Barreira:2013eea,Brax:2015dma}), and particularly in \cite{Izumi:2014loa,Reall:2014pwa,Minamitsuji:2015nca} properties of Killing horizons, such as black hole horizons in stationary spacetimes, were studied based on Lovelock theories that incorporate Gauss-Bonnet gravity theory, and also based on scalar-tensor theories with a non-minimally coupled scalar field. In Lovelock theories, 
gravitational wave may propagate superluminally depending on the background spacetime. However, it was shown that such superluminal propagation is prohibited on a Killing horizon \cite{Izumi:2014loa,Reall:2014pwa}, hence it becomes a boundary of causal contact in the sense that no wave can come out from it.
In the scalar-tensor theories with non-minimal coupling, however, it was shown that such a property is not guaranteed in general, and additional conditions must be imposed on the scalar field for a Killing horizon to be a causal edge~\cite{Minamitsuji:2015nca}. In this work, we will examine this issue on Killing horizons based on the bi-Horndeski theory.
We also exemplify wave propagation in this theory on a nontrivial background by taking the plane wave solution \cite{Babichev:2012qs}, which is an exact solution of the shift-symmetric Horndeski theory, as the background and examine how the causality is implemented on it.

Another issue we address in this work is the formation of shock (or caustics) in scalar-tensor theories that is caused by nonlinear self interaction of waves.
One of the simplest example of such shock formation is realized in Burgers' equation $u_{,t} + u\, u_{,x} = 0$, for which initially smooth wave profile is distorted in time evolution due to the nonlinear term $u\,u_{,x}$. Within finite time, the wave profile becomes double-valued and derivatives of $u$ diverge there, which may be interpreted as shock formation.
Such a phenomenon typically occurs when wave propagation speed depends on environment and also on its own amplitude. Wave obeying Burgers' equation and also gravitational wave obeying Lovelock theories \cite{Reall:2014sla} have this property, and indeed it can be shown that shock formation occurs for them.
In this work, we study such shock formation process based on the Horndeski theory, where we impose shift symmetry in scalar field to the theory for simplicity.
Such shock (or caustics) formation was studied for a probe scalar field with Horndeski-type action on flat spacetime in \cite{Babichev:2016hys,Mukohyama:2016ipl,deRham:2016ged} by constructing simple wave solutions.%
\footnote{Properties of caustics were studied also by \cite{Felder:2002sv,*Barnaby:2004nk,*Goswami:2010rs} in the DBI-type scalar field theories and by \cite{Bekenstein:2004ne,*Contaldi:2008iw,*Mukohyama:2009tp,*Setare:2010gy} in other theories.} We will take a different approach following \cite{Reall:2014sla}
that focuses on transport of discontinuity in second derivatives of dynamical fields. We will also check if gravitational wave in the Horndeski theory would suffer from shock formation. If it would, interesting implications may be obtained for gravitational wave observations which was recently realized by LIGO group~\cite{Abbott:2016blz}.

The organization of this paper is as follows. 
In section~\ref{Sec:characteristics}, we give a brief introduction on characteristics, which is a mathematical tool to study wave propagation, and based on it we derive the principal part and characteristic equation of the bi-Horndeski theory.
In section~\ref{Sec:causaledge}, we study properties of characteristic surfaces in the bi-Horndeski theory, particularly focusing on the causal structure in a spacetime with a Killing horizon.
As another application of the formalism developed in the previous sections, in section~\ref{Sec:planewave} we study wave propagation on the plane wave solution, which is an exact solution in the Horndeski theory.
In section~\ref{Sec:shock}, we turn to the issue of shock formation in the Horndeski theory. After reviewing the general formalism in section \ref{Sec:shock_general}, 
we apply it to the shift-symmetric Horndeski theory in section~\ref{Sec:shock-Horndeski}.
Then, we examine shock formation process on the plane wave solution and two-dimensionally maximally-symmetric dynamical spacetime in sections~\ref{Sec:Npp} and \ref{Sec:2-dim}, respectively. 
Background solutions in the latter example include
the Friedmann-Robertson-Walker (FRW) universe and and also spherically-symmetric static solutions. Using these backgrounds, we will study conditions for shock formation in waves propagating on them.
We then
conclude this work in section~\ref{Sec:summary} with discussions.
Some formulae necessary for this work are summarized in appendices. Field equations of the bi-Horndeski theory, and its relationship with those of other theories such as the Horndeski theory are summarized in appendices \ref{App:BiHorndeskiEoM} and \ref{App:toOtherTheories}. The integrability conditions of the bi-Horndeski theory is shown in appendix \ref{App:integrability} for reference. 
Appendices~\ref{App:ssHorndeski}, \ref{App:N} and \ref{App:Npp} are devoted to studies on wave propagation and shock formation in the shift-symmetric Horndeski theory.

We summarize the convention for indices used in this work in Table~\ref{Table:indices}.
Also we denote partial and covariant derivatives by comma and stroke, respectively ($v_{,a}\equiv \partial v / \partial x^a$, $v_{|a}\equiv \nabla_a v$).
The generalized Kronecker delta $\delta^{a_1\ldots a_n}_{b_1\ldots b_n} \equiv n! \delta^{\,a_1}_{[b_1}\cdots \delta^{a_n}_{b_n]}$ is used to describe the (bi-)Horndeski theory.
See the definitions in each section for more details.
\begin{table}[htbp]
\caption{Notation of indices in this work.}
\centering
\begin{tabular}{|l|l|}
\hline
$a,b,c,d,\ldots$, $q,r,s,t$ & Four-dimensional indices for $x^{a=0,1,2,3}$
\\ \hline 
$\mu,\nu,\ldots$ & Three-dimensional indices for $x^{\mu=1,2,3}$ on the hypersurface $\Sigma$ at $x^0=0$
\\ \hline
$\alpha,\beta,\ldots$ & Two-dimensional spatial indices for $x^{\alpha=2,3}$ in section~\ref{Sec:characteristics}, \ref{Sec:causaledge} and for angular directions  in section~\ref{Sec:2-dim}
\\ \hline
$i,j$ & Two-dimensional spatial indices of the null basis in section~\ref{Sec:planewave}
\\ \hline
$A, B, \ldots$ & Two-dimensional indices for $x^A=\tau,\chi$ in section~\ref{Sec:2-dim}
\\ \hline
$I, J, \ldots$ & Scalar field indices of the bi-Horndeski theory $\phi_{I=1,2}$; used also as generic indices (e.g.\ in Eq.~(\ref{fieldeq}))
\\ \hline
\end{tabular}
\label{Table:indices}
\end{table}

\section{Characteristics in bi-Horndeski Theory}
\label{Sec:characteristics}

In this section, we explain the method to examine the causal structure of the bi-Horndeski theory. The full expression of the field equations in this theory are shown in appendix~\ref{App:BiHorndeskiEoM}, and its relationship with the generalized multi-Galileon theory and the 
Horndeski theory with a single scalar field is summarized in appendix~\ref{App:toOtherTheories}.

\subsection{Characteristics}
\label{Sec:review}

As a preparation for the analysis on the bi-Horndeski theory, we give a short review on characteristics for a generic equation of motion.
Suppose that a vector of dynamical variables $v_I$ obeys a set of field equations
\begin{equation}
E_I(v, \partial v, \partial^2 v) = 0.
\label{fieldeq}
\end{equation}
To consider time evolution based on this equation, we introduce a three-dimensional hypersurface $\Sigma$ and a coordinate system $(x^a)=(x^0,x^{\mu})$. $\Sigma$ is given by $x^0 = 0$ and $x^{\mu}$ lies on $\Sigma$.
We use notation that Latin indices ($a,b,\ldots$) denote all the four dimensions and Greek indices ($\mu,\nu,\ldots$) denote only the three dimensions on $\Sigma$.
Now let us assume that $E$ is linear in $\partial_0^2 v$, which is the case in the Horndeski and bi-Horndeski theories. Then Eq.~(\ref{fieldeq}) is expressed as
\begin{equation}
\frac{\partial E_I}{\partial v_{J,00}} v_{J,00} + \cdots = 0,
\label{fieldeq2}
\end{equation}
where a comma denotes partial derivative ($v_{J,00}\equiv \partial^2 v_J/(\partial x^0)^2$) and the ellipses denote terms up to first order in derivatives with respect to $x^0$.
Equation~(\ref{fieldeq2}) can be solved to determine $v_{J,00}$ in terms of quantities with lower order $x^0$ derivatives as long as the coefficient matrix of the $v_{J,00}$ term, 
\begin{equation}
\frac{\partial E_I}{\partial v_{J,00}},
\label{Ppre}
\end{equation}
is not degenerate and invertible as a matrix acting on the vector $v_I$.
On the other hand, if it is not invertible then the value of $v_{J,00}$ cannot be fixed by Eq.~(\ref{fieldeq2}). In such a situation, we call $\Sigma$ characteristic.

A characteristic surface gives a boundary of causal domain and defines the maximum propagation speed allowed in the theory, as we can see by the following consideration~\cite{BA0392362X} (see also \cite{Izumi:2014loa,Reall:2014pwa}). Suppose that we have discontinuity in $v_{I,00}$ across $\Sigma$, while $v_a$ and $v_{a,0}$ are continuous there. Then this surface must be characteristic, because otherwise $v_{I,00}$ cannot be discontinuous as we argued above.
It implies that the discontinuity propagates on the characteristic surface $\Sigma$, and in this sense we may regard $\Sigma$ as the wave front corresponding to the discontinuity, which may be interpreted as wave in the high frequency limit.
Now let us consider time evolution from an initial time slice, and focus on a finite part of such a slice.
In the region enclosed by that part of the initial time slice and the characteristic surface emanating from the edge of that part, the time evolution will be uniquely specified by the initial data on that part of the initial time slice. It is because we can solve Eq.~(\ref{fieldeq2}) to fix the solution in such a region.
On the other hand, the solution outside this region cannot be fixed only by the initial data on that part of the initial time slice, because disturbances outside that part can propagate in this outer region along characteristic surfaces.
In this sense, for a region on an initial time slice, the boundary of the causal domain is given by the characteristic surface emanating from its edge.
In other words
the maximum propagation speed in a theory is determined by characteristic surfaces.

To express the coefficient matrix~(\ref{Ppre}) covariantly, we introduce a normal vector of the surface $\Sigma$, $\xi_a \equiv (dx^0)_a$, with which we can express the equation to determine characteristic surfaces as
\begin{equation}
P(x,\xi)\cdot r
\equiv
\frac{\partial E_I}{\partial v_{J,st}}\xi_s \xi_t r_J
=0.
\label{chareq}
\end{equation}
$\Sigma$ is characteristic if Eq.~(\ref{chareq}) has nontrivial solutions, which is realized when $\det P=0$, and the eigenvector $r$ for a vanishing eigenvalue
corresponds to a mode propagating on $\Sigma$.
$P$ is called the principal symbol of the field equation~(\ref{fieldeq2}), and $\det P=0$ is called the characteristic equation.

\subsection{Characteristic equation of bi-Horndeski theory}
\label{Sec:char-BiHondeski}

We apply the analysis shown in the previous section to the field equations of the bi-Horndeski theory in this section.
Clarifying the principal symbol of the field equations in sections~\ref{Sec:P-bH}, we introduce the characteristic equation and the principal symbol in this theory in section~\ref{Sec:Pwhole}.

\subsubsection{Equations of motion}
\label{Sec:P-bH}

We start our analysis from the metric part of the field equations.
It turns out that, in the bi-Horndeski theory,
the $(00)$ and $(0\mu)$ components of the gravitational equations do not contain second derivatives with respect to $x^0$ and only the $(\mu\nu)$ components has them. Hence it suffices to look at the $(\mu\nu)$ components of field equations, which are
\begin{align}
E^{\mu\nu}(\mathcal{L})&=\mathcal{A}^{\mu\nu,\rho\sigma}g_{\rho\sigma,00}+\mathcal{B}^{\mu\nu}_I\phi_{I}{}_{,00}+\mathcal{C}^{\mu\nu}
\end{align}
where $\mathcal{A}^{\mu\nu,\rho\sigma}, \mathcal{B}^{\mu\nu}_I$ and $\mathcal{C}^{\mu\nu}$ are given by,
denoting a covariant derivative by a stroke as $\nabla_a \phi^I \equiv \phi^I_{|a}$,
\begin{align}
\mathcal{A}^{\mu\nu,\rho\sigma}&=
-\left(\mathcal{F}+2\mathcal{W}\right)
g^{p(\mu}\delta^{\nu)0(\rho}_{pmf}g^{\sigma)f}g^{0m}
-2J_{IJ}g^{p(\mu}\delta^{\nu)c0(\rho}_{pdmf}g^{\sigma)f}g^{0m}\phi^{I}_{|c}\phi^{J|d}
-2K_Ig^{p(\mu}\delta^{\nu)c0(\rho}_{pdmf}g^{\sigma)f}g^{0m}\phi^{I|d}_{|c},
\\
\mathcal{B}^{\mu\nu}_{I}&=\tilde{B}_Ig^{l(\mu}\delta^{\nu)0}_{lm}g^{0m}
                              +D_{JKI}g^{l(\mu}\delta^{\nu)c0}_{ldm}g^{0m}\phi^{J}_{|c}\phi^{K|d}
                              +E_{JKLMI}g^{l(\mu}\delta^{\nu)ce0}_{ldfm}g^{0m}\phi^{J}_{|c}\phi^{K|d}\phi^{L}_{|e}\phi^{M|f}\notag \\
                              &\quad
+2\mathcal{F}_{,IJ}g^{l(\mu}\delta^{\nu)c0}_{ldm}g^{0m}\phi^{J|d}_{|c}
                              +4J_{JK,LI}g^{l(\mu}\delta^{\nu)ce0}_{ldfm}g^{0m}\phi^{J}_{|c}\phi^{K|d}\phi^{L|f}_{|e}\notag \\
                              &\quad 
+K_Ig^{l(\mu}\delta^{\nu)0ce}_{lmdf}g^{0m}R_{ce}{}^{df}
                              +2K_{I,JK}g^{l(\mu}\delta^{\nu)ce0}_{ldfm}g^{0m}\phi^{J|d}_{|c}\phi^{K|f}_{|e},
\label{B}
\\
\mathcal{C}^{\mu\nu}
&=
\mathcal{C}^{\mu\nu}\bigl( g_{\rho\sigma}, g_{\rho\sigma,0}, g_{\rho\sigma,\kappa},g_{\rho\sigma,\kappa0}, g_{\rho\sigma,\kappa\lambda}, \phi_I, \phi_I{}_{,0}, \phi_I{}_{,\rho},\phi_I{}_{,0\rho}, \phi_I{}_{,\rho\sigma} \bigr).
\end{align}
$\tilde B_I$ in Eq.~(\ref{B}) is defined as
\begin{equation}
\tilde{B}_I\equiv -2\mathcal{F}_{,I}-\mathcal{W}_{,I}+2\left( D_{JKI}+8J_{J[K,I]}\right) X^{JK}-8E_{JKLMI}X^{JK}X^{LM}.
\end{equation}

We next look at the equations of motion for the scalar fields, which are given as
\begin{align}
\mathcal{E}_{I}&=\mathcal{G}^{\mu\nu}_Ig_{\mu\nu,00}+\mathcal{H}_{IJ}\phi_{J,00}+\mathcal{I}_I,
\end{align}
where $\mathcal{G}^{\mu\nu}_{I}, \mathcal{H}_{IJ}$ and $\mathcal{I}_I$ are defined by
\begin{align}
\mathcal{G}^{\mu\nu}_I
&= \left( D_{IJK} -8 J_{J[K,I]}-8E_{LKJIM}X^{LM}\right)\delta^{c0(\mu}_{dml}g^{\nu)l}\phi^{J}_{|c}\phi^{K|d}g^{m0}
                            +2D_{IJK}X^{JK}\delta^{0(\mu}_{ml}g^{\nu)l}g^{m0}\notag \\
                            &\quad
+2E_{LJKIM}\delta^{ce0(\mu}_{dfmh}g^{\nu)h}\phi^{L}_{|c}\phi^{J|d}\phi^{K}_{|e}\phi^{M|f}g^{m0}
                            +4J_{IL,JK}g^{l(\mu}\delta^{\nu)ce0}_{ldfh}\phi^{J}_{|c}\phi^{L|d}\phi^{K|f}_{|e}g^{h0}
                           \notag \\
                            &\quad
-4I_{,I}g^{d(\mu}\delta^{\nu)0}_{df}g^{f0}+4\left( J_{IJ}-K_{(I,J)}+2J_{K(I,J)L}X^{KL}\right) \delta^{c0(\mu}_{bhf}g^{\nu)f}\phi^{J|b}_{|c}g^{h0}\notag \\
                            &\quad
+2K_{J,IK}\delta^{ce0(\mu}_{dfmh}g^{\nu)h}\phi^{J|d}_{|c}\phi^{K|f}_{|e}g^{m0}+ K_I\delta^{ce0(\mu}_{dfmh}g^{\nu)h}R_{ce}{}^{df}g^{m0},
\\
\mathcal{H}_{IJ}&=2B_{IJ}g^{00}-2B_{IK,LJ}\phi^{K|0}\phi^{L|0}
                          +2C_{J,I}g^{00}+4D_{K[I|J,|L]}\left( 2X^{KL}g^{00}+\phi^{K|0}\phi^{L|0}\right) \notag \\
                          &\quad+2D_{IKJ,LM}\delta^{c0}_{bf}\phi^{K}_{|c}\phi^{L|d}\phi^{M|b}_{|d}g^{f0}+2D_{IKL,MJ}\delta^{ce}_{bf}\phi^{K}_{|c}\phi^{L|f}_{|e}\phi^{M|0}g^{b0}
                          -4D_{I(JK)}\delta^{c0}_{bf}\phi^{K|b}_{|c}g^{f0}\notag \\
                          &\quad+2\left( E_{KLMNJ,I} -2E_{KLMIJ,N}\right)\delta^{ce0}_{dfh}\phi^{K}_{|c}\phi^{L|d}\phi^{M}_{|e}\phi^{N|f}g^{h0}\notag \\
                          &\quad-4E_{KLMIJ,NO}\delta^{ce0}_{bfh}\phi^{K}_{|c}\phi^{L|f}\phi^{M}_{|e}\phi^{N|l}\phi^{O|b}_{|l}g^{h0}-4E_{KLMIN,OJ}\delta^{ceg}_{bfh}\phi^{K}_{|c}\phi^{L|f}\phi^{M}_{|e}\phi^{N|h}_{|g}\phi^{O|0}g^{b0}\notag \\
                          &\quad+16E_{(K|LMI|J)}\delta^{ce0}_{bfh}\phi^{K|b}_{|c}\phi^{M}_{|e}\phi^{L|f}g^{h0}
                          +4G_{JK,I}\delta^{c0}_{df}g^{f0}\phi^{K|d}_{|c}\notag \\
                          &\quad+8\left( J_{KL,MJ,I}-J_{IK,MJ,L}\right) \delta^{ce0}_{dfh}\phi^{K}_{|c}\phi^{L|d}\phi^{M|f}_{|e}g^{h0}-12J_{I(J,KL)}\delta^{ce0}_{bfh}\phi^{K|b}_{|c}\phi^{L|f}_{|e}g^{h0}\notag \\
                          &\quad-4J_{KM,LN,IJ}\delta^{ceg0}_{dfhm}\phi^{K}_{|c}\phi^{M|d}\phi^{L|f}_{|e}\phi^{N|h}_{|g}g^{m0}-8J_{IK,LM,NJ}X^{KM}\delta^{eg0}_{fhm}g^{m0}\phi^{L|f}_{|e}\phi^{N|h}_{|g}\notag \\
                          &\quad-2J_{KL,IJ}\delta^{ceg0}_{dfhm}g^{m0}\phi^{K}_{|c}\phi^{L|d}R_{eg}{}^{fh}\notag \\
                          &\quad-2\left( -K_{(I,J)}+J_{IJ}+2J_{K(I,J)L}X^{KL}\right) \delta^{0eg}_{dfh}g^{0d}R_{eg}{}^{fh}-2K_{J,IK}\delta^{0egl}_{dfhm}g^{0d}\phi^{K|f}_{|e}R_{gl}{}^{hm}\notag \\
                          &\quad+4K_{J,KL,I}\delta^{ce0}_{dfh}\phi^{K|d}_{|c}\phi^{L|f}_{|e}g^{h0}-\frac43K_{L,JK,IM}\delta^{ceg0}_{dfhm}\phi^{M|d}_{|c}\phi^{L|f}_{|e}\phi^{K|h}_{|g}g^{m0},
\\
\mathcal{I}_{I}&=\mathcal{I}_{I}\bigl(g_{\rho\sigma}, g_{\rho\sigma,0}, g_{\rho\sigma,\kappa},g_{\rho\sigma,\kappa0}, g_{\rho\sigma,\kappa\lambda}, \phi_I, \phi_I{}_{,0}, \phi_I{}_{,\rho},\phi_I{}_{,0\rho}, \phi_I{}_{,\rho\sigma} \bigr).
\end{align}

\subsubsection{Characteristic equation}
\label{Sec:Pwhole}

Using the above expressions, 
the equations of motion are written as
\begin{align}
P
\cdot v_{,00} = S,
\end{align}
where
\begin{equation}
P
=\left( 
\begin{array}{@{}cc@{}}
\mathcal{A}^{\mu\nu,\rho\sigma} & \mathcal{B}^{\mu\nu}_{J}\\
\mathcal{G}^{\rho\sigma}_{I} & \mathcal{H}_{IJ}\\
\end{array}
\right),
\qquad
v
=\left( 
\begin{array}{@{}c@{}}
g_{\rho\sigma}\\
\phi_{J}\\
\end{array}
\right),
\qquad
S
=\left( 
\begin{array}{@{}c@{}}
\mathcal{C}^{\mu\nu}\\
\mathcal{I}_{I}\\
\end{array}
\right).
\end{equation}
Then the characteristics are found by solving
\begin{equation}
P\cdot r = 0,
\label{Pdotr=0}
\end{equation}
where $r=(r_{ab},r_J)$ is a vector made of a symmetric tensor $r_{ab}$ and a vector of scalars with two components $r_J$.
The characteristic equation is given by $\det P=0$, and eigenvectors for vanishing eigenvalues corresponds to the modes propagating on the characteristic surface $\Sigma$ as we argued in section~\ref{Sec:review}.

As we discuss in appendix~\ref{App:integrability}, the integrability conditions for equations of motion guarantee that the matrix $P$ is symmetric, i.e., 
\begin{equation}
\mathcal{A}^{\mu\nu,\rho\sigma}=\mathcal{A}^{\rho\sigma,\mu\nu},
\quad
\mathcal{B}^{\mu\nu}_I =\mathcal{G}^{\mu\nu}_I,
\quad
\mathcal{H}_{IJ}=\mathcal{H}_{JI}.
\label{Psymmetry}
\end{equation}
The integrability conditions have not been imposed to the field equations in~\cite{Ohashi:2015fma}, and then the principal symbol derived above does not have the symmetry~(\ref{Psymmetry}) in general.
We proceed without imposing these conditions in the analysis below, and we leave the full analysis with these conditions for future work.
In the next section, we find that some properties of causal structure in this theory can be read out despite this restriction.

\section{Causal edge in bi-Horndeski theory}
\label{Sec:causaledge}

In GR, a null surface is always a characteristic surface and hence it gives the boundary of causal domain. This property is lost in the bi-Horndeski theory and hence the causal structure in this theory becomes nontrivial. Particularly, a Killing horizon in a stationary spacetime may not be a characteristic surface in this theory, which means that the black hole region defined in a usual sense may be visible from outside due to the presence of superluminal modes.
In this section, as a first application of the formalism developed in the previous section, we clarify conditions for a null hypersurface to be characteristic.
We will find that, for a Killing horizon to become a boundary of causal domain, the scalar fields must satisfy some additional conditions similar to that found in Ref.~\cite{Minamitsuji:2015nca}.

\subsection{Null hypersurface}

We assume that a $x^0=\text{constant}$ surface $\Sigma$ is null, that is,
\begin{align}
g^{00}=0, \ \ \ 
g^{0\alpha}=0, \ \ \
g_{11}=0, \ \ \
g_{1\alpha}=0,
\label{nullcondition}
\end{align}
where $x^1$ is the null coordinate lying on $\Sigma$ and $x^{\alpha}$ ($\alpha=2,3$) are other spatial coordinates along $\Sigma$.
Under the conditions~(\ref{nullcondition}), the components of the principal symbol become
\begin{align}
\mathcal{A}^{11,11}&=\mathcal{A}^{11,1\alpha}=\mathcal{A}^{1\alpha,11}=0
\\
\mathcal{A}^{1\alpha,\beta\gamma}
&=
\mathcal{A}^{\beta\gamma,1\alpha}
=
g^{01} \left(
-2J_{IJ}g^{p(1}\delta^{\alpha)q0(\gamma}_{pr1f}g^{\delta)f}\phi^{I}_{|q}\phi^{J|r}
-2K_I g^{p(1}\delta^{\alpha)q0(\gamma}_{pr1f}g^{\delta)f}\phi^{I|r}_{|q}
\right)
\\
\mathcal{A}^{\alpha\beta,\gamma\delta}
&=
g^{01}\left(
-2J_{IJ}g^{p(\alpha}\delta^{\beta)q0(\gamma}_{pr1f}g^{\delta)f}\phi^{I}_{|q}\phi^{J|r}
-2K_I g^{p(\alpha}\delta^{\beta)q0(\gamma}_{pr1f}g^{\delta)f}\phi^{I|r}_{|q}
\right)
\\
\mathcal{A}^{11,\alpha\beta}
&=
\mathcal{A}^{\alpha\beta,11}
= -2 \mathcal{A}^{1\alpha,1\beta}
\notag \\ &=
g^{01}\left[
\left( \mathcal{F} + 2 \mathcal{W}\right) g^{01} g^{\alpha\beta}
-2J_{IJ} g^{p1}\delta^{1q0(\alpha}_{pr1f}g^{\beta)f}\phi^{I}_{|q}\phi^{J|r}
-2K_I g^{p1}\delta^{1q0(\alpha}_{pr1f}g^{\beta)f}\phi^{I|r}_{|q}
\right]
\end{align}
\begin{align}
\mathcal{B}^{11}_I
&=
g^{01}\biggl(
-\tilde{B}_I  g^{01}
+D_{JKI}g^{l1}\delta^{1c0}_{ld1} \phi^{J}_{|c}\phi^{K|d}
+E_{JKLMI}g^{l1}\delta^{1ce0}_{ldf1} \phi^{J}_{|c}\phi^{K|d}\phi^{L}_{|e}\phi^{M|f}
+2\mathcal{F}_{,IJ}g^{l1}\delta^{1c0}_{ld1} \phi^{J|d}_{|c}
\notag \\ & \qquad\quad
+4J_{JK,LI}g^{l1}\delta^{1ce0}_{ldf1} \phi^{J}_{|c}\phi^{K|d}\phi^{L|f}_{|e}
+K_Ig^{l1}\delta^{10ce}_{l1df} R_{ce}{}^{df}
+2K_{I,JK}g^{l1}\delta^{1ce0}_{ldf1} \phi^{J|d}_{|c}\phi^{K|f}_{|e}
\biggr)
\\
\mathcal{B}^{1\alpha}_I
&=
g^{01}\biggl(
D_{JKI}g^{l(1}\delta^{\alpha)c0}_{ld1} \phi^{J}_{|c}\phi^{K|d}
+E_{JKLMI}g^{l(1}\delta^{\alpha)ce0}_{ldf1} \phi^{J}_{|c}\phi^{K|d}\phi^{L}_{|e}\phi^{M|f}
+2\mathcal{F}_{,IJ}g^{l(1}\delta^{\alpha)c0}_{ld1} \phi^{J|d}_{|c}
\notag \\& \qquad\quad
+4J_{JK,LI}g^{l(1}\delta^{\alpha)ce0}_{ldf1} \phi^{J}_{|c}\phi^{K|d}\phi^{L|f}_{|e}
+K_Ig^{l(1}\delta^{\alpha)0ce}_{l1df} R_{ce}{}^{df}
+2K_{I,JK}g^{l(1}\delta^{\alpha)ce0}_{ldf1} \phi^{J|d}_{|c}\phi^{K|f}_{|e}
\biggr)
\\
 \mathcal{B}^{\alpha\beta}_I&=
-D_{JKI}g^{\alpha\beta}\phi^{J|0}\phi^{K|0}
+E_{JKLMI}g^{l(\alpha}\delta^{\beta)ce0}_{ldf1}g^{01}\phi^{J}_{|c}\phi^{K|d}\phi^{L}_{|e}\phi^{M|f}
-2\mathcal{F}_{,IJ}g^{\alpha\beta}\phi^{J|00}
\notag \\& \quad
+4J_{JK,LI}g^{l(\alpha}\delta^{\beta)ce0}_{ldf1}g^{01}\phi^{J}_{|c}\phi^{K|d}\phi^{L|f}_{|e}
+K_Ig^{l(\alpha}\delta^{\beta)0ce}_{l1df}g^{01}R_{ce}{}^{df}
+2K_{I,JK}g^{l(\alpha}\delta^{\beta)ce0}_{ldf1}g^{01}\phi^{J|d}_{|c}\phi^{K|f}_{|e}
\end{align}
\begin{align}
\mathcal{G}^{11}_I&= 
g^{01}\biggl[
- \left( D_{IJK} -8 J_{J[K,I]}-8E_{LKJIM}X^{LM}\right)
\delta^{c01}_{dl1} g^{1l}\phi^J_{|c}\phi^{K|d}
- 2D_{IJK}X^{JK}g^{01}
\notag \\ & \quad
-2E_{LJKIM}\delta^{ce01}_{dfh1}g^{1h}\phi^{L}_{|c}\phi^{J|d}\phi^{K}_{|e}\phi^{M|f}
-4J_{IL,JK}g^{l1}\delta^{10ce}_{1ldf} \phi^{J}_{|c}\phi^{L|d}\phi^{K|f}_{|e}
+4I_{,I}g^{01}
\notag \\ & \quad
-4\left( J_{IJ}-K_{(I,J)}+2J_{K(I,J)L}X^{KL}\right) \delta^{c01}_{bf1}g^{1f}\phi^{J|b}_{|c}
-2K_{J,IK}\delta^{ce01}_{dfh1}g^{1h}\phi^{J|d}_{|c}\phi^{K|f}_{|e}
- K_I\delta^{ce01}_{dfh1}g^{1h}R_{ce}{}^{df}
\biggr]
\\
\mathcal{G}^{1\alpha}_I&= 
g^{01}\biggl[
\left( D_{IJK} -8 J_{J[K,I]}-8E_{LKJIM}X^{LM}\right)\delta^{c0(1}_{d1l}g^{\alpha)l}\phi^{J}_{|c}\phi^{K|d}
+2E_{LJKIM}\delta^{ce0(1}_{df1h}g^{\alpha)h}\phi^{L}_{|c}\phi^{J|d}\phi^{K}_{|e}\phi^{M|f}
\notag \\&
\qquad\quad
+4J_{IL,JK}g^{l(1}\delta^{\alpha)ce0}_{ldf1}\phi^{J}_{|c}\phi^{L|d}\phi^{K|f}_{|e}
+4\left( J_{IJ}-K_{(I,J)}+2J_{K(I,J)L}X^{KL}\right) \delta^{c0(1}_{b1f}g^{\alpha)f}\phi^{J|b}_{|c}
\notag \\&
\qquad\quad
+2K_{J,IK}\delta^{ce0(1}_{df1h}g^{\alpha)h}\phi^{J|d}_{|c}\phi^{K|f}_{|e}
+ K_I\delta^{ce0(1}_{df1h}g^{\alpha)h}R_{ce}{}^{df}
\biggr]
\\
\mathcal{G}^{\alpha\beta}_I&=
-\left( D_{IJK} -8 J_{J[K,I]}-8E_{LKJIM}X^{LM}\right)
g^{\alpha\beta}\phi^{J|0}\phi^{K|0}
-4\left( J_{IJ}-K_{(I,J)}+2J_{K(I,J)L}X^{KL}\right) 
g^{\alpha\beta}\phi^{J|00}
\notag \\& \quad
+
g^{01}\biggl[
2E_{LJKIM}\delta^{ce0(\alpha}_{df1h}g^{\beta)h}\phi^{L}_{|c}\phi^{J|d}\phi^{K}_{|e}\phi^{M|f}
+4J_{IL,JK}g^{l(\alpha}\delta^{\beta)ce0}_{ldf1}\phi^{J}_{|c}\phi^{L|d}\phi^{K|f}_{|e}
\notag \\& \qquad\qquad
+2K_{J,IK}\delta^{ce0(\alpha}_{df1h}g^{\beta)h}\phi^{J|d}_{|c}\phi^{K|f}_{|e}
+ K_I\delta^{ce0(\alpha}_{df1h}g^{\beta)h}R_{ce}{}^{df}
\biggr]
\end{align}
\begin{align}
\mathcal{H}_{IJ}&=
2\left(
- B_{IK,LJ}+2D_{K[I|J,|L]}
\right)\phi^{K|0}\phi^{L|0} 
-2D_{IKJ,LM}\phi^{K|0}\phi^{L|d}\phi^{M|0}_{|d}
+ 4 \left(
D_{I(JK)} - G_{JK,I}
\right)\phi^{K|00}
\notag \\&\quad
+ g^{01}\biggl[
+2D_{IKL,MJ}\delta^{ce}_{1f}\phi^{K}_{|c}\phi^{L|f}_{|e}\phi^{M|0}
+2\left( E_{KLMNJ,I} -2E_{KLMIJ,N}\right)\delta^{ce0}_{df1}\phi^{K}_{|c}\phi^{L|d}\phi^{M}_{|e}\phi^{N|f}
\notag \\& \qquad\qquad
-4E_{KLMIJ,NO}\delta^{ce0}_{bf1}\phi^{K}_{|c}\phi^{L|f}\phi^{M}_{|e}\phi^{N|l}\phi^{O|b}_{|l}
-4E_{KLMIN,OJ}\delta^{ceg}_{1fh}\phi^{K}_{|c}\phi^{L|f}\phi^{M}_{|e}\phi^{N|h}_{|g}\phi^{O|0}
\notag \\& \qquad\qquad
+16E_{(K|LMI|J)}\delta^{ce0}_{bf1}\phi^{K|b}_{|c}\phi^{M}_{|e}\phi^{L|f}
+8\left( J_{KL,MJ,I}-J_{IK,MJ,L}\right) \delta^{ce0}_{df1}\phi^{K}_{|c}\phi^{L|d}\phi^{M|f}_{|e}
\notag \\& \qquad\qquad
-12J_{I(J,KL)}\delta^{ce0}_{bf1}\phi^{K|b}_{|c}\phi^{L|f}_{|e}
-4J_{KM,LN,IJ}\delta^{ceg0}_{dfh1}\phi^{K}_{|c}\phi^{M|d}\phi^{L|f}_{|e}\phi^{N|h}_{|g}
\notag \\& \qquad\qquad
-8J_{IK,LM,NJ}X^{KM}\delta^{eg0}_{fh1}\phi^{L|f}_{|e}\phi^{N|h}_{|g}
-2J_{KL,IJ}\delta^{ceg0}_{dfh1}\phi^{K}_{|c}\phi^{L|d}R_{eg}{}^{fh}
\notag \\& \qquad\qquad
-2\left( -K_{(I,J)}+J_{IJ}+2J_{K(I,J)L}X^{KL}\right) \delta^{0eg}_{1fh}R_{eg}{}^{fh}
-2K_{J,IK}\delta^{0egl}_{1fhm}\phi^{K|f}_{|e}R_{gl}{}^{hm}
\notag \\& \qquad\qquad
+4K_{J,KL,I}\delta^{ce0}_{df1}\phi^{K|d}_{|c}\phi^{L|f}_{|e}
-\frac43K_{L,JK,IM}\delta^{ceg0}_{dfh1}\phi^{M|d}_{|c}\phi^{L|f}_{|e}\phi^{K|h}_{|g}
\biggr].
\end{align}
From these equations, we find Eq.~(\ref{Pdotr=0}) has the following structure:
\begin{equation}
0 = P \cdot r 
= 
\left(
\begin{array}{@{}cccc@{}}
{\cal A}^{11,11}
&
2{\cal A}^{11,1\gamma}
&
{\cal A}^{11,\gamma \delta}
&
{\cal B}^{11}_J
\\
{\cal A}^{1\alpha,11}
&
2{\cal A}^{1\alpha,1\gamma}
&
{\cal A}^{1\alpha,\gamma \delta}
&
{\cal B}^{1\alpha}_J
\\
{\cal A}^{\alpha \beta,11}
&
2{\cal A}^{\alpha \beta,1\gamma}
&
{\cal A}^{\alpha \beta,\gamma \delta}
&
{\cal B}^{\alpha \beta}_J
\\
{\cal G}^{11}_I
&
2{\cal G}^{1\gamma}_I
&
{\cal G}^{\gamma \delta}_I
&
{\cal H}_{IJ}
\end{array}\right)
\left(
\begin{array}{@{}c@{}}
r_{11}
\\
r_{1\gamma}
\\
r_{\gamma \delta}
\\
r_J
\end{array}
\right)
=
\left(
\begin{array}{@{}cccc@{}}
0
&
0
&
{\cal A}^{11,\gamma \delta}
&
{\cal B}^{11}_J
\\
0
&
-{\cal A}^{11,\alpha\gamma}
&
{\cal A}^{1\alpha,\gamma \delta}
&
{\cal B}^{1\alpha}_J
\\
{\cal A}^{11,\alpha \beta}
&
2{\cal A}^{\alpha \beta,1\gamma}
&
{\cal A}^{\alpha \beta,\gamma \delta}
&
{\cal B}^{\alpha \beta}_J
\\
{\cal G}^{11}_I
&
2{\cal G}^{1\gamma}_I
&
{\cal G}^{\gamma \delta}_I
&
{\cal H}_{IJ}
\end{array}\right)
\left(
\begin{array}{@{}c@{}}
r_{11}
\\
r_{1\gamma}
\\
r_{\gamma \delta}
\\
r_J
\end{array}
\right).
\label{Pnull}
\end{equation}
Although three components ${\cal A}^{11,11}$, ${\cal A}^{1\alpha,11}$, ${\cal A}^{11,1\gamma}$ vanish identically, 
$P$
is invertible unless some of the remaining components happen to vanish, hence a null hypersurface is not characteristic in general.

\subsection{Killing Horizon with additional conditions}
\label{Sec:KH}

Next, we consider the case when $\Sigma$ is a Killing horizon, on which the metric satisfies~\cite{Izumi:2014loa}
\begin{align}
\partial_1 g_{\alpha\beta} =0, \ \ \
\partial_1^2 g_{\alpha\beta} =0, \ \ \
\partial_1\partial_{\gamma} g_{\alpha\beta}=0,
\label{KillingH}
\end{align}
which implies that the following components of the Riemann tensor vanish on the null surface $\Sigma$:
\begin{equation}
R_{1\alpha\beta\gamma} = R_{1\alpha1\beta} = 0.
\end{equation}
In Gauss-Bonnet and Lovelock theories, a Killing horizon becomes characteristic once these conditions are imposed~\cite{Izumi:2014loa,Reall:2014pwa}.
In Horndeski theory, however, it was noticed that these conditions are insufficient and some additional conditions on the scalar field must be imposed to make the Killing horizon characteristic~\cite{Minamitsuji:2015nca}.

What happens in the bi-Horndeski theory is similar to that for the Horndeski theory.
By explicit calculations, we can show that the conditions~(\ref{KillingH}) make some terms in $P$ with curvature tensors vanishing. 
Even when this occurs, however, no components in the principal symbol~(\ref{Pnull}) vanish completely and hence the Killing horizon $\Sigma$ will not be characteristic in general.

In Ref.~\cite{Minamitsuji:2015nca} it was noticed that a Killing horizon in the Horndeski theory become characteristic if the scalar field satisfies some additional conditions. 
We consider generalizations of these conditions\footnote{In some class of scalar-tensor theories, these conditions on scalar fields are automatically satisfied if spacetime is stationary~\cite{Smolic:2015txa,Graham:2014ina}.} given by
\begin{align}
\partial_1 \phi_I &=0, \ \ \ 
\partial_1^2 \phi_I =0, \ \ \ 
\partial_1\partial_{\alpha} \phi_I =0.
\label{phiconditions}
\end{align}
Under these conditions, 
the components of $P$ drastically simplify as
\begin{align}
\mathcal{A}^{11,11}&=\mathcal{A}^{11,1\alpha}=\mathcal{A}^{1\alpha,11}=\mathcal{A}^{1\alpha,\beta\gamma}=\mathcal{A}^{\beta\gamma,1\alpha}=\mathcal{A}^{\alpha\beta,\gamma\delta}=0
\\
\mathcal{A}^{11,\alpha\beta}&=\mathcal{A}^{\alpha\beta,11}
= -2 \mathcal{A}^{1\alpha,1\beta}
\notag \\&
=
(g^{01})^2\left[
\left(\mathcal{F}+2 \mathcal{W}\right)g^{\alpha\beta}
-2J_{IJ} \left( 2X^{IJ}g^{\alpha\beta}+\phi^{I}{}^{|\alpha}\phi^{J|\beta}\right)
+2K_I\left( \phi^{I|\sigma}_{|\sigma}g^{\alpha\beta}-\phi^{I|\alpha\beta}\right) 
\right]
\end{align}
\begin{align}
\mathcal{B}^{11}_I&=
(g^{01})^2
\biggl[
-\tilde{B}_I
+2D_{JKI}X^{JK}
-8E_{JKLMI}X^{J K}X^{M L}
-2\mathcal{F}_{,IJ}\phi^{J|\gamma}_{|\gamma}
\notag \\ & \qquad \qquad
-4J_{JK,LI}\delta^{\gamma\hat e}_{\delta\hat f}\phi^{J}_{|\gamma}\phi^{K|\delta}\phi^{L|\hat f}_{|\hat e}
-2K_{I,JK}\delta^{\gamma\hat e}_{\delta\hat f}\phi^{J|\delta}_{|\gamma}\phi^{K|\hat f}_{|\hat e}
- 2 K_I R_{\gamma\delta}{}^{\gamma\delta}
\biggr]
\\
\mathcal{B}^{1\alpha}_I&=\mathcal{B}^{\alpha\beta}_I =0
\\
\mathcal{G}^{11}_I&=
2(g^{01})^2
\biggl[
\left( D_{IJK} -8 J_{J[K,I]}-8E_{LKJIM}X^{LM}\right)X^{JK}
- D_{IJK}X^{JK}
- 8 E_{LJKIM} X^{JL} X^{KM}
\notag \\ & \qquad\qquad\quad
-2 J_{IL,JK}\delta^{\gamma\hat e}_{\delta\hat f} \phi^{J}_{|\gamma}\phi^{L|\delta}\phi^{K|\hat f}_{|\hat e}
+2I_{,I}
-2\left( J_{IJ}-K_{(I,J)}+2J_{K(I,J)L}X^{KL}\right) \phi^{J|\beta}_{|\beta}
\notag \\ & \qquad\qquad\quad
- K_{J,IK}\delta^{\gamma\hat e}_{\delta\hat f}\phi^{J|\delta}_{|\gamma}\phi^{K|\hat f}_{|\hat e}
- K_I R_{\gamma\delta}{}^{\gamma\delta}
\biggr]
\\
\mathcal{G}^{1\alpha}_I &= \mathcal{G}^{\alpha\beta}_I = 0
\\
\mathcal{H}_{IJ} &= 0.
\end{align}
Then, 
Eq.~(\ref{Pdotr=0}) reduces to
\begin{equation}
0=P\cdot r=
\left(
\begin{array}{@{}cccc@{}}
0
&
0
&
{\cal A}^{11,\gamma \delta}
&
{\cal B}^{11}_J
\\
0
&
-{\cal A}^{11,\alpha\gamma}
&
0
&
0
\\
{\cal A}^{11,\alpha \beta}
&
0
&
0
&
0
\\
{\cal G}^{11}_I
&
0
&
0
&
0
\end{array}\right)
\left(
\begin{array}{@{}c@{}}
r_{11}
\\
r_{1\gamma}
\\
r_{\gamma \delta}
\\
r_J
\end{array}
\right).
\end{equation}
The principal symbol $P$ is degenerate, and it gives $d$ conditions in total on an eigenvector $r$ with vanishing eigenvalue. Then the eigenvector will have $\frac12d(d-1) + 2 - d = \frac12d(d-3) + 2$ degrees of freedom, which coincides with the number of physical degrees of freedom of the bi-scalar-tensor theory. Therefore, the additional conditions~(\ref{phiconditions}) on the scalar fields guarantee the Killing horizon $\Sigma$ to be characteristic for all of the physical modes.

\section{Wave propagation on plane wave solution}
\label{Sec:planewave}

As another application of the technique shown in section~\ref{Sec:characteristics}, 
we analyze characteristic surfaces on the plane wave solution in the shift-symmetric Horndeski theory, which is an exact solution constructed in Ref.~\cite{Babichev:2012qs}.
For this purpose, we employ a covariant formalism provided in Ref.~\cite{Reall:2014pwa}, which is equivalent to a description of the equations of motion in the harmonic gauge and useful to find effective metrics that govern wave propagation.

Using this formalism, we will find that we can define effective metrics that differ from the physical metric on the plane wave solution, and characteristic surfaces are given by null hypersurfaces with respect to the effective metrics.
This result can be viewed as an generalization of that for Lovelock theories addressed in Ref.~\cite{Reall:2014pwa}.
Also, we will study shock formation phenomena on the plane wave solution later in section~\ref{Sec:Npp}, which heavily relies on contents of this section.

\subsection{Shift-symmetric Horndeski theory}
\label{Sec:EoM}

The shift-symmetric Horndeski theory is the most-general single scalar field theory whose arbitrary functions are invariant under constant shift in the scalar field.
Within this theory, Babichev constructed an exact solution describing plane wave of metric and scalar field propagating in a common null direction~\cite{Babichev:2012qs}. 
We will review the construction procedure of this solution later in section~\ref{Sec:BG} based on equations summarized in the current section.

The shift-symmetric Horndeski theory possesses four arbitrary functions $K(X)$, $G_{3,4,5}(X)$, where $X\equiv -\frac12 \phi_{|a}\phi^{|a}$, in its Lagrangian and equations of motion.
We use the following notation
\begin{equation}
K_{X} \equiv \frac{\partial K}{\partial X},
\qquad
G_{nX} \equiv \frac{\partial G_n}{\partial X},
\qquad
\square \phi = \phi_{|a}^{|a},
\end{equation}
where $n=3,4,5$. 

\subsubsection{Equations of motion and principal symbol }

The action and equations of motion of the shift-symmetric Horndeski theory are summarized in Ref.~\cite{Babichev:2012qs}. They can be reproduced from appendix B of Ref.~\cite{Kobayashi:2011nu} by making the arbitrary functions $K$, $G_n$ independent of $\phi$, and also from the equations of the bi-Horndeski theory given in appendix~\ref{App:BiHorndeskiEoM} if we specify the functions as shown in appendix~\ref{App:toHorndeski}.
Just to reproduce them here, the Lagrangian of the shift-symmetric Horndeski theory is given by ${\cal L} = \sum_{n=2}^5 {\cal L}_n$, where
\begin{equation}
\begin{aligned}
{\cal L}_2 &= K(X),
\quad
{\cal L}_3 = -G_{3}(X) \square \phi,
\quad
{\cal L}_4 = G_4(X) R + G_{4X}(X) 
\delta_{a_1 a_2}^{b_1 b_2} \phi^{|a_1}_{|b_1} \phi^{|a_2}_{|b_2},
\\
{\cal L}_5 &= G_5(X) G_{ab}\phi^{|ab} - \frac{1}{6}G_{5X}(X)
\delta_{a_1 a_2 a_3}^{b_1 b_2 b_3} \phi^{|a_1}_{|b_1} \phi^{|a_2}_{|b_2} \phi^{|a_3}_{|b_3}.
\end{aligned}
\end{equation}
Metric equation following from this Lagrangian is given by
\begin{equation}
\sum_{n=2}^5 {\cal G}^n_{ab}=0,
\label{geq}
\end{equation}
where
\begin{align}
{\cal G}^2{}_a^b &=
-\frac12 K_X \phi_{|a} \phi^{|b} - \frac12 K \delta_a^b
\\
{\cal G}^3{}^b_a &=
-\frac12G_{3X}
\left(
\delta_{a a_1 a_2}^{b b_1 b_2} \phi^{|a_1}_{|b_1} \phi^{|a_2} \phi_{|b_2}
+2X
\delta_{a a_1}^{b b_1} \phi^{|a_1}_{|b_1} 
\right)
\\
{\cal G}^4{}_a^b
 &=
 \frac12 \left(G_{4X} + 2X G_{4XX}\right)\delta_{a a_1 a_2}^{b b_1 b_2} \phi^{|a_1}_{|b_1} \phi^{|a_2}_{b_2}
 + \frac12 G_{4XX} \delta_{a a_1 a_2 a_3}^{b b_1 b_2 b_3} \phi^{|a_1}_{|b_1} \phi^{|a_2}_{|b_2} \phi^{|a_3}\phi_{|b_3}
 \notag \\ &\quad
 -\frac14 \left(G_4 - 2X G_{4X}\right) \delta_{a a_1 a_2}^{b b_1 b_2} R^{a_1 a_2}_{b_1 b_2}
 + \frac14 G_{4X} \delta_{a a_1 a_2 a_3}^{b b_1 b_2 b_3} R^{a_1 a_2}_{b_1 b_2}\phi^{|a_3}\phi_{|b_3}
 \\
{\cal G}^5{}_a^b
 &=
 -\frac1{24} \delta_{a a_1 a_2 a_3 a_4}^{b b_1 b_2 b_3 b_4}
 \left(
3 G_{5X} R^{a_1 a_2}_{b_1 b_2} + 2 G_{5XX} \phi^{|a_1}_{|b_1} \phi^{|a_2}_{|b_2}
 \right)
 \phi^{|a_3}_{|b_3} \phi^{|a_4}\phi_{|b_4}
 \notag \\ & \quad
 -\frac16\left(G_{5X}+X G_{5XX}\right)
\delta_{a a_1 a_2 a_3}^{b b_1 b_2 b_3} \phi^{|a_1}_{|b_1} \phi^{|a_2}_{|b_2} \phi^{|a_3}_{|b_3}
-\frac14 X G_{5X}
\delta_{a a_1 a_2 a_3}^{b b_1 b_2 b_3} R^{a_1 a_2}_{b_1 b_2} \phi^{|a_3}_{|b_3}
 \notag \\ &=
 -\frac16\left(G_{5X}+X G_{5XX}\right)
\delta_{a a_1 a_2 a_3}^{b b_1 b_2 b_3} \phi^{|a_1}_{|b_1} \phi^{|a_2}_{|b_2} \phi^{|a_3}_{|b_3}
-\frac14 X G_{5X}
\delta_{a a_1 a_2 a_3}^{b b_1 b_2 b_3} R^{a_1 a_2}_{b_1 b_2} \phi^{|a_3}_{|b_3}.
\end{align}
The above equations hold in any dimensions except for the second expression of ${\cal G}^5_{\mu\nu}$, 
which is simplified by eliminating the fifth-order generalized Kronecker delta ($\delta_{\mu a_1 a_2 a_3 a_4}^{\nu b_1 b_2 b_3 b_4}$) that identically vanishes in four dimensions.
The scalar equation of motion is given by
\begin{equation}
\sum_{n=2}^5 \nabla^a {\cal J}^n_a = 0,
\label{Jeq}
\end{equation}
where ${\cal J}^n_a$ is the current associated with the shift symmetry of the scalar field.
Their divergences are given by
\begin{align}
\nabla^a {\cal J}^2_a
&=
-K_X \square \phi + K_{XX} \phi_{|a_1}\phi^{|a_1}_{|a_2}\phi^{|a_2}
\\
\nabla^a {\cal J}^3_a
 &=
 \left( G_{3X} + X G_{3XX}  \right) \delta_{a_1 a_2}^{b_1 b_2}\phi^{|a_1}_{|b_1}\phi^{|a_2}_{|b_2}
+ \frac12 G_{3XX} \delta_{a_1 a_2 a_3}^{b_1 b_2 b_3}\phi^{|a_1}_{|b_1}\phi^{|a_2}_{|b_2}\phi^{|a_3}\phi_{|b_3}
- G_{3X} R^{a_1}_{b_1}\phi_{|a_1}\phi^{b_1}
\\
 \nabla^a {\cal J}^4_a
 &=
 -\frac16\left( 3 G_{4XX} + 4X G_{4XXX}  \right)
 \delta_{a_1 a_2 a_3}^{b_1 b_2 b_3}\phi^{|a_1}_{|b_1}\phi^{|a_2}_{|b_2}\phi^{|a_3}_{|b_3}
 -\frac13 G_{4XXX} \delta_{a_1 a_2 a_3 a_4}^{b_1 b_2 b_3 b_4}\phi^{|a_1}_{|b_1}\phi^{|a_2}_{|b_2}\phi^{|a_3}_{|b_3} \phi^{|a_4} \phi_{|b_4}
 \notag \\ & \quad
 -\frac12 \left(G_{4X} + 2X G_{4XX}\right) \delta_{a_1 a_2 a_3}^{b_1 b_2 b_3}\phi^{|a_1}_{|b_1}R^{a_2 a_3}_{b_2 b_3}
  -\frac12 G_{4XX} \delta_{a_1 a_2 a_3 a_4}^{b_1 b_2 b_3 b_4}\phi^{|a_1}_{|b_1}R^{a_2 a_3}_{b_2 b_3} \phi^{|a_4} \phi_{|b_4}
\\
\nabla^a {\cal J}_a^5
 &=
\frac1{12}\left(2 G_{5XX} + X G_{5XXX}\right)
\delta_{a_1 a_2 a_3 a_4}^{b_1 b_2 b_3 b_4}
\phi^{|a_1}_{|b_1} \phi^{|a_2}_{|b_2}
\phi^{|a_3}_{|b_3} \phi^{|a_4}_{|b_4}
 \notag \\ & \quad
+ \frac14 \left( G_{5X} + X G_{5XX} \right)
\delta_{a_1 a_2 a_3 a_4}^{b_1 b_2 b_3 b_4}
R^{a_1 a_2}_{b_1 b_2} \phi^{|a_3}_{|b_3} \phi^{|a_4}_{|b_4}
+ \frac1{16}X G_{5X} \delta_{a_1 a_2 a_3 a_4}^{b_1 b_2 b_3 b_4}
R^{a_1 a_2}_{b_1 b_2} R^{a_3 a_4}_{b_3 b_4}.
\label{divJ5}
\end{align}
We used Schouten identity (obtained by expanding
$\delta_{a_1 a_2 a_3 a_4 a_5}^{b_1 b_2 b_3 b_4 b_5}R^{a_1 a_2}_{b_1 b_2}
\phi^{|a_3}_{|b_3} \phi^{|a_4}_{|b_4} \phi^{|a_5} \phi_{|b_5}=0$)
that holds only in four dimensions to derive Eq.~(\ref{divJ5}).
To simplify the analysis, we introduce also the trace-reversed equations of motion $\tilde {\cal G}_{ab}$ by
\begin{equation}
\tilde {\cal G}^n_{ab} 
\equiv 
{\cal G}^n_{ab} - \frac12 {\cal G}^{n~c}_{~c} \,g_{ab}.
\label{tracereverse}
\end{equation}
The principal symbol of the equations of motion, which is the set of the metric equation~(\ref{tracereverse}) and the scalar field equation~(\ref{Jeq}),
is constructed by taking derivatives of these equations
with respect to partial derivatives of the dynamical variables $g_{qr,st}$ and $\phi_{,st}$.
We summarize the explicit expressions of these derivatives in appendix~\ref{App:ssHorndeski}.
Using them, the principal symbol $\tilde P$ based on the trace-reversed metric equation~(\ref{tracereverse}) and the scalar equation is then constructed as%
\begin{equation}
\tilde P(x,\xi)\cdot r
=
\left(
\begin{array}{@{}c@{}}
\bigl(\tilde P(x,\xi)\cdot r \bigr)_{ab}
\\ 
\bigl(\tilde P(x,\xi)\cdot r \bigr)_\phi
\end{array}
\right)
= 
\sum_{n=2}^5
\xi_s \xi_t
\left(
\begin{array}{@{}cc@{}}
\frac{\partial\tilde {\cal G}^n_{ab}}{\partial g_{qr,st}}
&
\frac{\partial\tilde {\cal G}^n_{ab}}{\partial \phi_{,st}}
\\ 
\frac{\partial \nabla^c{\cal J}_c^n}{\partial g_{qr,st}}
&
\frac{\partial \nabla^c{\cal J}_c^n}{\partial \phi_{,st}}
\end{array}
\right)
\left(
\begin{array}{@{}c@{}}
r_{qr}
\\ 
r_\phi
\end{array}
\right),
\label{ssHorndeskiP}
\end{equation}
where $r_{qr}$ and $r_\phi$ are a symmetric tensor and a scalar corresponding to waves of $g_{qr}$ and $\phi$ propagating on a characteristic surface $\Sigma$, respectively.
The principal symbol $\tilde P$ has a symmetry%
\footnote{For the original principal symbol that is not trace-reversed, this symmetry is simply given by
\begin{equation}
r^a_b (P\cdot r' )^b_a = (P\cdot r )^a_b r'{}^b_a.
\label{IPsymmetry2}
\end{equation}}
\begin{equation}
\bigl(\tilde P\cdot r, r'\bigr) = \bigl(r, \tilde P\cdot r'\bigr),
\label{IPsymmetry}
\end{equation}
where $r_{ab}$ and $r'_{ab}$ are symmetric tensors
and the inner product is defined by
\begin{equation}
\left(r,r'\right) \equiv 
\left(
g^{a(c}g^{d)b} - \frac12 g^{ab}g^{cd}
\right)
r_{ab}r'_{cd} + r_\phi r'_\phi
=
r^{ab}
r'_{ab} - \frac12 r^a_{~a} r'^b_{~~b} + r_\phi r'_\phi.
\label{IP}
\end{equation}
This symmetry of the inner product follows from the fact that the equations of motion are derived from a Lagrangian by the variational principle, and equivalent to the symmetry discussed in Eq.~(\ref{Psymmetry}).

\subsubsection{Gauge symmetry and transverse condition}
\label{Sec:Gauge}

We can check that the components of the principal symbol $\tilde P$ satisfy, for any vector $X^a$,
\begin{equation}
\frac{\partial {\cal G}^{n}_{ab}}{\partial g_{qr,st}} \xi_s \xi_t \xi_{(q}X_{r)}
=
\frac{\partial \nabla^a {\cal J}^{n}_a}{\partial g_{qr,st}} \xi_s \xi_t \xi_{(q}X_{r)}
=0,
\label{annihilate}
\end{equation}
which implies that a vector given by $r = (r_{ab},r_\phi) = (\xi_{(a}X_{b)},0)$ is annihilated by $\tilde P$ for arbitrary vector $X^a$, and hence
$\tilde P\cdot r$ is invariant under gauge transformation
\begin{equation}
r_{ab} \to r_{ab} + \xi_{(a}X_{b)}.
\end{equation}
This property follows from the diffeomorphism invariance of the theory.
We define the vector space of the equivalence classes with respect to this invariance as $V_\text{physical}$, following Ref.~\cite{Reall:2014pwa}.

We can check also that the gravitational part of the principal symbol satisfies 
\begin{equation}
\xi^a \biggl(
\frac{\partial \tilde{\cal G}^n_{ab}}{\partial g_{qr,st}} 
-\frac12 \frac{\partial \tilde {\cal G}^{nc}{}_{c}}{\partial g_{qr,st}} g_{ab}
\biggr)
\xi_s \xi_t
=
\xi^a
\biggl(
\frac{\partial \tilde{\cal G}^{n}_{ab}}{\partial \phi_{,st}}
-\frac12 
\frac{\partial \tilde{\cal G}^{nc}{}_{c}}{\partial \phi_{,st}} g_{ab}
\biggr)
\xi_s \xi_t
=0.
\label{divfree}
\end{equation}
This property originates from the generalized Bianchi identity in this theory
\begin{equation}
\nabla^a {\cal G}^n_{ab} = \frac12  \phi_{|b} \nabla^a {\cal J}^n_a,
\end{equation}
and the fact that the left-hand side of this equation cannot not have third derivatives since $\nabla^a{\cal J}_a^n$ is given by derivatives up to second order.
We say that a symmetric tensor $r_{ab}$ is transverse if it obeys
\begin{equation}
\xi^a r_{ab} - \frac12 \xi_b r^a_{~a} = 0,
\label{transverse}
\end{equation}
and call the vector space made of transverse symmetric tensors $V_\text{transverse}$.
Identity (\ref{divfree}) implies that the gravitational part of the principal symbol is transverse, that is,
\begin{equation}
\xi^a \bigl( \tilde P\cdot r\bigr){}_{ab} - \frac12 \xi_b \bigl( \tilde P\cdot r\bigr){}^a_{~a}=0.
\label{transverseP}
\end{equation}
Hence, $\tilde P$ can be regarded as a map from $V_\text{physical}$ into $V_\text{transverse}$.
Note that these two vector spaces share the same number of dimensions: seven in total, six and one from the metric and scalar sectors, respectively.

We hereby assume $G_4- 2XG_{4X}$ and $K_X$ are nonzero. 
Then, for convenience, we separate and normalize the equation $\tilde P \cdot r = 0$ as
\begin{equation}
0 =
\begin{pmatrix}
\frac{2}{G_4- 2XG_{4X}} & 0\\ 0 &\frac{1}{K_X}
\end{pmatrix}
\begin{pmatrix}
\bigl( \tilde P \cdot r\bigr)_{ab}\\
\bigl( \tilde P \cdot r\bigr)_\phi
\end{pmatrix}
=
\begin{pmatrix}
\bigl( \tilde P_0 \cdot r\bigr)_{ab}\\
\bigl( \tilde P_0 \cdot r\bigr)_\phi
\end{pmatrix}
+
\begin{pmatrix}
\bigl( \tilde {\cal R} \cdot r\bigr)_{ab}\\
\bigl( \tilde {\cal R} \cdot r\bigr)_\phi
\end{pmatrix},
\label{P0def}
\end{equation}
where $\bigl( \tilde P_0 \cdot r\bigr)_{ab}$ and $\bigl( \tilde P_0 \cdot r\bigr)_\phi$ are defined by
\begin{align}
\bigl(\tilde P_0\cdot r\bigr)_{ab} 
&\equiv
g_{ab'}
 \left(
\delta^{b'b_1b_2}_{aa_1 a_2}
-\delta^{b'}_a\delta^{b_1 b_2}_{a_1 a_2}
\right)\xi^{a_1} \xi_{b_1} r^{a_2}_{b_2} 
=
-
\left(
 \xi^2 r_{ab} - 2\xi^c r_{c(a}\xi_{b)} + r^c_c \xi_a \xi_b
\right)
\label{P0def2}
\\
\bigl(\tilde P_0\cdot r\bigr)_\phi
&=
- \xi^2 r_\phi.
\label{P0def3}
\end{align}
These terms are the first terms of Eqs.~(\ref{dtildeGdGpp}) and (\ref{ddivJdphipp}), respectively, with the coefficients removed by the normalization introduced in Eq.~(\ref{P0def}).
$\tilde P_0$ corresponds to the principal symbol of GR 
and a minimally-coupled scalar field described by $K(X)$. Both $\tilde P_0$ and $\tilde {\cal R}$ satisfy the symmetry with respect to the inner product~(\ref{IPsymmetry}) and the transverse condition~(\ref{transverseP}).
Also, both $\tilde P_0\cdot t$ and $\tilde {\cal R}\cdot t$ vanish for the gauge mode $r=(\xi_{(a}X_{b)},0)$.

$\tilde P$ may be regarded as a matrix acting on the seven physical components of $r$. Then, characteristic surfaces can be found by solving the characteristic equation $\det\tilde P(x,\xi)=0$ with respect to $\xi$, or equivalently by finding eigenvectors of $\tilde P$ with vanishing eigenvalues. Below, we will take the the latter approach to find characteristic surfaces and eigenvectors associated with them.

We first study a non-null characteristic surface.
For this purpose, 
rather than solving Eq.~(\ref{P0def}) directly, it is useful to solve an eigenvalue equation for $\tilde {\cal R}$ defined by Eq.~(\ref{P0def}) as a first step:
\begin{equation}
\tilde {\cal R}\cdot r = \lambda \, r.
\label{EVeq}
\end{equation}
Equation (\ref{annihilate}) implies that $r=(\xi_{(a}X_{b)},0)$ for any vector $X^a$ is an eigenvector with $\lambda=0$.
These eigenvectors correspond to gauge modes.
Since $\xi$ is not null, 
based on Eq.~(\ref{transverse})
we can uniquely decompose $r_{ab}$ into the physical and gauge parts as
\begin{equation}
r_{ab} = \tilde r_{ab} + \xi_{(a}X_{b)},
\end{equation}
so that $\tilde r_{ab}$ satisfies the transverse condition~(\ref{transverse}).
Then, the characteristic equation (\ref{P0def}) becomes
\begin{equation}
\tilde {\cal R}\cdot r 
= \lambda \, r 
= -\tilde P_0\cdot r = 
\xi^2
\, r
,
\label{lambda-xi2}
\end{equation}
in which terms in (\ref{P0def2}) other than the $\xi^2$ terms vanish due to the transverse condition.
Equation~(\ref{lambda-xi2}) has a nonzero eigenvector $r$ if $\xi$ is chosen to satisfy $\xi^2 = \lambda$.%
\footnote{Since $\lambda$ is a function of $\xi$, in general it is not trivial how $\xi$ can be chosen to satisfy $\xi^2=\lambda$. In the case of the plane wave background shown in section~\ref{Sec:effmetric}, it turns out that $\lambda \propto \left(\ell^a\xi_a\right)^2$ for a vector $\ell^a$, hence it is straightforward to find $\xi_a$ that satisfies $\xi^2 = \lambda$. When $\lambda$ is given by a more complicated function of $\xi$, a similar construction would not work and even the existence of solutions is not guaranteed.}
If such a $\xi$ exists, the surface normal to $\xi$ is a characteristic surface that is timelike (spacelike) if $\lambda > 0$ ($\lambda<0$). Physically, it corresponds to subluminal (superluminal) propagation of wave whose profile is proportional to $r$.

Next, let us study a null characteristic surface.
In this case it is useful to introduce a null basis $\{e_0,e_1,e_{i=2,3}\}$, where $e_0^a = \xi^a$ and $e_1$ are null vectors satisfying $e_0\cdot e_1 = 1$ and $e_i$ are spacelike orthonormal vectors ($e_i\cdot e_j = \delta_{ij}$) orthogonal to $e_{0,1}$. The transverse condition~(\ref{transverse}) constrains the eigenvector as $r_{00} = r_{ii} = r_{0i} = 0$, and the same components of $\tilde{\cal R}\cdot r$ vanish identically.
Then, the nontrivial components of the equation $\tilde P \cdot r = 0$, or equivalently Eq.~(\ref{P0def}), 
are given by
\begin{subequations}
\label{nulleq}
\begin{align}
r_{00} + \bigl(\tilde{\cal R}\cdot r\bigr)_{01} &= 0
\label{nulleq01}
\\
\bigl(\tilde {\cal R}\cdot r\bigr)_{ij} &= 0
\label{nulleqij}
\\
r_{0i} + \bigl(\tilde {\cal R}\cdot r\bigr)_{1i} &= 0
\label{nulleq1i}
\\
- r_{ii} + \bigl(\tilde {\cal R}\cdot r \bigr)_{11} &= 0
\label{nulleq11}
\\
\bigl(\tilde {\cal R}\cdot r\bigr)_\phi &= 0,
\label{nulleqphi}
\end{align}
\end{subequations}
where only the traceless part of Eq.~(\ref{nulleqij}) is nontrivial since its trace part $\bigl(\tilde {\cal R}\cdot r\bigr)_{ii} = 0$ is satisfied identically.
It can be shown that $r_{1a}$ do not appear in the above equations, and they correspond to gauge modes.
Then, we have seven physical independent variables, and (\ref{nulleq}) comprise seven equations for them.
Generically there are no non-vanishing solutions since the number of the unknown variables are the same as that of the equations. If solutions do not exist, there are no null characteristic surfaces.
In some special cases, the above equation happen to have nontrivial solutions and there are null characteristic surfaces correspondingly. We will see such examples below.

\subsection{Characteristics on the plane wave background}
\label{Sec:WaveOnWave}

Using the formalism in the previous section,
we analyze wave propagation and causal structure on the plane wave solution constructed by Ref.~\cite{Babichev:2012qs}.
After briefly describing the background solution in section~\ref{Sec:BG}, we solve the characteristic equation on this background to find the structure of causal cones on this background in section~\ref{Sec:PonBG}. 

\subsubsection{Plane wave solution}
\label{Sec:BG}

The first step to construct the plane wave solution in the shift-symmetric Horndeski theory is to introduce the pp-wave ansatz given by%
\footnote{This ansatz differs from the one used in Ref.~\cite{Babichev:2012qs} by a sign flip in the $dudv$ term, which corresponds to changing the direction of $u$ or $v$ coordinate.}
\begin{equation}
ds^2 = F(u,x,y) du^2 + 2dudv + dx^2 + dy^2,
\qquad
\phi = \phi(u).
\label{ansatz}
\end{equation}
To describe this solution, we introduce a null basis $\ell^a$, $n^a$ and $m_i{}^a$ with $i=x,y$ satisfying
\begin{equation}
\ell^a n_a = 1, \quad m_{ia} m_{j}{}^{a} = \delta_{ij}, 
\quad 
\ell^2 = n^2 = 
\ell^a m_{ia} = n^a m_{ia} = 0.
\label{nullbasis}
\end{equation}
We set $\ell$ to the null direction of the background spacetime~(\ref{ansatz}), that is, $\ell_a = (du)_a$.
Then we find that 
curvature tensor components in this spacetime vanish except for
\begin{equation}
R_{uiuj} = - \frac12 F_{,ij}~,
\qquad
R_{uu} 
= - \frac12 \Delta F,
\end{equation}
where
$F_{,ij}\equiv \partial^2 F/\partial x^i \partial x^j$ and $\Delta F \equiv F_{,ii}$.
Derivatives of $\phi$ and the curvature tensor are then expressed covariantly as
\begin{equation}
\phi_{|a} = \phi' \ell_a,
\quad
\phi_{|ab} = \phi'' \ell_a \ell_b,
\quad
R_{a_1 a_2}^{b_1 b_2}=-2 \ell_{[a_1}\ell^{[b_1}F_{a_2]}^{b_2]},
\quad
R_{ab} = - \frac12 \Delta F\, \ell_a\ell_b,
\label{derivatives}
\end{equation}
where $\phi' \equiv d\phi/du$.
Because $\phi^{|a}$ vanishes when contracted with $\phi_{|a}a$, $\phi_{|ab}$ or $R_{abcd}$, and also because $\square \phi = R= 0$, the equations of motion are simplified drastically under the ansatz~(\ref{ansatz}).
Also, Eq.~(\ref{derivatives}) implies that
\begin{equation}
X = -\frac12 \phi^{|a}\phi_{|a} = 0,
\end{equation}
hence only the values of arbitrary functions $K(X)$, $G_n(X)$ and their derivatives evaluated at $X=0$ appear in equations.

Plugging the ansatz~(\ref{ansatz}) into the equations of motion,
it turns out that the scalar field equation~(\ref{Jeq})
is satisfied for arbitrary $\phi(u)$, and the metric equation~(\ref{geq}) reduces to
\begin{equation}
-\frac12 K_X(0) \phi_{|a} \phi_{|b} + G_4(0)R_{ab} - \frac12 K(0)g_{ab} = 0.
\end{equation}
This equation can be satisfied only when $K(0)=0$, i.e., 
when the cosmological constant vanishes, 
and
\begin{equation}
\Delta F = - \frac{K_X(0)}{G_4(0)} \phi'^2 \equiv -\kappa \, \phi'^2
.
\label{Feq}
\end{equation}
$\kappa$ defined by this equation becomes positive if we assume $K(0), G_4(0)>0$, which are no-ghost condition for a canonical scalar coupled to GR.
The general solution of Eq.~(\ref{Feq}) is
\begin{equation}
F = - \frac{1}{4}\kappa\phi'^2 \left(x^2+y^2\right)   + F_h \, ,
\label{Ffunc}
\end{equation}
where $F_h$ is the general homogeneous solution satisfying $\Delta F_h = 0$.
A simple example of the homogeneous regular solution is given by
\begin{equation}
F_h = a_{ij}(u) x^i x^j,
\label{amatrix}
\end{equation}
where $a_{ij}(u)$ is a symmetric traceless matrix.

\subsubsection{Principal symbol on the plane wave solution}
\label{Sec:PonBG}

On the background of the plane wave solution described above,
the components of the principal symbol~(\ref{ssHorndeskiP}) are given by
\begin{align}
r_{qr}\sum_{n=2}^5 \frac{\partial \tilde {\cal G}^n_{ab}}{\partial g_{qr,st}}\xi_s \xi_t 
&=
\frac12 G_4\left(- \xi^2 r_{ab}+ 2\xi^c r_{c(a}\xi_{b)} - \xi_a \xi_b r^c_{~c}\right)
\notag \\ & ~~
+ \frac12 G_{4X}\phi'^2\biggl\{
2(\xi\cdot \ell)\ell^c r_{c(a}\xi_{b)}- (\xi\cdot \ell)^2 r_{ab}-\ell^c \ell^d r_{cd } \xi_a \xi_b-\left(\xi^c \xi^d r_{cd} - \xi^2 r^c_{~c}\right)\ell_a \ell_b
\notag \\ & \qquad\qquad\quad~~
-\frac12\left(2 (\xi\cdot \ell)\ell^c\xi^d r_{cd}-(\xi\cdot\ell)^2 r^c_{~c}-\xi^2 \ell^c\ell^d r_{cd}\right)g_{ab}
\notag \\ & \qquad\qquad\quad~~
+ 2\left(\ell^c\xi^d r_{cd} \xi_{(a}\ell_{b)}+ (\xi\cdot\ell)\xi^c r_{c(a}\ell_{b)}- \xi^2 \ell^c r_{c(a}\ell_{b)} - (\xi\cdot\ell) r^c_{~c} \xi_{(a}\ell_{b)} \right)
\biggr\}
\\
\sum_{n=2}^5 \frac{\partial\tilde{\cal G}^n_{ab}}{\partial\phi_{st}}\xi_s \xi_t
&=
\Bigl( 
-\frac12 G_{3X}\phi'^2 + G_{4X}\phi'' 
\Bigr)\left(  2(\xi\cdot \ell) \xi_{(a}\ell_{b)}- \xi^2  \ell_a \ell_b
\right)
\\
r_{qr}\sum_{n=2}^5 \frac{\partial \nabla^a{\cal J}^n_a}{\partial g_{qr,st}}
\xi_s \xi_t 
&=
\Bigl( -\frac12 G_{3X} \phi'^2 + G_{4X} \phi'' \Bigr)\left(2(\xi\cdot \ell)\xi^c \ell^d r_{cd}- \xi^2 \ell^c \ell^d r_{cd}-(\xi\cdot \ell)^2 r^c_{~c}\right)
\\
\sum_{n=2}^5 \frac{\partial \nabla^a{\cal J}^n_a}{\partial\phi_{st}}\xi_s \xi_t
&= 
- K_X \xi^2 + \left(K_{XX}\phi'^2 - 2G_{3X}\phi'' - G_{4X} \Delta F\right)\left(\xi\cdot \ell\right)^2,
\end{align}
where $\xi\cdot \ell \equiv \xi^a \ell_a$.

\subsubsection{Characteristic surfaces on the plane wave solution}
\label{Sec:effmetric}

In this section, 
we construct characteristic surfaces on the plane wave solution background
based the formalism of section~\ref{Sec:EoM}.
It is accomplished by finding eigenvalues $\lambda$ of the eigenvalue equation~(\ref{EVeq}) and then solving $\lambda = \xi^2$,
which is equivalent to Eq.~(\ref{lambda-xi2}), to fix $\xi$.

We start from solving the eigenvalue equation~(\ref{EVeq}) by finding eigenvectors satisfying this equation.
As we observed in section~\ref{Sec:Gauge}, 
$r=(\xi_{(a}X_{b)},0)$ for any vector $X^a$ is an eigenvector with $\lambda=0$.
Adding to that, $r=(\ell_{(a}X_{b)},0)$ gives $\lambda=0$ for any $X^a$, hence we have found seven eigenvectors with vanishing eigenvalues in total, since the vector $r=(\ell_{(a}\xi_{b)},0)$ is contained in the both kinds of the eigenvectors.
In the null basis~(\ref{nullbasis}), we may freely choose $n_a$ keeping the other null vector $\ell_a=(du)_a$ 
invariant.
Using this arbitrariness, in our analysis
we choose $n^a$ as the null vector made of a linear combination of $\ell^a$ and $\xi^a$, that is, $n^a = (\xi\cdot \ell)^{-2}(-\frac12 \xi^2 \ell^a + (\xi\cdot \ell)\xi^a)$.
Then the eigenvectors with vanishing eigenvalues can be taken as $r=\bigl(\ell_{(a}X_{b)},0\bigr)$ and $\bigl(n_{(a}X_{b)},0\bigr)$.

Eigenvectors with non-vanishing eigenvalues must be orthogonal to these eigenvectors with respect to the inner product~(\ref{IP}), from which we find that the eigenvector takes the form
\begin{equation}
r = \left(
2r_{\ell n}\ell_{(a}n_{b)} + r_{ij}m_{ia}m_{jb}, r_\phi
\right),
\end{equation}
where $r_{ij}$ is a symmetric traceless tensor.
The $r_{\ell n}$ part belongs to the Kernel of $\tilde{\cal R}$, then the $\ell n$ component of the eigenvalue equation becomes an equation to fix $r_{\ell n}$ in terms of the other components as
\begin{equation}
\frac{2}{G_4}
\left( -\frac12 G_{3X} \phi'^2 + G_{4X} \phi'' \right)(\xi\cdot\ell)^2 r_\phi = \lambda \, r_{\ell n}.
\end{equation}
Other than this one, only the $ij$ and $\phi$ components remain nontrivial and given by%
\footnote{This property is due to the fact that $\tilde {\cal R}\cdot r$ is orthogonal to $r=\bigl(\ell_{(a}X_{b)},0\bigr)$ and $\bigl(n_{(a}X_{b)},0\bigr)$, that is, 
}
\begin{align}
\left(
\begin{array}{@{}c@{}}
\bigl(\tilde {\cal R}\cdot r\bigr)_{ij}
\\
\bigl(\tilde{\cal R}\cdot r\bigr)_\phi
\end{array}
\right)
=
\left(
\begin{array}{@{}c@{}}
- \frac{G_{4X}}{G_4}
\phi'^2 (\xi\cdot\ell)^2 r_{ij}
\\
\frac{1}{K_X}
\left(
K_{XX}\phi'^2 - 2 G_{3X}\phi'' - G_{4X} \Delta F
\right)(\xi\cdot\ell)^2 r_\phi
\end{array}
\right)
=
\lambda
\left(
\begin{array}{@{}c@{}}
r_{ij}
\\
r_\phi
\end{array}
\right).
\end{align}
This eigenvalue equation is solved by
\begin{equation}
 (r_{\ell n},r_{ij},r_\phi)=(0, r_{ij},0),
\qquad 
\lambda = - \frac{G_{4X}}{G_4} \phi'^2 (\xi\cdot\ell)^2,
\label{EVgrav}
\end{equation}
and also by
\begin{equation}
 (r_{\ell n},r_{ij},r_\phi)=(\tilde r_{\ell n},0,r_\phi),
\qquad 
\lambda = 
\frac{1}{K_X}
\left(
K_{XX}\phi'^2 - 2 G_{3X}\phi'' - G_{4X} \Delta F
\right)(\xi\cdot\ell)^2,
\label{EVscalar}
\end{equation}
where
\begin{equation}
\tilde r_{\ell n} = 
\frac{2 K_X}{G_4} \, 
\frac{ -\frac12 G_{3X} \phi'^2 + G_{4X} \phi''}
{K_{XX}\phi'^2 - 2 G_{3X}\phi'' - G_{4X} \Delta F} 
\, r_\phi.
\end{equation}
The first solution~(\ref{EVgrav}) has two modes because $r_{ij}$ is a traceless symmetric tensor.
We call the first solution~(\ref{EVgrav}) and the second one~(\ref{EVscalar}) the tensor and scalar modes, respectively.

We have found ten eigenvectors up to this point, and
in principle there may be one more eigenvector since $\tilde {\cal R}$ can be regarded as a matrix for eleven variables which consist of ten metric components and one scalar field.
To search for the last eigenvector, 
it is useful to note that a general eigenvector may be expanded as
\begin{equation}
r = \left( r_{ab},r_\phi \right)
=\left(
2\ell_{(a}X_{b)} + 2n_{(a}Y_{b)}
+ 
(\hat r_{ij} + \alpha\, \delta_{ij})
m_{ia}m_{jb}, r_\phi
\right),
\label{finalEV}
\end{equation}
where $X^a$ and $Y^a$ are arbitrary vectors and $\hat r_{ij}$ is a symmetric traceless tensor.
The nonzero components of the eigenvalue equation for this vector turn out to be
\begin{align}
\bigl(\tilde {\cal R}\cdot r\bigr)_{\ell n}
&=
- 
\frac{(\xi\cdot\ell)^2}{G_4}
\left[
G_{4X} \phi'^2  \alpha 
 +
\left(
G_{3X} \phi'^2 -2 G_{4X} \phi''
\right)r_\phi 
\right] 
= \lambda \left(
\ell\cdot X + n \cdot Y
\right)
\label{Rln}
\\
\bigl(\tilde {\cal R}\cdot r\bigr)_{ij}
&=
-
(\xi\cdot\ell)^2 
\frac{G_{4X}}{G_4}\phi'^2 \hat r_{ij} 
= 
\lambda \left( \hat r_{ij} + \alpha \, \delta_{ij} \right)
\label{Rij}
\\
\bigl(\tilde {\cal R}\cdot r\bigr)_\phi
&=
\frac{(\xi\cdot\ell)^2 }{K_X}
\left[
\left(G_{3X} \phi'^2-2 G_{4X} \phi''\right)
\alpha
+ \left(
K_{XX}\phi'^2 - 2 G_{3X}\phi'' - G_{4X} \Delta F
\right)r_\phi 
\right]
=\lambda \,  r_\phi.
\label{Rphi}
\end{align}
If $\lambda \neq 0$, Eq.~(\ref{Rij}) forces $\alpha = 0$ and the vector~(\ref{finalEV}) becomes a linear combination of the eigenvectors found in the previous step and is not a new one.
Then setting $\lambda = 0$, we find that Eq.~(\ref{Rij}) implies $\hat r_{ij}=0$,
and Eqs.~(\ref{Rln}) and (\ref{Rphi}) give only a trivial solution $\alpha = r_\phi=0$ unless these equations happen to be degenerate with each other. 
Hence, generically the ten vectors found in the previous step exhaust all of nontrivial eigenvectors, and the characteristic equation forces the last eigenvector to be a trivial one.
A special case where Eqs.~(\ref{Rln}) and (\ref{Rphi}) become degenerate with each other is when the characteristic surface is parallel to the null direction $\ell$ and then $\xi\cdot\ell = 0$. We will examine this case later.

From the results above and Eq.~(\ref{lambda-xi2}),
$\xi$ for the tensor and scalar modes satisfy
\begin{align}
\text{tensor}:&\quad
- \frac{G_{4X}}{G_4} \phi'^2 (\xi\cdot\ell)^2
=  \xi^2
\\
\text{scalar}:&\quad
\frac{1}{K_X}
\left(
K_{XX}\phi'^2 - 2 G_{3X}\phi'' - G_{4X} \Delta F
\right)(\xi\cdot\ell)^2 = \xi^2.
\end{align}
Then, we may define effective inverse metrics $G^{ab}_\omega$ 
for which $\xi$ of these modes are null as
\begin{equation}
0=G^{ab}_\omega \xi_a \xi_b \equiv \left(
g^{ab} + \omega\, \ell^a \ell^b
\right)
\xi_a \xi_b
,
\label{geff_inv}
\end{equation}
where
\begin{align}
\text{tensor}:&\quad
\omega = 
\frac{G_{4X}}{G_4}\phi'^2
\label{omegatensor}
\\
\text{scalar}:&\quad
\omega = - 
\left(\frac{K_{XX}}{K_X} + \frac{G_{4X}}{G_4}\right)\phi'^2
+ \frac{2G_{3X}}{K_X}\phi''.
\label{omegascalar}
\end{align}
We plugged in Eq.~(\ref{Feq}) to simplify the expression~(\ref{omegascalar}).
Using $G^{ab}_\omega$, a characteristic surface $\Sigma$ is obtained as 
a surface whose normal is null with respect to $G^{ab}_\omega$.
We can also construct effective metrics for tangent vectors of $\Sigma$
as 
\begin{equation}
G^\omega_{ab} = g_{ab} - \omega \ell_a \ell_b.
\label{geff}
\end{equation}
The characteristic cones are superluminal (the normal $\xi$ being timelike) if $\omega>0$.
We can also see that the (maximum) propagation speeds of the two tensor modes are the same while that of the scalar mode is different from them.

The above derivation does not work when 
$\xi$ is null, which can be separated into two cases where $\xi$ is not parallel to $\ell$ and when it is parallel.
We examine each of these two cases below.

In the first case where $\xi$ is not parallel to $\ell$, we may take $\xi = n$ by appropriately choosing $n$. Then, 
we may solve Eq.~(\ref{nulleq}) for null characteristic surfaces taking $e_0 = n$ and $e_1 = \ell$, which is equivalent to swapping 0 and 1 in that equation.
Plugging the general ansatz for an eigenvector~(\ref{finalEV}) into this equation, we obtain
\begin{subequations}
\begin{gather}
2 X\cdot n + \frac2{G_4}\left(-\frac12 G_{3X}\phi'^2 + G_{4X}\phi''\right) r_\phi = 0
\label{nulleq01_PW}
\\
\frac{G_{4X}\phi'^2}{G_4}\hat r_{ij} = 0
\label{nulleqij_PW}
\\
X\cdot e_i = 0
\label{nulleq0i_PW}
\\
\alpha = 0
\\
\left\{
\left( \frac{K_{XX}}{K_X} + \frac{G_{4X}}{G_4} \right) \phi'^2
- \frac{2G_{3X}}{K_X} \phi''
\right\}
r_\phi = 0.
\label{nulleqphi_PW}
\end{gather}
\end{subequations}
Unless the coefficients in Eqs.~(\ref{nulleqij_PW}) or (\ref{nulleqphi_PW}) happen to vanish, these equations do not have nontrivial solutions hence there are no null characteristic surfaces.
In the special case where the coefficient of Eq.~(\ref{nulleqphi_PW}) vanishes, $r_\phi$ can be nonzero, then Eqs.~(\ref{nulleq01_PW}) and (\ref{nulleq0i_PW}) will fix $X^a$ in terms of $r_\phi$. Also $\hat r_{ij}$ can be nonzero if the coefficient of Eq.~(\ref{nulleqij_PW}) happen to vanish. These correspond to the scalar and tensor modes, respectively.

In the second case where $\xi$ is parallel to $\ell$,
we can check that $\tilde {\cal R}\cdot r$ identically vanishes, hence the characteristic equation~(\ref{P0def}) reduces to $\tilde P_0\cdot r=0$, which is equivalent to that in GR with a minimally-coupled scalar field.
Then we immediately see that the number of the physical propagating modes is given by two plus one, which comes from the metric and scalar sectors, respectively.
Since $\xi\propto \ell$ is a null vector with respect to the effective metric~(\ref{geff}), we may conclude that any null vector (not only the one parallel to $\ell$) with respect to $G_\omega^{ab}$ gives a characteristic surface.

Before closing this section, let us make some comments on physical features of the effective metrics~(\ref{geff}) and the characteristic surfaces derived from them.
When $\phi'= \phi''=0$, $\omega$ vanishes and then all the modes propagate at the speed of light.
This situation is realized not only on the flat background but also on the purely gravitational plane wave background, for which $\phi$ is constant but $F=F_h$ is nontrivial. The effect of nontrivial $F=F_h$ appears only in the physical metric but not in the deviation of the effective metrics from the physical one.

\begin{figure}[htbp]
\centering
\includegraphics[width=8.5cm]{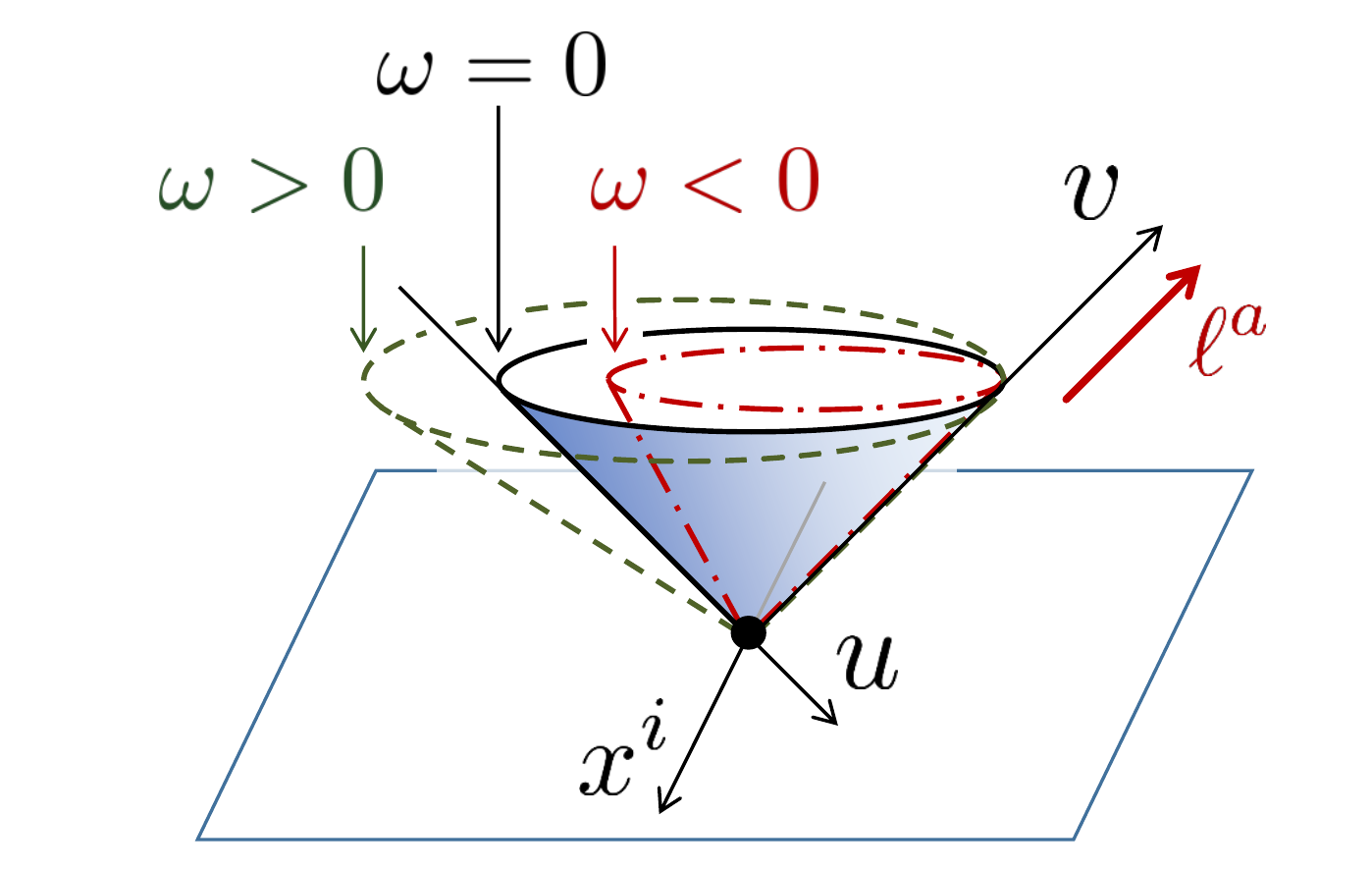}
\caption{A schematic of characteristic cones for waves on the plane wave solution at $F(u,x,y)=0$, where $-u$ and $v$ are taken toward the future direction.
The null cone with respect to the physical metric~(\ref{ansatz}) is given by solid black line, and the characteristic cones obtained from the effective metrics~(\ref{geff}) are shown by green dashed curve and red dot-dashed curve for $\omega > 0$ and $\omega<0$, respectively.
All the cones are aligned in the direction parallel to $\ell^a$ and they split in other directions if $\omega\neq 0$, hence the characteristic surfaces form a nested set of cones in general.
The $u$-axis is aligned to the physical null cone since $F=0$ in this figure, while it is not the case when $F\neq 0$.
\hfill
\label{Fig:cones}}
\end{figure}

The characteristic cones for the tensor and scalar modes do not coincide with each other in general, while they always do along $\ell$, because $\ell$ is null with respect to $G_\omega^{ab}$ for any $\omega$. Therefore the characteristic surfaces form nested cones that touch with each other along $\ell$, as shown in Fig.~\ref{Fig:cones}.
This feature is similar to that in Lovelock theories on type N spacetime background~\cite{Reall:2014pwa}.

As long as $\omega$ in Eqs.~(\ref{omegatensor}) and (\ref{omegascalar}) are finite, 
propagation speeds of waves are finite and we can define causality as usual despite the propagation becomes superluminal if $\omega > 0$. The only difference from GR is that the causality is not defined with respect to the light cone but to the largest cone, which is realized for the largest $\omega$.
Since the effective metrics~(\ref{geff}) are always Lorentzian, the hyperbolicity of the field equation is maintained and the initial value problem is guaranteed to be well-posed.
See Ref.~\cite{Reall:2014pwa} for further discussions on the hyperbolicity in theories with modifications.

\section{Shock formation in shift-symmetric Horndeski theory}
\label{Sec:shock}

Based on the technique summarized in sections~\ref{Sec:characteristics} and \ref{Sec:planewave}, we examine shock formation process in the Horndeski theory in this section.
Some previous works~\cite{Babichev:2016hys,Mukohyama:2016ipl,deRham:2016ged} studied such shock (caustics) formation in generalized Galileon theories focusing on the simple wave solution of the scalar field. We re-examine the problem of shock formation taking the gravitational effect into account. 

For this purpose, we focus on propagation of discontinuity in second derivatives of the scalar field and also the metric.
Such a shock formation process based on transport of second-order discontinuity was studied in \cite{Reall:2014sla} for Lovelock theories in higher dimensions, and it was found that gravitational wave in these theories suffers from shock formation generically.
We will examine if this kind of phenomena could occur for gravitational wave in Horndeski theory, and also check what would happen for scalar field wave and shock formation in them when the gravitational sector is taken into account.
For simplicity, we will focus on Horndeski theory with a single scalar field and particularly the shift-symmetric version of it, as we did in section~\ref{Sec:planewave}.

We first review the formalism of shock formation for a generic equation of motion in section~\ref{Sec:shock_general}, following Ref.~\cite{Reall:2014sla}. We will apply this formalism to the shift-symmetric Horndeski theory in section~\ref{Sec:shock-Horndeski}, and examine conditions to avoid the shock formation without specifying the background solution in section~\ref{Sec:N=0}.

To study properties of shock formation in this theory more explicitly, we focus on some examples of background solutions in the following sections.
In section~\ref{Sec:Npp}, we take the plane wave solution studied in section~\ref{Sec:planewave} as the background solution, and check if this solution suffers from the shock formation.
Another typical class of solutions in the Horndeski theory is solutions whose two-dimensional angular part of the metric is maximally symmetric. 
For example, isotropic homogeneous cosmological solutions such as the FRW universe and also (dynamical) spherically-symmetric solutions belong to this class of solutions.
We study shock formation on such dynamical solutions with two-dimensional maximally-symmetric part in their metrics in section~\ref{Sec:2-dim}.

\subsection{General formalism of shock formation}
\label{Sec:shock_general}

In this section, we introduce a formalism for propagation of discontinuity in second derivatives based on a general equation of motion~(\ref{fieldeq}).
This formalism 
was introduced in Ref.~\cite{anile2005relativistic} and 
was employed by Ref.~\cite{Reall:2014sla} to analyze shock formation process in Lovelock theories. We reproduce a part of the derivation explained therein to get our analysis oriented and to fix the notation.

We will employ the coordinates $(x^a)=(x^0,x^\mu)$ introduced in section~\ref{Sec:review}, where a characteristic surface $\Sigma$ lies on $x^0=0$.
We assume that the equation of motion has the following structure:
\begin{equation}
P_{IJ}(v_{,\mu\nu}\, , v_{,0}\,,v_{,\mu}\,, v \, , x) 
v_{J,00}
+ b_I(v_{,0\mu}\,,v_{,\mu\nu}\,,v_{,0}\,,v_{,\mu}\,,v\,,x) = 0.
\label{fieldeqv2}
\end{equation}
Here we assumed that $P_{IJ}$ is independent of $v_{,0\mu}$, which is the case in the shift-symmetric Horndeski theory at least.
On the characteristic surface $\Sigma$, $\det P = 0$ is satisfied and hence there are eigenvectors of $P$ with vanishing eigenvalues:
\begin{equation}
r_I P_{IJ} = P_{IJ} r_J = 0,
\label{EVdef}
\end{equation}
where we assumed that $P_{IJ}$ is symmetric in its indices hence the left and right eigenvectors of $P$ coincide with each other.

\begin{figure}[htbp]
\centering
\includegraphics[width=10.5cm]{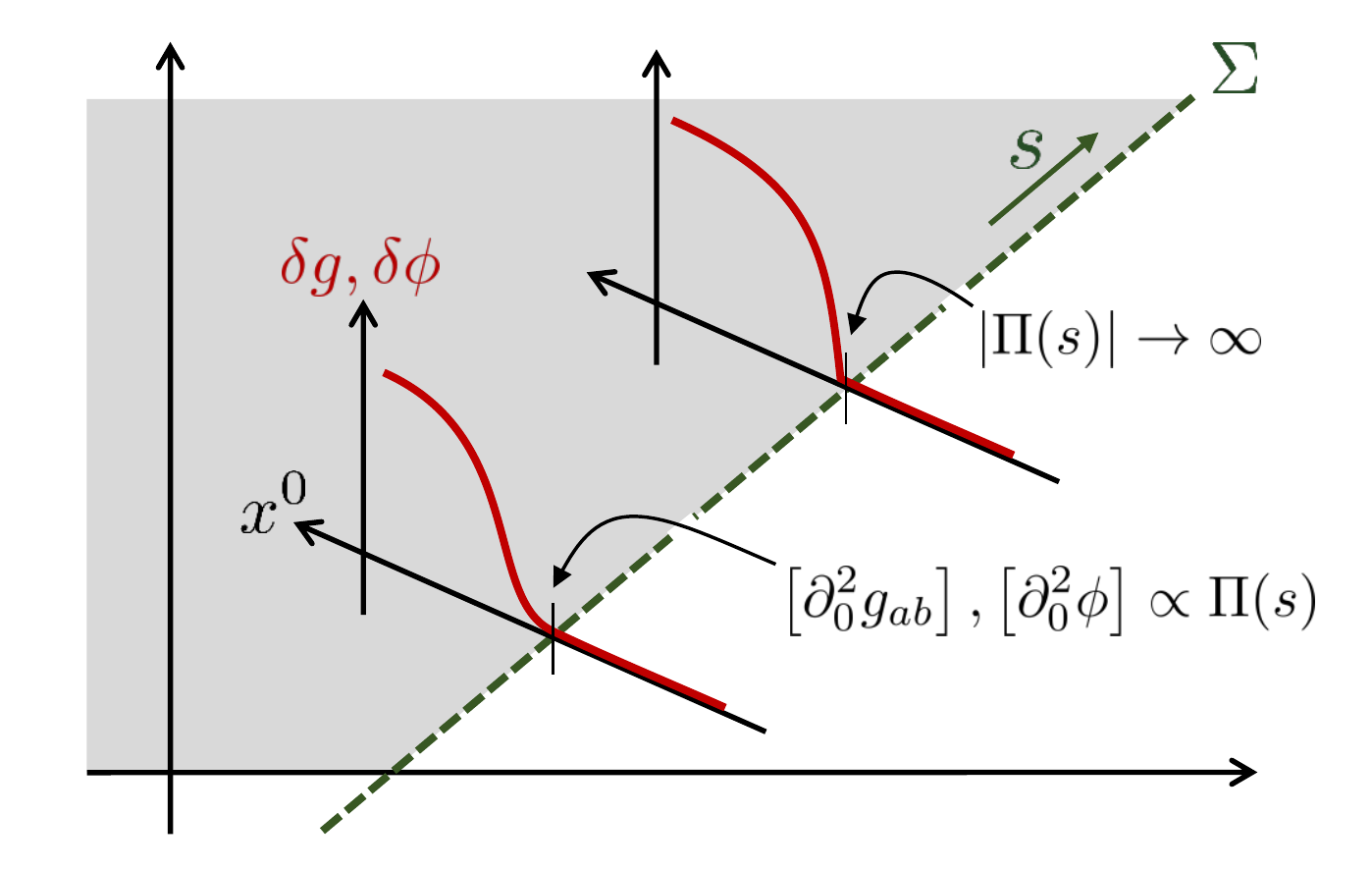}
\caption{Propagation of discontinuity in second derivatives of the metric and scalar field, where red curves show profile of difference between the ``background solution'' without discontinuity and the solution to which the discontinuity is added.
Solution in the future region of $\Sigma$ (shaded region) is influenced by the discontinuity, while that in the past region is not.
The discontinuity propagates on $\Sigma$, and its amplitude $\Pi(s)$ evolves as it propagates along a bicharacteristic curve parameterized by $s$. When ${\cal N} \neq 0$, $\Pi(s)$ diverges at finite $s$ if the initial amplitude $\Pi(0)$ is chosen appropriately. At this point, second derivatives of fields blow up and single derivatives become discontinuous. We call this phenomenon a shock formation in this work.
\hfill
\label{Fig:disc}}
\end{figure}

Now let us consider time evolution from an initial time slice that intersects with $x^0=0$, and assume that the dynamical variable $v$ has a discontinuity in its second derivative with respect to $x^0$ at the locus of $x^0=0$ on the initial time slice. 
This discontinuity will propagate on $\Sigma$, and the solution on the past side of $\Sigma$ will not be influenced by the discontinuity. Hence we may regard wave of the discontinuity to propagate into the ``background solution'', which is a solution to the unperturbed equation of motion (see Fig.~\ref{Fig:disc}).

Since the discontinuous part of (\ref{fieldeqv2}) is given by $P_{IJ}\left[v_{J,00}\right] = 0$, where a quantity in square brackets denotes its discontinuous part, 
comparing with Eq.~(\ref{EVdef})
we find that $\left[v_{I,00}\right]$ must be proportional to an eigenvector $r_{I}$ and hence
\begin{equation}
\left[v_{I,00}\right] = \Pi(x^\mu)\, r_I,
\label{Pidef}
\end{equation}
where $\Pi(x^i)$ is the proportional constant, which may be regarded as amplitude of the discontinuity.

In the following, we focus on how the amplitude of discontinuity $\Pi(x^\mu)$ changes as it propagates on $\Sigma$.
A transport equation of $\Pi$ can be constructed by firstly taking $x^0$ derivative of Eq.~(\ref{fieldeqv2}), acting $r^I$ on it to remove third derivatives with respect to $x^0$, and finally picking up discontinuous part of the resultant equation. Leaving the details of these steps to Ref.~\cite{Reall:2014sla}, we find that the final outcome of these steps is given by
\begin{equation}
{\cal K}^\mu \Pi_{,\mu} + {\cal M} \, \Pi + {\cal N}\,\Pi^2  = 0,
\label{dotPieqpre}
\end{equation}
where
\begin{align}
{\cal K}^\mu &=r_I r_J \frac{\partial b_I }{\partial v_{J,0\mu}} 
\label{Kdef-orig}
\\
{\cal M} &= r_I
\biggl\{
\frac{\partial b_I}{\partial v_{J,0\mu}} r_{J,\mu}
+ \left(
  \frac{\partial P_{IJ}}{\partial v_{K,\mu\nu}}v_{K,0\mu\nu}
+ \frac{\partial P_{IJ}}{\partial v_{K,\mu}} v_{K,0\mu}
+ \frac{\partial P_{IJ}}{\partial v_{K}} v_{K,0}
+ \frac{\partial P_{IJ}}{\partial x^0}
+ \frac{\partial b_I}{\partial v_{J,0}}
\right) r_J
+ 2\frac{\partial P_{IJ}}{\partial v_{K,0}}
\left(v_{(J|,00}\right)^- r_{|K)} 
\biggr\}
\label{Mdef-orig}
\\
{\cal N} &= r_I r_J r_K \frac{\partial P_{IJ}}{\partial v_{K,0}},
\label{Ndef-orig}
\end{align}
where $\left(v_{J,00}\right)^- \equiv \lim_{x^0\to -0} v_{J,00}$.
The coefficients in Eq.~(\ref{dotPieqpre}) depend only on the field values at $x^0 \to -0$, that is, the background solution on the past side of the characteristic surface.
The discontinuity propagates along the integral curve generated by ${\cal K}^\mu$, which can be found by integrating
\begin{equation}
\frac{dx^\mu}{ds} = {\cal K}^\mu(x^\nu),
\label{sdef}
\end{equation}
where we have introduced a parameter $s$ along the integral curve $x^\mu = x^\mu(s)$ that becomes zero on the initial time slice.
It can be shown that this integral curve coincides with a bicharacteristic curve, which is a geodesic curve with respect to the effective metric and along which waves on characteristic surface propagate~\cite{Reall:2014sla}. 
Then, denoting $\dot \Pi \equiv d \Pi(s)/ds$,
Eq.~(\ref{dotPieqpre}) may be written as
\begin{equation}
\dot \Pi + {\cal M} \, \Pi + {\cal N} \, \Pi^2 = 0.
\label{dotPieq}
\end{equation}
An equation equivalent to Eq.~(\ref{dotPieq}) is obtained also for propagation of weakly nonlinear high frequency waves, whose frequency is sufficiently large compared to the background time dependence~\cite{ChoquetBruhat:1969,Hunter-Keller,anile2005relativistic,choquet2009general,Reall:2014sla}.
This equation is nonlinear in $\Pi$ as long as ${\cal N}$ does not vanish. In such a case, the theory is called genuinely nonlinear and suffers from shock formation as we see below. There are certain theories for which ${\cal N}$ identically vanishes and the above equation becomes linear, in which case the theory is called exceptional or linearly degenerate~\cite{anile2005relativistic,Lax:1954gzy,*Lax:1957hec,*1976pitm.book.....J,doi:10.1063/1.1664860}.
For example, GR coupled to a canonical scalar field is an exceptional theory.

Time evolution of the amplitude $\Pi(s)$ is described by the general solution of Eq.~(\ref{dotPieq}), which is given by
\begin{equation}
\Pi(s) = \frac{\Pi(0) e^{-\Phi(s)}}{
1 + \Pi(0)\int_0^s {\cal N}(s') e^{-\Phi(s')} ds'
} \, ,
\label{Pisol}
\end{equation}
where
\begin{equation}
\Phi(s) \equiv \int_0^s {\cal M}(s') ds'.
\label{Phidef}
\end{equation}
When ${\cal N}=0$, $\Pi(s)$ obeying (\ref{Pisol}) diverges only when $\Phi(s)$ and hence ${\cal M}(s)$ do so.
Since ${\cal M}(s)$ is determined only by information of the background solution and the characteristic surface on it, $\Phi$ diverges only when the background solution or $\Sigma$ is not regular.
This happens when bicharacteristic curves on $\Sigma$ form a caustic on it by crossing with each other, where the amplitude of wave may diverge due to focusing effect.
To distinguish it from the shock generated by nonlinear effect due to nonzero $\cal N$, sometimes this type of shock formation is called a linear shock~\cite{anile2005relativistic}.

When ${\cal N}\neq 0$, $\Pi(s)$ may diverge even when $\Phi(s)$ is regular, that is, there are no caustics on $\Sigma$. Such a divergence is realized when the denominator of (\ref{Pisol}) vanishes as $s$ increases from zero.
As long as ${\cal N}(s)$ and $\Phi(s)$ are regular functions, 
we can always tune the signature and magnitude of $\Pi(0)$ to make $\Pi(0)\int_0^s {\cal N}(s') e^{-\Phi(s')} ds'$ cross $-1$ at finite $s$, because there will be a sufficiently small region of $s$ in which $\int_0^s {\cal N}(s') e^{-\Phi(s')} ds'$ behaves as a monotonic function of $s$.
Hence, when ${\cal N}\neq 0$, it is guaranteed that $\Pi(s)$ can blow up if the initial amplitude $\Pi(0)$ is sufficiently large, where the divergence is expected to occur roughly at $s \sim-\bigl(\Pi(0)\,{\cal N}e^{-\Phi}\bigr)^{-1} $.
In some cases, such a divergence of $\Pi(s)$ may happen even when $\Pi(0)$ is arbitrarily small, as we see an example in section~\ref{Sec:Npp}.

At the moment when $\Pi(s)$ diverges, second derivatives on the future side become infinite hence first derivatives become discontinuous at $\Sigma$.
We call it a shock formation in this work. This phenomenon originates from the nonlinear effect due to nonzero ${\cal N}$ as we observed above, and also it can be seen that it occurs when two different characteristic surfaces collide with each other in time evolution~\cite{Reall:2014sla,majda1984compressible,johnnonlinear,christodoulou2007formation}.

\subsection{Shock formation in shift-symmetric Horndeski theory}
\label{Sec:shock-Horndeski}

As shown in the previous section, shock formation in the discontinuity of second derivatives may occur when $\cal N$ defined by Eq.~(\ref{Ndef-orig}) does not vanish.
In the following, we examine properties of $\cal N$ and shock formation process in the shift-symmetric Horndeski theory.

We first show the expression of $\cal N$ in the shift-symmetric Horndeski theory on general background in section~\ref{Sec:N}, then examine sufficient conditions for ${\cal N}=0$ in section~\ref{Sec:N=0}.
To evaluate $\cal N$ explicitly, we need to specify the background solution.
We do it for the plane wave solution in section~\ref{Sec:Npp}, and also for dynamical solution whose two-dimensional angular part of metric is maximally symmetric in section~\ref{Sec:2-dim}.

\subsubsection{Expression of $\cal N$ on general background}
\label{Sec:N}

As shown in Eq.~(\ref{Ndef-orig}), the coefficient of the nonlinear term $\cal N$ is given by a derivative of the principal symbol of the field equation.
In the shift-symmetric Horndeski theory, it is given by
\begin{align}
{\cal N}
= 
\left(
r_{ef}\frac{\partial}{\partial g_{ef,0}}
+ 
r_\phi \frac{\partial}{\partial \phi_{,0}}
\right)
\left\{
\sum_{n=2}^5
\bigl(
\begin{array}{@{}cc@{}}
r^{ab} & r_\phi
\end{array}
\bigr)
\left(
{
\begin{array}{@{}cc@{}}
\frac{\partial {\cal G}^n_{ab}}{\partial g_{cd,00}}
&
\frac{\partial {\cal G}^n_{ab}}{\partial \phi_{,00}}
\\ 
\frac{\partial \nabla^h{\cal J}_h^n}{\partial g_{cd,00}}
&
\frac{\partial \nabla^h{\cal J}_h^n}{\partial \phi_{,00}}
\end{array}
}
\right)
\left(
{
\begin{array}{@{}c@{}}
r_{cd}
\\ 
r_\phi
\end{array}
}
\right)
\right\}
,
\label{Ndef}
\end{align}
where we have taken $\xi_a = (dx^0)_a$ and denoted $\xi_c \xi_d \frac{\partial}{\partial g_{ab,cd}} \equiv \frac{\partial}{\partial g_{ab,00}}$ and $\xi_c \xi_d \frac{\partial}{\partial \phi_{,cd}} \equiv \frac{\partial}{\partial \phi_{,00}}$.
The derivatives
with respect to $g_{ef,0}$ and $\phi_{,0}$
act only on the components of the two-by-two matrix.
For the terms appearing in this expression, we can confirm that the following relation holds:
\begin{align}
r_{ef}\frac{\partial}{\partial  g_{ef,0}} \frac{\partial \nabla^h {\cal J}_h^n}{\partial \phi_{,00}}
&= \frac{\partial}{\partial\phi_{,0}}  \frac{\partial {\cal G}^n_{ab}}{\partial \phi_{,00}}r^{ab}
= \frac{\partial}{\partial\phi_{,0}} \frac{\partial \nabla^h {\cal J}_h^n}{\partial g_{cd,00}}r_{cd} \, ,
\\
r_{ef}\frac{\partial}{\partial g_{ef,0}}
\frac{\partial {\cal G}^n_{ab}}{\partial \phi_{,00}} r^{ab}
&=
r_{ef}\frac{\partial}{\partial g_{ef,0}}
 \frac{\partial \nabla^h {\cal J}_h^n}{\partial g_{cd,00}}r_{cd} 
=
\frac{\partial}{\partial\phi_{,0}}
\frac{\partial {\cal G}^n_{ab}}{\partial g_{cd,00}} r^{ab} r_{cd} \, .
\end{align}
Using them, $\cal N$ may be expressed as
\begin{equation}
{\cal N} = 
\sum_{n=2}^5\biggl\{
r^{ab} r_{cd}r_{ef}
\frac{\partial}{\partial g_{ef}}
\frac{\partial {\cal G}^n_{ab}}{\partial g_{cd,00}} 
+
r_\phi{}^3
\frac{\partial}{\partial\phi_{,0}}  \frac{\partial \nabla^h {\cal J}^n_h}{\partial \phi_{,00}}
+ 3 \left(
r_\phi r_{ef}\frac{\partial}{\partial  g_{ef,0}}
\frac{\partial {\cal G}^n_{ab}}{\partial \phi_{,00}} r^{ab}
+
r_\phi{}^2
r_{ef}\frac{\partial}{\partial g_{ef,0}} \frac{\partial \nabla^h {\cal J}_h^n}{\partial \phi_{,00}}
\right)
\biggr\}
.
\label{Ndef2}
\end{equation}
We summarize explicit formula of the terms in (\ref{Ndef2}) in appendix~\ref{App:N}.
$\cal N$ on a general solution is given by a summation of the expressions therein.
It does not vanish in general, hence we may expect that shock formation occurs generically in this theory, while there are some cases where ${\cal N}=0$ as we see below.

\subsubsection{Sufficient conditions for ${\cal N}=0$}
\label{Sec:N=0}

To evaluate $\cal N$ explicitly, we first need to specify the background solution, and then find characteristic surfaces and eigenvectors associated with them.
We will follow such a procedure taking some examples of the background solutions later in sections~\ref{Sec:Npp} and \ref{Sec:2-dim}.
Before studying such examples, in this section
let us examine sufficient conditions for $\cal N$ to vanish without specifying the background solution.
Among various theories, we find that the {\it k}-essence~\cite{ArmendarizPicon:1999rj,ArmendarizPicon:2000ah} coupled to GR stands out as a special theory where the scalar field decouples from metric sector, and that the scalar field version of the DBI model~\cite{Born410,*Born425,*Dirac57} turns out to be the unique nontrivial theory to make ${\cal N}=0$ on the general background.

For the {\it k}-essence coupled to GR ($G_3 = G_5 = 0$, $G_4=\text{{const.}}$ with general $K(X)$), non-vanishing parts of the (trace-reversed) principal symbol are given by%
\footnote{The analysis in this section is unchanged even when $G_3$ and $G_5$ are promoted to non-zero constants, for which the Lagrangians ${\cal L}_{3,5}$ become total derivatives and do not contribute to the dynamics of the theory.
}
\begin{align}
\xi_s \xi_t \frac{\partial \nabla^\alpha {\cal J}_{\alpha}^2}{\partial \phi_{,st}}
&=
-K_X\left(g^{ab} - \frac{K_{XX}}{K_X}\phi^{|a} \phi^{|b}\right)\xi_a \xi_b,
\label{Pphi-K}
\\
\xi_s \xi_t r_{qr}
\frac{\partial \tilde{\cal G}^4{}_a^b}{\partial g_{qr,st}} 
&=
\frac12 G_4
\left(
\delta_{aa_1 a_2}^{bb_1 b_2}
- \delta_a^b \delta_{a_1 a_2}^{b_1 b_2}
\right)
\xi^{a_1} \xi_{b_1} r^{a_2}_{b_2}
=
\frac12 G_4\left(
- \xi^2 r_a^b - r^c_c \xi_a\xi^b
+ \xi_c r^c_a \xi^b 
+ \xi^c r_c^b \xi_a
\right).
\label{Pg-K}
\end{align}
There are no terms that mix the scalar and metric parts, and in this sense the scalar part decouples from the metric part in this analysis.
Characteristic surfaces can be found by equating these expressions with zero and solving for $\xi$.

In the metric part~(\ref{Pg-K}), if $\xi \neq 0$ we find that $r$ is forced to have a form $r_{ab} = \xi_{(a}X_{b)}$, which is pure gauge. Hence $\xi$ must be null to find physical modes, then Eq.~(\ref{Pg-K}) becomes a constraint which reduces the number of degree of freedom by four. Hence there will be two physical degrees of freedom in $r = (r_{ab},r_\phi) = (r_{ab},0)$, which corresponds to the usual tensor modes in GR.

In the scalar part~(\ref{Pphi-K}), the factor on the right-hand side gives an effective metric for the scalar mode as
\begin{equation}
\left(g^{ab} - \frac{K_{XX}}{K_X}\phi^{|a} \phi^{|b}\right)\xi_a \xi_b = 0.
\label{Kscalar}
\end{equation}
A surface 
whose normal vector
$\xi$ satisfies this equation is characteristic, and the eigenvector is simply given by $r = (r_{ab},r_\phi) = (0,r_\phi)$.

Let us evaluate $\cal N$ for these modes next. 
For the {\it k}-essence coupled to GR, 
only~(\ref{d2J2dphippdphip}) contributes among the terms appearing in $\cal N$, hence
\begin{equation}
{\cal N} =
r_\phi{}^3
\frac{\partial}{\partial\phi_{,0}}  \frac{\partial \nabla^h {\cal J}^2_h}{\partial \phi_{,00}}
=
3 \phi^{|0} K_{XX} \xi^2 - (\phi^{|0})^3 K_{XXX},
\end{equation}
where $\phi^{|0} = \xi_a \phi^{|a}$.
For the tensor modes, this $\cal N$ vanishes identically and hence shock will not form.
For the scalar mode, normalizing $r_\phi=1$ and using (\ref{Kscalar}), we find
\begin{equation}
{\cal N} = 
\frac{\left(\phi^{|0}\right)^3}{K_X}
\left( 3 K_{XX}{}^2 - K_X K_{XXX} \right).
\label{N_K}
\end{equation}
The condition ${\cal N}=0$ to avoid shock formation is equivalent to $3 K_{XX}{}^2 - K_X K_{XXX}=0$ (assuming $\phi^{|0}\neq 0$), which
may be viewed as a differential equation of $K(X)$. 
A trivial solution is $K\propto X$, which corresponds to a canonical scalar field.
The general solution other than this one is given by
\begin{equation}
K
 = -\lambda \sqrt{c \pm X} + \Lambda,
\label{K-DBI}
\end{equation}
where $\lambda, c, \Lambda$ are constants.
This is the Lagrangian of the scalar DBI model, which reduces to the cuscuton model (${\cal L}\propto \sqrt{X}$) for $c=0$ or in the limit $X\to\infty$. Hence, among the theories described by the {\it k}-essence coupled to GR, the scalar DBI model (\ref{K-DBI}) is singled out as the theory free from shock formation.

This behavior is the same as that of plane-symmetric simple wave solutions for probe scalar field studied by Ref.~\cite{Babichev:2016hys,Mukohyama:2016ipl}, where the scalar DBI model turned out to be the theory free from caustics formation.%
\footnote{See \cite{boillat2006recent,*1959PThPS...9...69T,*Whitham_chap17,*Barbashov:1966frq,*doi:10.1142/9789812708908_0003,*Deser:1998wv} for earlier discussions on exceptional theories for a scalar field on flat spacetime, in which the canonical scalar and the scalar DBI were found as such theories. Also the DBI model for a probe vector field is shown to be exceptional~\cite{doi:10.1063/1.1665231}.}
As discussed above, even in our setup the scalar and metric part decouples if $G_{3,5}(X)$ and non-constant part of $G_4(X)$ are set to zero. Having this decoupling, it seems natural that the result obtained from our setup coincides with those for the probe scalar field.

For the theories other than the {\it k}-essence coupled to GR,
it seems difficult to find characteristics and eigenvectors since the scalar and metric parts do not decouple in the characteristic equation.
However, there is a sufficient condition to realize
${\cal N}=0$ even in more general theories, which
is to have $\phi^{|0} = \Gamma^0_{ab} = 0$ on a characteristic surface. 
Imposing this condition to the expressions in appendix~\ref{App:N}, it can be checked that
all the terms appearing in $\cal N$ vanish identically.
A flat spacetime with constant $\phi$ is an example where this condition is satisfied, hence shock formation in the sense of section~\ref{Sec:shock_general}
does not occur for wave propagating into such background. This is consistent with the results of Refs.~\cite{Babichev:2016hys,Mukohyama:2016ipl,deRham:2016ged} for simple waves of a probe scalar field, for which $\phi$ and its first derivative are not zero where caustics form.
Another less trivial example of ${\cal N}=0$ is the Killing horizon with $\phi^{|0}=0$ imposed additionally, which was discussed in section~\ref{Sec:KH} in the context of the bi-Horndeski theory. On the Killing horizon $\Gamma^0_{ab}=0$ is satisfied by virtue of the condition~(\ref{KillingH}), then $\cal N$ vanishes if $\phi^{|0}=0$ is satisfied as well.

\subsection{Shock formation on plane wave solution}
\label{Sec:Npp}

To study properties of shock formation for more complicated choices of $G_n(X)$ on generic background solutions, it seems that we need to look into explicit examples of background solutions and study wave propagation on them. Such a study using explicit background solutions is the main subject of this and the next sections.

The first example is the plane wave solution examined in section~\ref{Sec:WaveOnWave},
where we follow the analysis of Ref.~\cite{Reall:2014sla} for shock formation on the plane wave solution in Lovelock theories.
To prepare for the analysis, we introduce coordinates adapted to geodesics in this theory in section~\ref{Sec:Npp_geometry}.
Using them, we examine shock formation on this solution in section~\ref{Sec:Npp-result}.
We will find that the tensor modes or the gravitational wave, and also the scalar mode propagating along $\ell$ do not suffer from the shock formation, while the scalar mode propagating in the opposite direction forms a shock in general.

\subsubsection{Geometry of characteristic surfaces}
\label{Sec:Npp_geometry}

Characteristic surfaces on the plane wave background is given by null hypersurfaces with respect to effective metrics~(\ref{geff}), which can be transformed as
\begin{align}
G^\omega_{\mu\nu}dx^\mu dx^\nu
&=
(F-\omega) du^2 + 2 du dv + dx^2 + dy^2
=
Fdu^2 + 2 du dv' + dx^2 + dy^2,
\label{geff2}
\end{align}
where we have introduced a new coordinate $v' \equiv v - \frac{1}{2}\omega \, u$.
For simplicity, we consider plane-fronted wave propagating from a surface $u=v'=0$,
and focus on the propagation in the (negative) $u$ direction, which is opposite from the direction along $\ell^a$ in Fig.~\ref{Fig:cones}.%
\footnote{For the wave propagation along $\ell^a$, i.e.\ when $\xi_a = \ell_a$, it can be checked that $\phi^{|0}$ and $\Gamma^0_{ab}$ vanish on the plane wave solution background. Then $\cal N$ vanishes in this case, as argued at the end of section~\ref{Sec:N=0}.
Hence the shock formation does not occur for the wave propagating along $\ell^a$.}
Such wave propagates 
along the null geodesics of the effective metric~(\ref{geff2}).
Parameterizing the coordinates on a geodesic by affine parameter $\lambda$, the geodesic equation $\frac{d^2}{d\lambda^2}x^a + \Gamma^a_{bc}\frac{dx^b}{d\lambda}\frac{dx^c}{d\lambda}=0$ for the effective metric~(\ref{geff2}) is given by
\begin{equation}
\ddot u = 0,
\qquad
\ddot v' + \frac12 \dot u \left(
\dot u F_{,u} + 2 \dot x^\alpha F_{,x^\alpha} 
\right) = 0,
\qquad
\ddot x^\alpha - \frac12 \dot u \,F_{,x^\alpha} = 0,
\end{equation}
which are the $a = u, v'$ and $x^\alpha(=x,y)$ components of the equation, respectively.
The $u$ component implies that we may take $u= \lambda$.
Below, we assume for simplicity that $\phi'$ is a nonzero constant and $\phi''=0$. 
Also, we assume that $F(x^\mu)$ is given by
Eq.~(\ref{Ffunc}) whose homogeneous part is Eq.~(\ref{amatrix}) with $a_{xx}=-a_{yy} = A$ and $a_{xy}=0$ for a constant $A$, that is, 
\begin{equation}
F = 
- \frac{\kappa}{2}\phi'^2 \left(x^2+y^2\right) 
+ A\,\left(x^2-y^2\right).
\label{Fexp}
\end{equation}
In this case, the $x^\alpha$ components of the geodesic equations can be solved by
\begin{equation}
x^\alpha = \eta^\alpha \cosh\bigl(
\sqrt{A_\alpha} \lambda
\bigr),
\end{equation}
where
$\eta^I$ is the initial position $x^I$ of a geodesic at $\lambda=0$,
and $A_\alpha$ are given by
\begin{equation}
A_{x} = A - \frac14 \kappa \phi'^2,
\qquad
A_{y} = - A - \frac14 \kappa \phi'^2.
\label{alphaxy}
\end{equation}
Assuming $\kappa>0$, it is guaranteed that one of $A_x$ and $A_y$ becomes negative at least for any choice of $A$.
The $v$ component can then be integrated as
\begin{equation}
v = -\frac14 \sum_{\alpha=x,y} \left(\eta^\alpha\right)^2
\sinh\bigl( 2 \sqrt{A_\alpha} \lambda \bigr).
\end{equation}
Introducing Gaussian null coordinates 
adapted to the characteristic surface and geodesics
\begin{equation}
\eta^\alpha = \frac{x^\alpha}{
\cosh\bigl(\sqrt{A_\alpha} \lambda\bigr)
},
\qquad
x^0 = 
v - \frac12 \omega \, u
+ \frac14
\sum_{\alpha=x,y} \sqrt{A_\alpha} \left(\eta^\alpha\right)^2
\sinh\bigl( 2 \sqrt{A_\alpha} u \bigr),
\end{equation}
the physical metric~(\ref{ansatz}) becomes
\begin{equation}
ds^2 = \omega du^2 + 2 du dx^0 
+ \sum_{\alpha=x,y} \cosh^2\bigl( \sqrt{A_\alpha} u\bigr) \left(d\eta^\alpha\right)^2.
\label{gppmod}
\end{equation}
The characteristic surface is at $x^0=0$ and its normal is given by $\xi = dx^0$.
This metric becomes singular at $u = \pm \pi / 2 \sqrt{-A_\alpha}$ for $A_\alpha<0$, which corresponds to a caustic of the null geodesics.
The only nonzero components of $\Gamma^0_{\alpha\beta}$ on $x^0=0$ are 
\begin{equation}
\Gamma^0_{\alpha\alpha} = -\frac12 \sqrt{A_\alpha} \sinh\bigl(2\sqrt{A_\alpha}u\bigr)
\qquad (\text{no sum on $\alpha$}).
\label{Gamma-gpp}
\end{equation}
This quantity is proportional to extrinsic curvature of the surface $x^0=0$ when it is not null.

\subsubsection{$\cal N$ on the plane wave solution}
\label{Sec:Npp-result}

$\cal N$ on the plane wave solution can be obtained by plugging the background solution~(\ref{gppmod}), (\ref{Gamma-gpp}) with $\phi=\phi(u)$ into the general expression~(\ref{Ndef2}), whose explicit expressions are given in appendix~\ref{App:N}.
We summarize the explicit formula obtained from this procedure in appendix~\ref{App:Npp}.

For the tensor mode~(\ref{EVgrav}), the only term that could contribute to $\cal N$ 
is the pure metric term~(\ref{d2G5dgpdgpp}), which is given by
\begin{equation}
r^{ab} r_{cd} r_{ef}\frac{\partial}{\partial  g_{ef,0}}
\frac{\partial {\cal G}^5_{ab}}{\partial g_{cd,00}} 
=
 \frac{\phi^{|0}}{4}
 X G_{5X}
 \delta_{a a_1 a_2 a_3}^{b b_1 b_2 b_3} 
\xi^a \xi_b r^{a_1}_{b_1} r^{a_2}_{b_2} r^{a_3}_{b_3}.
\label{d2G5dgpdgpp_maintext}
\end{equation}
This term becomes zero because $X=0$ for the plane wave solution and also $ \delta_{a a_1 a_2 a_3}^{b b_1 b_2 b_3} 
\xi^a \xi_b r^{a_1}_{b_1} r^{a_2}_{b_2} r^{a_3}_{b_3}$ vanishes identically if Eq.~(\ref{EVgrav}) is plugged in.
Hence, $\cal N$ vanishes for the tensor modes on the plane wave background, or in other words gravitational wave on this background does not suffer from the shock formation.

For the scalar mode~(\ref{EVscalar}), $\cal N$ is given by
\begin{equation}
{\cal N} =
  C_+ \, t_+(u)
+ C_- \, t_-(u)
+ C_0,
\label{Nscalar_pp}
\end{equation}
where
\begin{equation}
t_\pm(u) \equiv 
\sqrt{A_x}\tanh\bigl(\sqrt{A_x}u\bigr) 
\pm 
\sqrt{A_y}\tanh\bigl(\sqrt{A_y}u\bigr),
\label{tpmdef}
\end{equation}
and $C_{\pm,0}$ are constants given by
\begin{align}
C_+ &=
\phi'^2
\left\{
-\frac{1}{2G_4} \left(2 G_{3X}G_{4X} + K_X G_{5X}\right)
+ \frac{2K_{XX}G_{3X}}{K_X}
- G_{3XX}
\right\}
\\
C_- &=-2A \, G_{5X} 
\\
C_0 &=
\phi'^3
\left\{
\frac{3}{G_4}\left(
-G_{3X}{}^2 - K_X G_{4XX} + K_{XX}G_{4X} 
\right)
+ \frac{3 K_{XX}{}^2}{K_X} - K_{XXX}
\right\}
.
\end{align}
We have also taken a normalization $r_\phi=1$.
This expression can be obtained following the calculation procedure explained above.
Since we have already found geodesics and the coordinates adapted to it, as summarized in section~\ref{Sec:Npp_geometry}, there is an alternative method to derive $\cal N$
and also the entire part of the transport equation~(\ref{dotPieqpre}). 
In this method, we assume the field variables are given by
\begin{equation}
g_{ab} = \bar g_{ab} + \frac{1}{2} \, (x^0)^2 \, \Theta(x^0) \, \Pi\bigl(u,\eta^\alpha\bigr) \, r_{ab} \, ,
\qquad
\phi = \bar \phi  + \frac{1}{2} \, (x^0)^2 \, \Theta(x^0) \, \Pi\bigl(u,\eta^\alpha\bigr) \, r_\phi \, ,
\label{discansatz}
\end{equation}
where $\bar g_{ab}, \bar\phi$ are the background solutions and $\Theta(x^0)$ is a step function.
The above $g_{ab}$ and $\phi$ correctly give discontinuities in their second derivatives at $x^0=0$ as prescribed by Eq.~(\ref{Pidef}).
The transport equation~(\ref{dotPieqpre}) is obtained by evaluating the equation of motion at $x^0=0$ using Eq.~(\ref{discansatz}), although it is not a correct solution in $x^0>0$.

With the aid of computer algebra, we can follow this alternative procedure to find the transport equation of $\Pi(u,\eta^\alpha)$ to be given by
\begin{equation}
{\cal K}\, \Pi_{,u} + {\cal M} \,   \Pi + {\cal N}  \,  \Pi^2 = 0,
\label{transport_PP}
\end{equation}
where $\cal N$ is given by Eq.~(\ref{Nscalar_pp}), and the other coefficients turn out to be
\begin{equation}
{\cal K} =  -2 K_X, \qquad {\cal M} = - K_X t_+(u).
\end{equation}
Then, the transport equation takes the form of Eq.~(\ref{dotPieq}) once 
the parameter $s$ along the bicharacteristic curve is introduced following (\ref{sdef}) as
\begin{equation}
s = -\frac{1}{2K_X} \, u \, .
\end{equation}

Now let us assume $A < 0$ for definiteness, which implies $A_x<0$ and $|A_x|>|A_y|$.
Then Eq.~(\ref{tpmdef}) becomes
\begin{equation}
t_\pm = 
\sqrt{-A_x} \tan\left(\frac\pi2 \frac{s}{s_*}\right)
\mp 
\sqrt{A_y} \tanh\left(\frac\pi2 \frac{s}{s_*}\right),
\end{equation}
where $s_* \equiv \frac{\pi}{4K_X \sqrt{-A_x}}$, and also $e^{-\Phi}$ appearing in the general solution of $\Pi(s)$~(\ref{Pisol}) is calculated as
\begin{equation}
e^{-\Phi} = \left\{
\cos\left(\frac\pi2 \frac{s}{s_*}\right)\cosh\left(2K_X \sqrt{A_y} s \right)
\right\}^{-1/2}.
\end{equation}
This quantity diverges for $s\to s_*-$ as $\propto (s_* -s)^{-1}$.
Also, ${\cal N} e^{-\Phi}$ behaves for $s\to s_*-$ as
\begin{equation}
{\cal N}e^{-\Phi} \simeq
\frac{\left(C_+ + C_-\right) \sqrt{-A_x}}{\cosh^{1/2}\left(\frac\pi2 \sqrt{\frac{A_y}{-A_x}}\right)}
\left( \frac{s_*-s}{s_*} \right)^{-3/2}
\equiv
{\cal N}_0
\left( \frac{s_*-s}{s_*} \right)^{-3/2}.
\label{integrand}
\end{equation}
This quantity diverges for $s\to s_*-$, hence the denominator of $\Pi(s)$ given by Eq.~(\ref{Pisol}) can be zero at finite $s=s_0$ (such that $0<s_0<s_*$) no matter how small $|\Pi(0)|$ is.
Hence, $\Pi(s)$ can  diverge and the shock formation occurs at $s=s_0$ for an arbitrarily small $\Pi(0)$ in this example.

Assuming that ${\cal N}e^{-\Phi}$ is well approximated by Eq.~(\ref{integrand}),
$s_0$ may be estimated as follows.
Using Eq.~(\ref{integrand})
the integral in the denominator of (\ref{Pisol}) is estimated as 
\begin{equation}
\int_0^s {\cal N}(s') e^{-\Phi(s')} ds' \simeq 2 s_* {\cal N}_0\left(\frac{s_*-s}{s_*}\right)^{-1/2}.
\end{equation}
If this approximation is valid, $s_0$ will be approximated by
\begin{equation}
s_0 \simeq \left\{
1 - \left(2\Pi(0)s_* {\cal N}_0\right)^2
\right\} s_*.
\label{s0estimate}
\end{equation}
The approximation used in Eq.~(\ref{integrand}) does not work if $C_+=C_-=0$.
In this case, the denominator is given by $1 + \Pi(0) C_0 \int_0^s  e^{-\Phi(s')} ds'$.
The integral $\int_0^s  e^{-\Phi(s')} ds'$ diverges at $s=s_*$, hence
the denominator vanishes and $\Pi(s)$ diverges
at $s<s_*$ if $\Pi(0)$ is taken so that $\Pi(0) C_0 < 0$.
Hence the shock formation can occur even in this case.

In this example, $\Pi(0)$ diverges at $s=s_*$ even when $\cal N$ happens to vanish, 
because 
${\cal N}=0$ implies 
$\Pi(s) = \Pi(0) e^{-\Phi}$ and $e^{-\Phi} \propto (s*-s)^{-1/2}\to \infty$ for $s\to s_*-$.
This divergence occurs at the caustics of geodesics on $\Sigma$ and caused by the focusing effect in wave propagation.

Before closing this section, we briefly examine conditions to realize ${\cal N}=0$.
If $K$ and $G_n$ satisfy
\begin{equation}
\begin{gathered}
\frac{3}{G_4}\left(
-G_{3X}{}^2 - K_X G_{4XX} + K_{XX}G_{4X} 
\right)
+ \frac{3 K_{XX}{}^2}{K_X} - K_{XXX}
=0,
\\
-\frac{1}{2G_4} \left(2 G_{3X}G_{4X} + K_X G_{5X}\right)
+ \frac{2K_{XX}G_{3X}}{K_X}
- G_{3XX}
=0,
\qquad
A\, G_{5X}=0
\label{Npp0cond}
\end{gathered}
\end{equation}
at $X=0$, $\cal N$ vanishes at any $u$ hence shock due to the nonlinear effect does not form.
The scalar DBI model coupled to GR ($K = \lambda( 1 - \sqrt{1 + c\,X}), G_3=G_5=0, G_4=\text{const.}$) and also the canonical scalar ($K \propto X, G_3=G_5=0$) with general $G_4(X)$ satisfy this condition, while it is not satisfied in more generic theories.
For example, 
the pure Galileon coupled to GR
\begin{equation}
K= a_2 X,
\quad
G_3 = a_3 X,
\quad
G_4 = c + a_4 X,
\quad
G_5 = a_5 X
\qquad
\left(a_2\neq 0,~c\neq 0\right)
\end{equation}
does not satisfy (\ref{Npp0cond}) and 
makes $\cal N$ a nonzero function of $u$
unless $a_3 = a_5 \,A = 0$.
Also, the DBI Galileon
\begin{equation}
K=a_2  ( 1 - \sqrt{1 + c\,X}),
\quad
G_3 = a_3\log(1+c\,X),
\quad
G_4 = a_4 (1+c\,X)^{-1/2},
\quad
G_5 = a_5 (1+c\,X)^{-3/2}
\end{equation}
with $a_2, a_4, c\neq 0$ does not satisfy (\ref{Npp0cond}) and it results in nonzero $\cal N$ given by
\begin{equation}
{\cal N} =
3  c\,  a_5 \, A \,t_-(u)
+ 
c^2
\left(
\frac{a_3}{2} - \frac{3 a_2 a_5}{8a_4}
\right)
\phi'^2
t_+(u)
+ 3c^2
\left(
\frac{c\,a_2}{4} -\frac{a_3{}^2}{a_4}
\right)
\phi'^3.
\end{equation}

\subsection{Shock formation on two-dimensionally maximally-symmetric dynamical solutions}
\label{Sec:2-dim}

We now focus on another example of simple background solutions, in which spacetime is dynamical and has two-dimensional angular part that is maximally symmetric.
Wave propagation on such solutions have been studied in Refs.~\cite{Izumi:2014loa,Minamitsuji:2015nca} in the context of the Gauss-Bonnet gravity in higher dimensions and a scalar-tensor theory with a scalar field coupled to gravity non-minimally.

We first summarize basics of these solutions in section~\ref{Sec:2D_basics}.
It turns out that gravitational wave on these solutions can be studied without specifying the background explicitly if the wave front is parallel to background symmetry direction. 
We summarize the results for such gravitational wave in section~\ref{Sec:2D_GW}. We will find that the gravitational wave is free from shock formation in this case, which is the same behavior as the gravitational wave on the plane wave background.

For the scalar field wave, we need more careful analysis as shown in section~\ref{Sec:2D_scalar}.
Based on the procedure in this section, we study homogeneous isotropic solutions, which we simply call the FRW universe in this work, in section~\ref{Sec:2D_FRW}. We find that basic properties of the solution shown in Ref.~\cite{Kobayashi:2011nu}, such as propagation speeds, are correctly reproduced from our analysis. Then we study properties of shock formation on this solution.
Last, 
in section~\ref{Sec:2D_sph}
we look at another simple example, that is static spherically-symmetric solutions and waves with spherically-symmetric wave front on them.
We will see that theories other than the scalar DBI model typically suffer from  shock formation, even in the limit to treat the scalar field as a probe field on flat spacetime.

In section~\ref{Sec:Npp} for the plane wave solution, we firstly clarified the structure of geodesics on a characteristic surface and then derived the full expression of the transport equation~(\ref{transport_PP}).
Based on this expression, we gave an estimate~(\ref{s0estimate}) on the time parameter $s$ on the bicharacteristic curve at which the shock formation occurs.
In the following sections, we skip deriving geodesics and the full expression of the transport equation, and evaluate only the coefficient of the nonlinear term ${\cal N}$ in the transport equation.
As we argued based on the general solution of $\Pi(s)$~(\ref{Pisol}), it is guaranteed that shock formation occurs for sufficiently large $\Pi(0)$ when ${\cal N}$ is nonzero, assuming that ${\cal K}^\mu$ is a regular non-vanishing function and  ${\cal M}$ is regular as well.
To follow this argument, we need to know the expressions of all of ${\cal K}^\mu$, ${\cal M}$ and ${\cal N}$ in principle.
However, without knowing the precise expressions of ${\cal K}^\mu$ and ${\cal M}$, we may still say that shock formation based on (\ref{Pisol}) does not occur if ${\cal N}$ identically vanishes, and also we may expect that shock would form if ${\cal N} \neq 0$ assuming ${\cal K}^\mu$ and ${\cal M}$ satisfies the above-mentioned properties. In the following, we take this attitude and check whether ${\cal N}$ vanishes or not, understanding that nonzero ${\cal N}$ suggests shock formation to occur while it is avoided when ${\cal N}=0$.

\subsubsection{Two-dimensionally maximally-symmetric dynamical solutions}
\label{Sec:2D_basics}

Solutions with metric whose two-dimensional spatial part is maximally symmetric can be expressed in general as
\begin{equation}
\begin{aligned}
 ds^2 &= 
f(\tau,\chi)
\left(-d\tau^2 + d\chi^2\right)
 + \rho^2(\tau,\chi)\gamma_{\alpha\beta} dx^\alpha dx^\beta
\equiv
f
\eta_{AB} dx^A dx^B
 + \rho^2 \gamma_{\alpha\beta} dx^\alpha dx^\beta,
\\
  \phi &= \phi(\tau,\chi),
\end{aligned}
  \label{maxsym}
\end{equation}
where
$x^A=\tau,\chi$ and $\gamma_{\alpha\beta}$ is the metric of the two-dimensional subspace with constant curvature $k =0,\pm1$ spanned by $x^\alpha$.
Non-vanishing components of the curvature tensor of this solution ares given by
\begin{equation}
R^{A_1 A_2}_{B_1 B_2}
=
\frac{
\left(\partial_\tau^2 - \partial_\chi^2\right)\log f
}{2f}
\delta^{A_1 A_2}_{B_1 B_2}
\equiv R_1 \delta^{A_1 A_2}_{B_1 B_2},
\quad
R^{A_1 \alpha_2}_{B_1 \beta_2} \equiv
R_2{}^{A_1}_{B_2} \delta^{\alpha_2}_{\beta_2},
\quad
R^{\alpha_1 \alpha_2}_{\beta_1 \beta_2} =
\frac{
k\,f+\rho_{,\tau}^2 - \rho_{,\chi}^2
}{f\rho^2}
\equiv R_3 \delta^{\alpha_1 \alpha_2}_{\beta_1 \beta_2},
\end{equation}
where
\begin{equation}
\left(
R_2{}^A_B 
\right)
=
\frac{1}{2f^2\rho}
\begin{pmatrix}
- \left(
f_{,\tau}\rho_{,\tau}+f_{,\chi}\rho_{,\chi}-2f\rho_{,\tau\tau}
\right)
& 
- 2\left(
f_{,(\tau}\rho_{,\chi)}-f\rho_{,\tau\chi}
\right)
\\
2\left(
f_{,(\tau}\rho_{,\chi)}-f\rho_{,\tau\chi}
\right)
&
f_{,\tau}\rho_{,\tau}+f_{,\chi}\rho_{,\chi}-2f\rho_{,\chi\chi}
\end{pmatrix}.
\end{equation}
Also the nonzero components of $\phi_{|ab}$ are given by
\begin{equation}
\left(\phi_{|AB} \right)
= -\frac1{2f}
\begin{pmatrix}
f_{,\tau}\phi_{,\tau}+f_{,\chi}\phi_{,\chi}-2f\phi_{,\tau\tau}
& 
2\left(
f_{,(\tau}\phi_{,\chi)}-f\phi_{,\tau\chi}
\right)
\\
2\left(
f_{,(\tau}\phi_{,\chi)}-f\phi_{,\tau\chi}
\right)
&
f_{,\tau}\phi_{,\tau}+f_{,\chi}\phi_{,\chi}-2f\phi_{,\chi\chi}
\end{pmatrix}
,
\quad
\phi_{|\alpha\beta} =
\frac{\rho}{f}
\left(
-\rho_{,\tau}\phi_{,\tau} + \rho_{,\chi}\phi_{,\chi}
\right) \gamma_{\alpha\beta}.
\end{equation}
We can evaluate the principal symbol and the quantities appearing in $\cal N$ using these formula.

Below, we focus on wave whose wave front shares the same symmetry as background spacetime, that is, we assume that the wave front is given by a $\chi$-constant surface and $\xi_\mu$ has only $(\tau,\chi)$ components.
For example, plane wave in flat FRW universe and wave with spherically-symmetric wave front around a spherically-symmetric dynamical star fulfill such an assumption.

\begin{figure}[htbp]
\centering
  \begin{minipage}[c]{7.5cm}
    \centering
    \includegraphics[keepaspectratio,width=5.5cm]{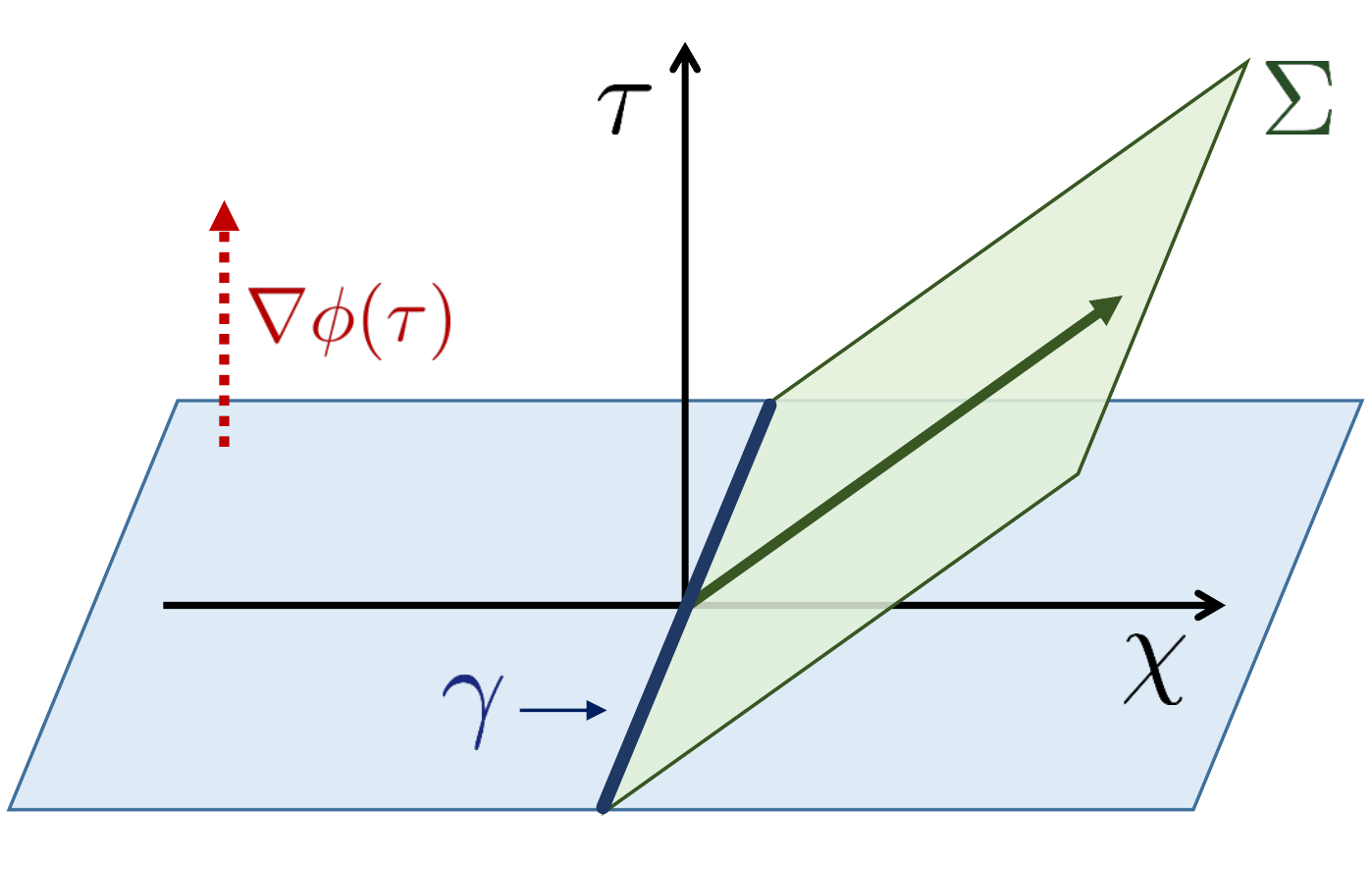}
    \subcaption{FRW universe}
\label{Fig:FRW}
  \end{minipage}
  \begin{minipage}[c]{7.5cm}
    \centering
    \includegraphics[keepaspectratio,width=5.5cm]{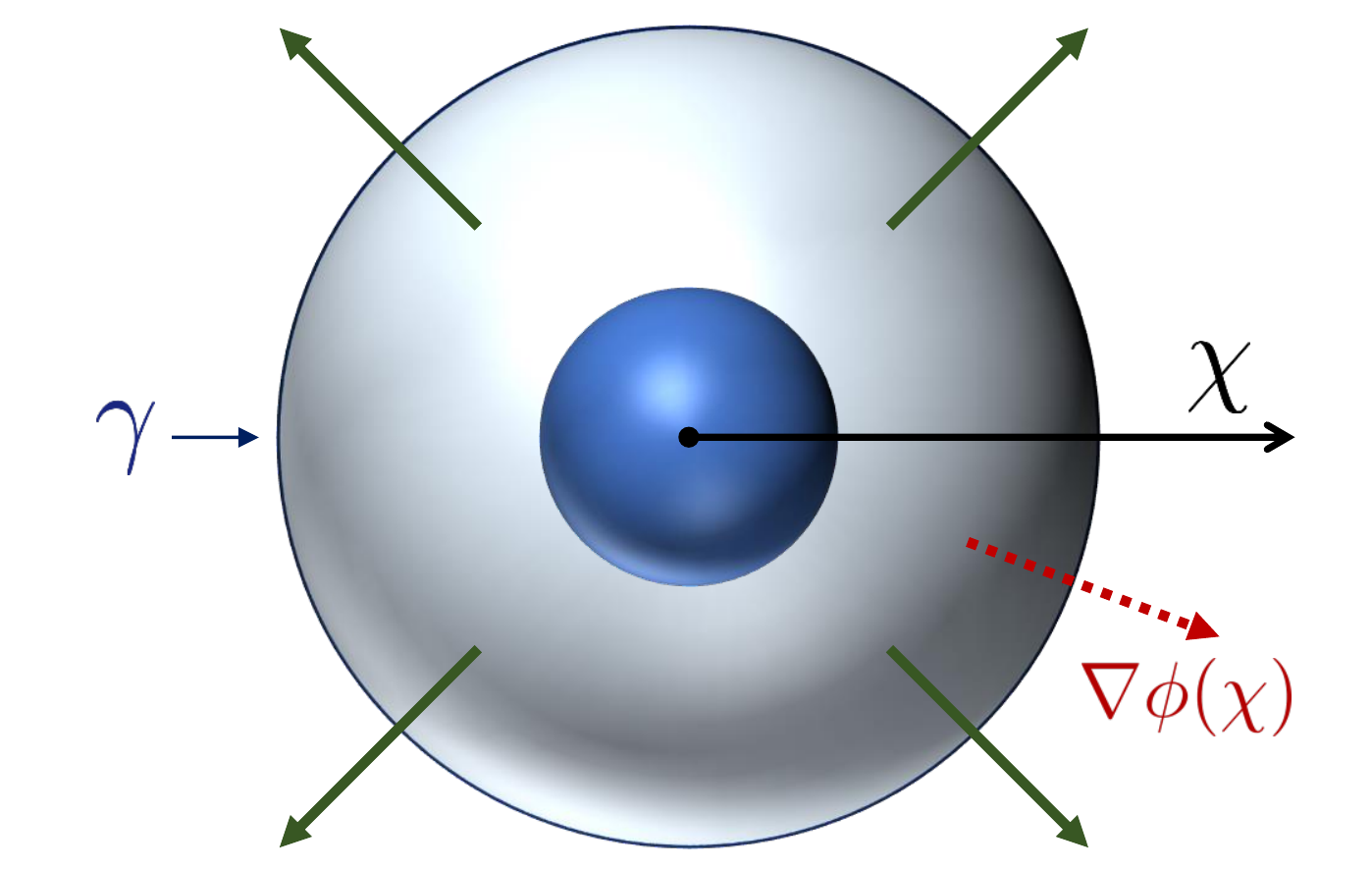}
    \subcaption{Spherically-symmetric static solution}
\label{Fig:sph}
  \end{minipage}
  \caption{Examples of two-dimensionally maximally-symmetric dynamical solutions and wave propagation therein. 
Panel~\subref{Fig:FRW} shows a flat FRW universe and wave with plane-symmetric wave front, and panel~\subref{Fig:sph} shows a spherically-symmetric static background solution and wave with spherically-symmetric wave front.
Gradient of the background scalar field is aligned to the symmetry direction of the background spacetime in these examples, while they may point toward different directions at the level of the general ansatz~(\ref{maxsym}).
\hfill
\label{Fig:2D-MS}}
\end{figure}

This assumption enable us to work out the characteristic analysis for the tensor modes without specifying explicit form of the background solution, and it is the main target in section~\ref{Sec:2D_GW}.
For the scalar mode, to simplify the analysis summarized in section~\ref{Sec:2D_scalar}, we will consider two explicit examples of background solutions that satisfy the above assumption.
The first one is plane wave propagating on homogeneous isotropic solutions (Fig.~\ref{Fig:FRW}), and the second one is wave with spherically-symmetric wave  front on spherically-symmetric static background solutions (Fig.~\ref{Fig:sph}), which are studied in sections~\ref{Sec:2D_FRW} and \ref{Sec:2D_sph}, respectively.

\subsubsection{Gravitational wave}
\label{Sec:2D_GW}

Characteristic surfaces on the background~(\ref{maxsym})
can be found by solving the eigenvalue equation~(\ref{EVeq}) following the procedure of section~\ref{Sec:EoM}.
To accomplish it for gravitational wave, we focus on a vector given by
\begin{equation}
r = (r_{ab}, r_\phi) = (r^\text{(T)}_{ab},0),
\label{rT}
\end{equation}
where $r^\text{(T)}_{ab}$ is a traceless tensor which has components only in the angular directions, that is, $r^\text{(T)}_{AB}=0$.
By explicit calculations, we can check that this vector is actually an eigenvector of $\tilde {\cal R}$ as follows.
The scalar and mixed parts of $\tilde {\cal R}$ vanish for this vector, hence only the metric part shown in appendix~\ref{App:ssHorndeski_trrev} remains nontrivial and is given by
\begin{equation}
\xi_s \xi_t r^\text{(T)}_{qr}\frac{\partial \tilde{\cal G}^4{}_a^b}{\partial g_{qr,st}} 
=
\frac12 G_{4X} \, 
\delta^{B_1 B_2}_{A_1 A_2} \xi^{A_1}\xi_{B_1} \phi^{|A_2}\phi_{|B_2} 
\,
r^\text{(T)}
{}^b_a,
\quad
\xi_s \xi_t r^\text{(T)}_{qr}
\frac{\partial \tilde{\cal G}^5{}_a^b}{\partial g_{qr,st}} 
=
-\frac12 X G_{5X} \,
\delta^{B_1 B_2}_{A_1 A_2} \xi^{A_1}\xi_{B_1} \phi^{|A_2}_{|B_2} \,
r^\text{(T)}{}^b_a,
\end{equation}
hence we have $\tilde {\cal R} \cdot r = \lambda \, r$ with
\begin{equation}
\lambda = 
\frac{1}{G_4-2X G_{4X}}
 \delta^{B_1 B_2}_{A_1 A_2} \xi^{A_1}\xi_{B_1} 
\left(
G_{4X} \phi^{|A_2}\phi_{|B_2} 
- X G_{5X}\phi^{|A_2}_{|B_2}
\right).
\end{equation}
From this expression, we find that 
characteristics are determined by%
\footnote{%
This effective metric for gravitational wave coincides with that of \cite{Bettoni:2016mij}, though $G_5$ was not taken into account in their analysis. Also, the propagation speed obtained from (\ref{gT}) coincides with that derived in \cite{Kobayashi:2011nu} when the background solution is set to the FRW universe.}
\begin{equation}
\left\{
\left(G_4 - 2X G_{4X}\right) \delta^B_A
- \delta^{B B_1}_{A A_1} 
\left(
 G_{4X} \phi^{|A_1} \phi_{|B_1}
- X G_{5X} \phi^{|A_1}_{|B_1}
\right)\right\} \xi^{A} \xi_{B} = 0,
\label{gT}
\end{equation}
where the expression in the curly brackets is the effective metric for gravitational wave on the background~(\ref{maxsym}).

Propagation speed of the gravitational wave can be read out from the effective metric~(\ref{gT}).
When the background solution is homogeneous and isotropic, the propagation speed $c_T=|\xi_\tau / \xi_\chi|$ in the frame for which $\phi=\phi(\tau)$ coincides with that shown in \cite{Kobayashi:2011nu}, which is given by $c_T = \sqrt{{\cal F}_T / {\cal G}_T}$ with
\begin{equation}
{\cal F}_T \equiv 2\bigl(G_4 - X\ddot \phi G_{5X}\bigr),
\qquad
{\cal G}_T \equiv 2 \bigl( G_4 -2X G_{4X} - H X \dot \phi G_{5X} \bigr).
\label{FGT}
\end{equation}

Let us check if shock formation could occur for gravitational wave~(\ref{rT}) by evaluating the coefficient $\cal N$ in the discontinuity transport equation.
The only term that could be nonzero is
$ r^{ab} r_{cd}r_{ef}\frac{\partial}{\partial g_{ef,0}} \frac{\partial {\cal G}^5_{ab}}{\partial g_{cd,00}}$ given by Eq.~(\ref{d2G5dgpdgpp}),
which is proportional to $\delta_{a a_1 a_2 a_3}^{b b_1 b_2 b_3}\xi^a \xi_b {r^\text{(T)}}^{a_1}_{b_1}{r^\text{(T)}}^{a_2}_{b_2}{r^\text{(T)}}^{a_3}_{b_3}$.
However, 
$r^\text{(T)}$
is a traceless tensor living in the two-dimensional angular part of the spacetime and then Eq.~(\ref{d2G5dgpdgpp}), which involves three $r^\text{(T)}$ tensors contracted with a single generalized Kronecker delta, identically vanishes. Hence $\cal N$ is zero and shock formation does not occur for gravitational wave on the background~(\ref{maxsym}).

\subsubsection{Scalar field wave}
\label{Sec:2D_scalar}

Next, we study wave that involves the scalar field and examine if it suffers from shock formation.
Recent studies about a probe scalar field on flat background clarified that shock generically forms in the Horndeski theory while it is avoided in the DBI-Galileon theory \cite{Babichev:2016hys,Mukohyama:2016ipl,deRham:2016ged}.
We re-examine these results using our formalism for transport of discontinuity in second derivatives of fields.

The first step is to find a characteristic surface and an eigenmode corresponding to scalar field wave.
On the background~(\ref{maxsym}) and for wave that inherits the symmetry of the background, we may assume the eigenvector $r$ has a structure given by 
\begin{equation}
r = (r_{ab},r_\phi),
\qquad
r_{AB} = r_\eta \, \eta_{AB}
+  \frac{2r_\gamma}{\xi^2} \xi_A \xi_B,
\quad
r_{A\alpha} = 0,
\quad
r_{\alpha\beta} = r_\gamma\, \gamma_{\alpha\beta},
\label{rS}
\end{equation}
where $r_\phi$, $r_\eta$ and $r_\gamma$ are functions of $\tau,\chi$.
The term involving $\xi_A \xi_B$ is the gauge part added so that $r$ satisfies the transverse condition~(\ref{transverse}).

For the ansatz~(\ref{rS}), 
we may parameterize the eigenvector by $(r_\eta, r_\gamma, r_\phi)$, 
and then the eigenvalue equation~(\ref{EVeq}) should have three eigenvalues in general.
Plugging the ansatz~(\ref{rS}) into (\ref{EVeq}), we can confirm that two eigenvalues are proportional to $\xi^2$ and another one is a nontrivial function of $\xi$. This nontrivial eigenvalue corresponds to a physical scalar mode propagating on the characteristic surface, and the propagation speed is given  in terms of $\xi$ as $c_S=|\xi_\tau / \xi_\chi|$.

Expression of the nontrivial eigenvalue and eigenvector are generically lengthy and not illuminating.
There are some cases in which their expressions become simple and $\cal N$ can be calculated explicitly.
We examine such cases realized for simple background solutions below.

\subsubsection{Shock formation in FRW universe}
\label{Sec:2D_FRW}

Based on the ansatz~(\ref{maxsym}), a solution describing homogeneous and isotropic universe is realized by
\begin{equation}
f = \rho^2 = a(\tau)^2, 
\quad
\phi = \phi(\tau),
\label{FRWansatz}
\end{equation}
where $\tau$ in this case is the conformal time from which a standard time coordinate may be defined by $dt = a(\tau)d\tau$.
We will use the Hubble parameter in terms of $t$, $H(t) \equiv a_{,t}/a$, and also the notation $\dot \phi \equiv \phi_{,t}$ and $X = \dot\phi^2/2$  below.
$tt$ and $\chi\chi$ components of the background equation of motion~(\ref{geq}) give the modified Friedmann equations
\begin{equation}
- K + 2X K_{X} 
+ 6HX\dot \phi G_{3X}
-6H^2 \left( G_4 - 4XG_{4X} + 4 X^2 G_{4XX} \right)
+ 2H^3 X \dot\phi \left(5G_{5X} + 2X G_{5XX}\right) = 0,
\label{Feq1}
\end{equation}
\begin{multline}
{\cal E} \equiv
K + 2\bigl(3H^2 + 2\dot H\bigr)\left(G_4 - 2X G_{4X}\right)
-4H \bigl(H^2 + \dot H \bigr) X \dot\phi G_{5X}
\\
-2 \ddot\phi \left[
X G_{3X} 
+2 H\dot\phi\left(G_{4X}+2X G_{4XX}\right)
+H^2 X\left(3G_{5X}+2XG_{5XX}\right)
\right]
=0.
\label{Feq2}
\end{multline}

Properties of the background solution~(\ref{FRWansatz}) and its perturbations are studied by Ref.~\cite{Kobayashi:2011nu}.
Particularly, propagation speed of the scalar mode is given by $c_S = \sqrt{{\cal F}_S / {\cal G}_S}$, where
\begin{equation}
\Sigma \equiv X {\cal E}_{,X} + \frac12 H {\cal E}_{,H},
\quad
\Theta \equiv -\frac16 {\cal E}_{,H},
\quad
{\cal F}_S \equiv \frac1a \frac{d}{dt} \left( \frac{a}{\Theta}{\cal G}_T^2\right) - {\cal F}_T,
\quad
{\cal G}_S \equiv \frac{\Sigma}{\Theta^2} {\cal G}_T^2 + 3 {\cal G}_T.
\end{equation}
This $c_S$ coincides with the propagation speed obtained from the eigenvalue obtained above once the Friedmann equations~(\ref{Feq1}), (\ref{Feq2}) are imposed.

Expressions of the propagation speed and the eigenvector $r$ become lengthy in general.
One exception is the case discussed in section~\ref{Sec:N=0}, where
 $G_{3,4,5}$ are constants while $K(X)$ is kept general. In this case it follows that
\begin{equation}
(r_{ab},r_\phi) = (0,r_\phi),
\qquad
c_S^2 = \frac{K_X}{K_X + 2X K_{XX}},
\label{cS2_K}
\end{equation}
and also we can check that ${\cal N}=0$ is achieved by the scalar DBI model coupled to GR, as we have seen in section~\ref{Sec:N=0}.

For general $K(X)$ and $G_n(X)$, the propagation speed and the eigenvector takes more complicated form.
A case that gives relatively simple $c_S$ and $r$ is when
\begin{equation}
K = -\lambda\sqrt{1 + c\,X},
\quad
G_4 = a_4 \sqrt{1+ \tilde c\,X},
\quad
G_{3,5} = 0,
\label{KG4model}
\end{equation}
where $\lambda, c,a_4, \tilde c$ are constants that satisfies $c\neq \tilde c$.\footnote{When $c=\tilde c$, ${\cal F}_S$ and ${\cal G}_S$ vanishes and then the quadratic Lagrangian for scalar perturbation shown in \cite{Kobayashi:2011nu} vanishes identically, which indicates that the theory is in the strong coupling regime.}
In this case, using the Friedmann equations (\ref{Feq1}), (\ref{Feq2}) we can simplify the propagation speed and the eigenvector as
\begin{equation}
c_S{}^2 = 1 + c\, X,
\qquad
\left(r_\eta, r_\gamma, r_\phi\right) \propto
\left(
\frac{\tilde c \left(1+c\, X\right)^{3/4}}{c \, \dot\phi},
\frac{\tilde c\,  \dot\phi}{\left(1+c\,X\right)^{1/4}},
\sqrt{\frac{6a_4}{\lambda}} \left(1+\tilde c \, X\right)^{1/4}
\right),
\end{equation}
and we can check that $\cal N$ identically vanishes in this case.
Other choices of $G_4$ such as $G_4 \propto \left(1+X\right)^{-1/2}$ typically result in ${\cal N} \neq 0$. Hence, it seems that the choice (\ref{KG4model}) is special among other choices of $K(X)$ and $G_4(X)$ in the sense that it leads to a cosmological solution free from shock formation.

For other choices of $K(X)$ and $G_n(X)$, $\cal N$ becomes nonzero generically.
For example, a choice given by $K=G_5 = 0$ with constant $G_4$ and general $G_3(X)$ results in%
\footnote{To deal with this case, we need to solve $\tilde P\cdot r= 0$ directly rather than Eq.~(\ref{P0def}) which is normalized with respect to $K_X$.}
\begin{equation}
c_S{}^2 = \frac{G_{3X}\left(23 G_{3X} + 16X G_{3XX}\right)}{3\left(5G_{3X}+ 4XG_{3XX}\right)^2},
\quad
\left( r_\eta, r_\gamma, r_\phi \right) \propto
\Bigl(
\left(c_S{}^2+1\right)G_{3X},
\left(c_S{}^2 - 1\right)X G_{3X},
\left(c_S{}^2 - 1\right)G_4
\Bigr).
\end{equation}
where Eqs.~(\ref{Feq1}) and (\ref{Feq2}) are used to simplify the expressions.
Using them, $\cal N$ is calculated as
\begin{multline}
{\cal N} \propto
\dot\phi X \Bigl[
-55G_{3X}{}^3 - 120 X^3 G_{3XX}{}^3 
+ 4X^2 G_{3X} G_{3XX}\left(
-55 G_{3XX} + 14 X G_{3XXX}
\right)
\\
+ 2 X G_{3X}{}^2\left(
-59 G_{3XX} + 38 X G_{3XXX}
\right)
\Bigr].
\label{NG3}
\end{multline}
For a generic choice of $G_3$, this expression does not vanish unless $\dot\phi=0$ and hence shock would form. In principle, we can find $G_3(X)$ that realizes ${\cal N}=0$ by equating Eq.~(\ref{NG3}) with zero and solving it as an ODE for $G_3(X)$. It seems difficult to obtain a closed form of $G_3(X)$ obtained in this way. Also it can be confirmed that $\cal N$ does not vanish for some simple choices such as $G_3\propto \log(1+c\,X)$ and $(1+c\, X)^n$ with integer $n$.

\subsubsection{Shock formation on spherically-symmetric static solutions}
\label{Sec:2D_sph}

Another example of a simple solution described by~(\ref{maxsym}) is
a spherically-symmetric static solution
\begin{equation}
f = f(\chi),
\quad
\rho=\rho(\chi),
\quad
\phi = \phi(\chi),
\label{functions_sph}
\end{equation}
where $\gamma_{\alpha\beta}$ is taken as the metric on $S^2$ with unit radius.
The metric functions and the scalar field in Eq.~\ref{functions_sph} are fixed by integrating $tt$ and $\theta\theta$ components of the metric equation~(\ref{geq}) and the scalar equation~(\ref{Jeq}) regarding them as second-order ODEs on $f(\chi)$, $\rho(\chi)$ and $\phi(\chi)$.
The $rr$ component of the metric equation is given by $f,\rho,\phi$ and their first derivatives, and it can be regarded as a constraint on the variables.
Below, we use the above second-order equations of motion to eliminate second derivatives of background fields from various expressions.

An eigenvector can be parameterized as (\ref{rS}) even in this case.
Also, as argued in section~\ref{Sec:N=0}, the {\it k}-essence coupled with GR is an example for which various quantities are easily derived.
In this case, the eigenvector is given by $(r_\eta,r_\gamma, r_\phi)\propto(0,0,1)$, and the propagation speed is given by
\begin{equation}
c_S{}^2 = \frac{K_{X} + 2XK_{XX}}{K_X},
\label{cS2Ksph}
\end{equation}
which is the reciprocal of the propagation speed in the FRW universe~\cite{ArmendarizPicon:2005nz}.
$\cal N$ in this case is given by~(\ref{N_K}), and the scalar DBI model turns out to be the unique theory 
other than the canonical scalar field
that makes $\cal N$ vanishing.

$\cal N$ does not vanish for a generic choice of $G_n(X)$, as it was the case of the FRW universe background. In that case, there was a nontrivial example~(\ref{KG4model}) that realizes ${\cal N}=0$.
Let us check if the shock formation could occur for this choice (\ref{KG4model}) when the background is spherically symmetric and static.
In this case, using the background equations we can show that
\begin{equation}
\begin{aligned}
c_S{}^2 &= \frac{1}{1+c\, X},
\qquad
\left(r_\eta,r_\gamma,r_\phi\right)\propto
\left(
\tilde r_\eta, -2 c\tilde c a_4 \phi' X \sqrt{1+\tilde c \, X} \rho'^2 ,
2c a_4 X \left(1+\tilde c \, X\right)^{3/2}f\rho\rho' 
\right),
\\
\frac{\tilde r_\eta}{\tilde c \, \phi'} &\equiv 
f(1+c \, X)^{1/2}(1+\tilde c \, X)^{3/2}\left(
\lambda \sqrt{1 + \tilde c \, X }\rho^2 - 2 a_4 \sqrt{1+c\,X}
\right)
-2 a_4 (1+c\, X)
\sqrt{1+\tilde c\,X}
\rho'^2.
\end{aligned}
\end{equation}
We can check that propagation speeds of the modes propagating in the positive and negative $r$ directions coincide with each other.
Using these expressions we can confirm that $\cal N$ does not vanish in this case.
Hence, the theory with (\ref{KG4model}) suffers from shock formation on a spherically-symmetric static background realized within this theory, contrarily to the case of the FRW universe background. 
Other choices such as $K=G_5=0$ with constant $G_4$ and generic $G_3$ give nonzero $\cal N$.

We can also check what happens when the scalar field is treated as a probe field and its gravitational backreaction is neglected. In this case, the background spacetime becomes flat ($f=1,\rho=\chi$), and the background scalar field is determined by the scalar equation~(\ref{Jeq}).
The principal symbol and $\cal N$ are given by their pure scalar part 
evaluated with a flat metric,
and using these expressions we can find a characteristic surface and check if shock formation could occur on it.
For the {\it k}-essence model, the propagation speed is given by (\ref{cS2Ksph}) and ${\cal N}=0$ is realized only for the scalar DBI model. For other models in which only one of the arbitrary functions $G_{3,4}(X)$ is non-zero, propagation speeds are given by%
\footnote{Instead of deriving propagation speed on spherically-symmetric static background, we could derive it for simple wave solutions by imposing $\phi_{,\tau\tau}\phi_{,\chi\chi} - \phi_{,\tau\chi}{}^2=0$ and taking the ratio of eigenvalues of the kinetic matrix of the theory following \cite{deRham:2016ged}. This method gives $c_S{}^2 = n\, G_{nX} / \bigl(n\,G_{nX} + 2X G_{nXX}\bigr)$ in the pure $G_n$ model for $n=2,3,4$ regarding $G_2=K$, and the kinetic matrix vanishes identically in the pure $G_5$ model. This propagation speed is the one measured in the frame comoving with $\phi$ where the gradient of $\phi$ is aligned to the time coordinate. When the gradient of $\phi$ is spacelike, the propagation speed is given by its reciprocal as shown in Eq.~(\ref{cS2Ksph}).}
\begin{align}
\text{pure $G_3$ model}:&\quad c_S^2 = \frac43 \frac{G_{3X} + X G_{3XX}}{G_{3X}}\\
\text{pure $G_4$ model}:&\quad c_S^2 = \frac{3G_{4XX} + 2 X G_{4XXX}}{3G_{4XX}}.
\end{align}
In the pure $G_5$ model, the scalar field equation becomes trivial for any static configuration of $\phi(\chi)$. Hence, a discontinuity in second derivative added at the initial time can remain static
unlikely to the other cases where such a discontinuity propagates at finite speed.
In the pure $G_3$ 
theory,
$\cal N$ does not vanish for a generic choice of the arbitrary function $G_3(X)$. We can still find some exceptions for which $\cal N$ vanishes on nontrivial backgrounds, such as $G_4 \propto \sqrt{c+X}$ and $G_4 \propto (1+X)^2$, where the former corresponds to the example (\ref{KG4model}) in which the metric part of the theory is taken into account.

\section{Summary and discussion}
\label{Sec:summary}

In this paper, 
we studied properties of wave propagation and causality defined by it, and also of the shock formation process in scalar-tensor theories.
For these studies we 
especially focused on 
the Horndeski theory, which is the most general scalar-tensor theory with one scalar field whose Euler-Lagrange equation is up to second order in derivatives, and its generalization with two scalar fields developed in \cite{Ohashi:2015fma}, which we called the bi-Horndeski theory in this work.
The latter theory reduces to the former and also to the generalized multi-Galileon theory by setting the arbitrary functions appropriately, as shown in appendix~\ref{App:toOtherTheories}.

About the wave propagation, in section~\ref{Sec:characteristics} 
we showed that characteristic surfaces in the bi-Horndeski theory are not null in general, that is,
propagation speeds of gravitational wave and scalar field wave can be faster or slower than the light speed defined by physical metric. We also showed that a Killing horizon in this theory is not a causal edge in general, hence there may be superluminal modes that can propagate across a Killing horizon.
However, it turned out that such propagation across a Killing horizon is prohibited if the scalar fields share the Killing symmetry on the horizon. These properties may be regarded as natural generalizations of those found for scalar-tensor theories with a non-minimally coupled scalar field studied in Ref.~\cite{Minamitsuji:2015nca}.
In \cite{Barausse:2011pu} the causal structure in Einstein-aether and Ho\v{r}ava--Lifshitz theories were studied, and it was noticed that the universal horizon, which acts as the causal edge for any mode, is orthogonal to the background vector field in those theories. This property seems similar to that found in this work if the gradient of the background scalar field is identified as such a vector field. It would be interesting to examine this possible correspondence and its implications.

We also studied wave propagation and causal structure on a nontrivial background taking the plane wave solution constructed by Ref.~\cite{Babichev:2012qs} as an example.
This exact solution has a null direction to which waves of metric and scalar field propagate. 
We found that the causal cones for metric and scalar perturbations on this background form a nested set of cones that are aligned along the background null direction, as shown in Fig.~\ref{Fig:disc} of section~\ref{Sec:planewave}.
As long as a the background solution is regular, there will be a characteristic cone with the largest but finite opening, which will define the causality on this background solution.

In section~\ref{Sec:shock}, we focused on another phenomenon associated with wave propagation, that is, the shock formation caused by nonlinear self interaction of waves.
For this purpose, we generalized the formalism of \cite{Reall:2014sla}  for the transport of discontinuity in second derivatives of the metric in Lovelock theories to the Horndeski theory with shift symmetry in the scalar field.
It turned out that the {\it k}-essence coupled to GR can be studied without specifying the background solution using this formalism, and the results are summarized in section~\ref{Sec:N=0}.
For this class of theory, we found that shock may form for scalar field wave while not for gravitational wave.
We have also shown that, 
Among the theories described by the {\it k}-essence, the scalar DBI model besides the canonical scalar field turned out to be the unique nontrivial theory that is free from shock formation. 
Such a property was found by previous works~\cite{Babichev:2016hys,Mukohyama:2016ipl,deRham:2016ged,boillat2006recent,*1959PThPS...9...69T,*Whitham_chap17,*Barbashov:1966frq,*doi:10.1142/9789812708908_0003,*Deser:1998wv} for simple waves of a probe scalar field on flat spacetime. 
Our result implies that this property persists even on nontrivial background that involves non-vanishing scalar field and spacetime curvature.

To study
properties of the shock formation on various backgrounds,
we focused on the plane wave solution and also dynamical solutions in which two-dimensional angular part of the spacetime is maximally symmetric
in sections~\ref{Sec:Npp} and \ref{Sec:2-dim}, respectively.
On the plane wave background 
studied 
in section~\ref{Sec:Npp}, we found that shock formation does not occur for wave propagating in the background null direction, while it occurs for the scalar field wave propagating in the opposite direction.
The gravitational wave, however, does not suffer from shock formation even when it is propagating in this direction.
On the two-dimensionally maximally-symmetric dynamical solutions,
we studied waves whose wave front is aligned to the background symmetry direction in section~\ref{Sec:2-dim}.
It turned out that shock formation occurs for scalar field wave in general, while it does not for gravitational wave. Taking some typical background solutions such as the FRW solutions and spherically-symmetric static solutions, we examined conditions for shock formation. We found that theories other than the scalar DBI model
generically suffer from shock formation on these backgrounds.

In any example studied in this work, the propagation modes corresponding to gravitational wave were always free from shock formation. If this feature persists on any background solution, we may conclude that the gravitational wave in this class of theory is more well-behaved compared to the scalar field, which typically suffers from shock formation. On the other hand, if we could find some background solutions or nontrivial shape of wavefront for which shock formation occurs even for gravitational wave, 
it 
might give interesting implications to
gravitational wave observations which was realized recently~\cite{Abbott:2016blz}. Such studies on shock formation in gravitational wave with nontrivial wave front on nontrivial background would be one possible future extension of this work.

There are some other open issues and future directions related to this work.
First one is about the time evolution after shock formation.
When a shock forms, the derivatives of fields diverge there and the theory would break down unless higher-order corrections to the theory ameliorate the singular behavior, or unless we accept such a field configuration as a weak solution of the theory.
Also, a shock formation would correspond to a naked singularity formation unless it is covered by an event horizon or resolved by corrections to the theory, as argued in \cite{Reall:2014sla,Mukohyama:2016ipl}.
It would be interesting to study time evolution after shock formation by taking higher-order corrections into account or by regarding a shock as a weak solution.
Next one is about extending the analysis in this work to more general theories.
In this work
we used the shift-symmetric Horndeski theory for a single scalar field in the analysis of shock formation
just for simplicity, 
although 
it is desirable to 
conduct the study using more broader class of theories.
It would be fruitful to extend the shock formation analysis, and also the analysis on wave propagation, for the bi-Horndeski theory and other more general theories mentioned in section~\ref{Sec:intro}.
Last, the phenomena discussed in this work could be important in time evolution that involves nonlinear dynamics of gravity and scalar fields. 
Investigations on wave propagation and shock formation in
scalar-tensor theories with such nonlinear dynamics
might provide useful implications in the context of cosmology and astrophysics.

\begin{acknowledgments}
We thank 
Tsutomu Kobayashi,
Hideo Kodama,
Kazuya Koyama,
Hayato Motohashi,
Shinji Mukohyama,
Ryo Namba,
Harvey S.\ Reall,
Alexander Vikman,
Yota Watanabe
and
Masahide Yamaguchi
for useful discussions and comments.
The work of N.T.\ was supported in part by JSPS KAKENHI Grant Numbers 15H03658 and 16H06932.
\end{acknowledgments}

\appendix

\section{Field equation of the most general bi-scalar-tensor theory}
\label{App:BiHorndeskiEoM}

In this appendix, we present the equations of motion for the most general bi-scalar-tensor theory. See Ref.~\cite{Ohashi:2015fma} for more details.

\subsection{Gravitational field equations}

The gravitational field equations for the most general bi-scalar-tensor theory are given by
\begin{align}
\calG{}^{a}_{b}&=
A\delta^{a}_{b}
+\left( -2 \mathcal{F}_{,I,J} -4\mathcal{W}_{,I,J} + A_{,IJ}
+2 D_{IKJ,L} X^{KL}
-16 E_{KIMNJ,L}X^{KL}X^{MN}
-16 J_{K[I,L],J} X^{KL}
\right)\phi^{(I|a} \phi^{J)}_{|b}
\notag \\
&
+\left[
-2 \mathcal{F}_{,I} -4 \mathcal{W}_{,I}
+2\left( D_{JKI} + 8 J_{J[K,I]}\right) X^{JK}
-8 E_{JKLMI} X^{JK} X^{LM}
\right] \delta^{ac}_{bd}\phi^{I |d}_{|c}
\notag \\&
+ D_{IJK}\delta^{ace}_{bdf}\phi^I_{|c} \phi^{J |d}\phi^{K|f}_{|e}
+ E_{IJKLM}\delta^{aceg}_{bdfh}\phi^{I}_{|c} \phi^{J |d}\phi^K_{|e} \phi^{L|f}\phi^{M |h}_{|g}
+\left( \frac{1}{2}\mathcal{F}+\mathcal{W}\right) \delta^{ace}_{bdf}R_{ce}^{~~\,df}
+ \mathcal{F}_{,IJ} 
\delta^{ace}_{bdf}\phi^{I |d}_{|c} \phi^{J|f}_{|e}
\notag \\&
+
J_{IJ}\delta^{aceg}_{bdfh}\phi^I_{|c}\phi^{J |d}R_{eg}^{~~\,fh}
+ 2 J_{IJ,KL}\delta^{aceg}_{bdfh}\phi^I_{|c} \phi^{J|d}\phi^{K |f}_{|e}\phi^{L|h}_{|g}
+K_{I}\delta^{aceg}_{bdfh}\phi^{I|d}_{|c}R_{eg}^{~~\,fh}
+\frac23 K_{I,JK}
\delta^{aceg}_{bdfh}\phi^{I|d}_{|c} \phi^{J|f}_{|e}\phi^{K|h}_{|g} ,
\label{EoM}
\end{align}
where $A, D_{IJK}, E_{IJKLM}, J_{IJ}, K_{I}$
are arbitrary functions of $\phi^I$ and $X^{IJ}$, $\mathcal{W}$ are arbitrary function of $\phi$.
These functions are subject to
\begin{equation}
 D_{IJK}=D_{JIK},  
\qquad
E_{IJKLM} =-E_{KJILM}=-E_{ILKJM} =E_{JILKM},
\qquad
 J_{IJ} = J_{JI},
\end{equation}
and $\mathcal{F}$ is defined as
\begin{align}
\mathcal{F}=\int dX^{IJ}\left( 2J_{IJ}-2K_{I,J}+J_{KI,JL}X^{KL}\right) .
\end{align}

\subsection{Scalar field equations}

The scalar field equations for the most general bi-scalar-tensor theory are given by
\begin{align}
{\cal E}_I = 2 {\cal Q}_I + \delta^{cegl}_{bdhm} \left(
             - \gamma_{JIKL} \phi^{K|b} \phi^{J}_{|c} \phi^{L|d}_{|e} 
             R_{gl}^{~~\,hm} 
             + \frac23 \sigma_{JIKLMN} \phi^{J}_{|c} \phi^{M|b}
             \phi^{K|d}_{|e} \phi^{L|h}_{|g} \phi^{N|m}_{|l}
             \right),
\end{align}
where
\begin{equation}
\gamma_{IJKL} = -2J_{IJ,KL} + H_{IKJL},
\qquad
\sigma_{IJKLMN} = H_{IJKL,MN} - H_{IMNL,JK}.
\label{gammadef}
\end{equation}
The explicit form of ${\cal Q}_I$ is given by
\begin{align}
{\cal Q}_{I}\equiv {\cal Q}_I^{(A)}+{\cal Q}_I^{(B)}+{\cal Q}_I^{(C)}+{\cal Q}_I^{(D)}+{\cal Q}_I^{(E)}+{\cal Q}_I^{(G)}+{\cal Q}_I^{(H)}+{\cal Q}_I^{(I)}+{\cal Q}_I^{(J)}+{\cal Q}_I^{(K)}+{\cal Q}_I^{(L)},
\end{align}
where
\begin{align}
\mathcal{Q}_I^{(A)}&=
A_{,I}, 
\\
\mathcal{Q}_I^{(B)}&=-2 B_{IJ,K} X^{JK}
-B_{IJ,KL}\phi^{K}_{|c}\phi^{L}{}^{|cb}\phi^{J}_{|b}
+ B_{IJ} \phi^J_{|b}{}^{|b},\\
\mathcal{Q}_I^{(C)}&= 
C_{J,I} \phi^J_{|c}{}^{|c},
\\
\mathcal{Q}_I^{(D)} &=
D_{JKL,I}\delta^{ce}_{df}\phi^J_{|c} \phi^K{}^{|d}\phi^L_{|e} {}^{|f}
+2D_{IJK,L}X^{JL}\phi^{K}_{|c}{}^{|c}
+D_{IJK,L}\phi^{J}{}^{|c}\phi^{K}_{|cd}\phi^L{}^{|d} 
\notag \\
& \quad
+D_{IJK,LM} \delta^{ce}_{bf}\phi^{L}_{|d}\phi^{M}{}^{|db}\phi^J_{|c} \phi^K_{|e}{}^{|f}  
-D_{IJK}\delta^{ce}_{bf}\phi^J_{|c}{}^{|b} \phi^K_{|e}{}^{|f}
-\frac{1}{2}D_{IJK}\delta^{ce}_{bf}\phi^J_{|c} \phi^K{}^{|l}R_{el}^{~~\,bf}, 
\\
\mathcal{Q}_I^{(E)} &=
E_{JKLMN,I}\delta^{ceg}_{dfh}\phi^J_{|c} \phi^K{}^{|d}\phi^L_{|e} \phi^M{}^{|f}\phi^N_{|g}{}^{|h}
-2E_{LJKIM,N}\delta^{ceg}_{dfh}\phi^L_{|c} \phi^J{}^{|d}\phi^K_{|e} \phi^N{}^{|f}\phi^M_{|g}{}^{|h}
\notag \\ &\quad
-2 E_{LJKIM,NO}\delta^{ceg}_{bfh}\phi^{N}_{|l}\phi^{O}{}{}^{|lb}\phi^L_{|c} \phi^K_{|e} \phi^J{}^{|f}\phi^M_{|g}{}^{|h}
+4E_{LJKIM}\delta^{ceg}_{bfh}\phi^L_{|c}{}^{|b} \phi^K_{|e} \phi^J{}^{|f}\phi^M_{|g} {}^{|h} 
\notag \\ &\quad
+ E_{LJKIM}\delta^{ceg}_{bfh}\phi^L_{|c} \phi^K_{|e} \phi^J{}^{|f}\phi^M{}^{|l}R_{gl}^{~\,bh},
\\
\mathcal{Q}_I^{(G)}&=
G_{JK,I}
\delta^{ce}_{df}\phi^J_{|c}{}^{|d} \phi^K_{|e}{}^{|f}, 
\\
\mathcal{Q}_I^{(H)} 
&=
H_{JKLM,I} \delta^{ceg}_{dfh}\phi^J_{|c}\phi^K{}^{|d} \phi^L_{|e}{}^{|f}\phi^M_{|g}{}^{|h} 
+2H_{IJKL,M}X^{JM}\delta^{eg}_{fh} \phi^K_{|e}{}^{|f}\phi^L_{|g}{}^{|h}
+2
H_{IJKL,M}\delta^{cg}_{fh}\phi^J_{|c} \phi^M{}^{|e}\phi^K_{|e}{}^{|f}\phi^L_{|g}{}^{|h}
\notag \\& \quad
-
H_{IJKL}\delta^{ceg}_{bfh}\phi^J_{|c}{}^{|b} \phi^K_{|e}{}^{|f}\phi^L_{|g}{}^{|h}
-
H_{IJKL}\delta^{ceg}_{bfh}\phi^J_{|c} \phi^K_{|e}{}^{|f}\phi^L{}^{|l}R_{gl}^{~~\,bh}
+H_{IJKL,MN}\delta^{ceg}_{bfh}\phi^{M}_{|l}\phi^{N}{}^{|lb}\phi^J_{|c} \phi^K_{|e}{}^{|f}\phi^L_{|g}{}^{|h},
\\
\mathcal{Q}_I^{(I)}&=
I_{,I}
\delta^{ce}_{df}R_{ce}^{~~\,df}, 
\\
\mathcal{Q}_I^{(J)} 
&=
J_{JK,I}\delta^{ceg}_{dfh}\phi^J_{|c}\phi^K{}^{|d}R_{eg}^{~~\,fh}
+2J_{IJ,K}X^{JK}\delta^{eg}_{fh}R_{eg}^{~~\,fh}
+2
J_{IJ,K}
\delta^{cg}_{fh}\phi^J_{|c}\phi^{K}{}^{|e}R_{eg}^{~~\,fh} 
\notag \\&\quad
+J_{IJ,KL}\delta^{ceg}_{bfh}\phi^{K}_{|d}\phi^{L}{}^{|db}\phi^J_{|c}R_{eg}^{~~\,fh}
-
J_{IJ}\delta^{ceg}_{bfh}\phi^J_{|c}{}^{|b}R_{eg}^{~~\,fh},
\\
\mathcal{Q}_I^{(K)}&=
K_{J,I}
\delta^{ceg}_{dfh}\phi^J_{|c}{}^{|d}R_{eg}^{~~\,fh}
-\frac12 K_{J,IK}
\delta^{cegl}_{dfhm}\phi^K_{|c}{}^{|d}\phi^J_{|e}{}^{|f}R_{gl}^{~~\,hm} 
-\frac{1}{8}K_I\delta_{dfhm}^{cegl}R_{ce}^{~~\,df}R_{gl}^{~~\,hm},
\\
\mathcal{Q}_I^{(L)}&=
L_{JKL,I}\delta^{ceg}_{dfh}\phi^J_{|c}{}^{|d}\phi^K_{|e}{}^{|f}\phi^L_{|g}{}^{|h} 
-\frac14 L_{LJK,IM}
\delta^{cegl}_{dfhm}\phi^M_{|c}{}^{|d}\phi^L_{|e}{}^{|f}\phi^J_{|g}{}^{|h}\phi^K_{|l}{}^{|m},
\end{align}
and the coefficient functions appearing in ${\cal Q}_I$ are defined as
\begin{align}
B_{IJ}&=
- 2  \left( \mathcal{F} + 2 \mathcal{W} \right)_{,I,J} 
+ A_{,IJ}
+ 2 D_{(I|K|J),L} X^{KL}
- 16 E_{K(I|MN|J),L} X^{KL} X^{MN}
- 8 \left( J_{K(I,J),L} - J_{KL,I,J} \right) X^{KL},
\label{coeff_top}
\\
C_{I}&=
-2\left(\mathcal{F} + 2\mathcal{W}\right)_{,I}
+ 2 \left( D_{JKI} + 8 J_{J[K,I]} \right) X^{JK}
- 8 E_{JKLMI} X^{JK} X^{LM}, 
\\
G_{IJ}
&=
2J_{IJ} - 2 K_{(I,J)} + 4 J_{K(I,J)L} X^{KL},
\\
H_{IJKL} &= 2 J_{IJ,KL},
\\
K_{[I,J]} &= -2 J_{K[I,J]L} X^{KL},
\\
K_{I,JK} &= K_{J,IK},
\\
L_{IJK} &= \frac23 K_{(I,JK)},
\\
I&=\frac{1}{2}\mathcal{F}+\mathcal{W}.
\label{coeff_bottom}
\end{align}

\section{Bi-scalar-tensor theory to generalized multi-Galileon and Horndeski theory}
\label{App:toOtherTheories}

In this section,
we show explicitly the relationship 
of the most general bi-scalar-tensor theory  
with
the generalized multi-Galileon theory and Horndeski theory.

\subsection{Generalized multi-Galileon theory}
\label{App:toMultiGalileon}

The Lagrangian of the generalized multi-Galileon theory is given as
\begin{align}
\frac{1}{\sqrt{-g}}\mathcal{L}&=G_2-G_3{}_I \phi_I{}_{|a}^{|a} 
+G_4R+G_4{}_{,IJ}\left( \phi^{I|a}_{|a} \phi^{J|b}_{|b} -\phi^I_{|ab} \phi^{J|ab}\right) \notag \\
                  &\quad
+G_{5I}G^{ab}\phi^I_{|ab}-\frac{1}{6}G_{5I,JK}\left( 
\phi^{I|a}_{|a}\phi^{J|b}_{|b}\phi^{K|c}_{|c}-3 \phi^{I|a}_{|a}\phi^{J|c}_{|b}\phi^{K|b}_{|c}
+2 \phi^{I|b}_{|a} \phi^{J|c}_{|b} \phi^{K|a}_{|c} \right),
\end{align} 
where $G_2, G_3{}_I, G_4$ and $G_5{}_I$ are arbitrary functions of
$\phi^I$ and $X^{IJ}$, and $G^{ab}$ is the Einstein tensor.  The
functions $G_3{}_{I}{}_{,JK}, G_4{}_{,IJ,KL}, G_5{}_{I,JK}$ and
$G_5{}_{I,JK,LM}$ must be totally symmetric in all of their
indices, $I,J,K,L$ and $M$ in order for the field equations to be of
second order.  

We can move from the the most general bi-scalar-tensor theory to the generalized multi-Galileon theory by setting the functions as (see Ref.~\cite{Ohashi:2015fma} for details)
\begin{equation}
\begin{aligned}
A&=-\frac{1}{2}G_2+G_{3(I,J)}X^{IJ}-2G_{4,I,J}X^{IJ},
\quad
D_{IJK}=0,
\quad
E_{IJKLM}=0,
\\
J_{IJ}&=\frac{1}{4}\left( G_{4IJ}-G_{5(I,J)}\right),
\quad
K_I=-\frac{1}{4}X^{JK}G_{5IJK}.
\end{aligned}
\end{equation}

\subsection{Horndeski theory}
\label{App:toHorndeski}

The Lagrangian of the Horndeski theory is written as
\begin{equation}
\frac{1}{\sqrt{-g}}\mathcal{L} =G_2-G_3{} \phi{}_{|a}^{|a} 
+G_4R+G_4{}_{,X}\left( \phi^{|a}_{|a} \phi^{|b}_{|b} -\phi_{|ab} \phi^{|ab}\right)
+G_{5}G^{ab}\phi_{|ab}
-\frac{G_{5,X}}{6}\left( 
\phi^{|a}_{|a}\phi^{|b}_{|b}\phi^{|c}_{|c}-3 \phi^{|a}_{|a}\phi^{|c}_{|b}\phi^{|b}_{|c}
+2 \phi^{|b}_{|a} \phi^{|c}_{|b} \phi^{|a}_{|c} \right),
\end{equation}
We can move from the the most general bi-scalar-tensor theory to the Horndeski theory by setting the functions as
\begin{equation}
\begin{aligned}
\phi^I  &\rightarrow \phi,
\quad
A\rightarrow -\frac{1}{2}G_2+G_{3,\phi}X-2G_{4,\phi \phi}X,
\quad
D_{IJK}\rightarrow0,
\quad
E_{IJKLM}\rightarrow0,
\\
J_{IJ}&\rightarrow \frac{1}{4}\left( G_{4,X}-G_{5,\phi }\right),
\quad
K_I\rightarrow -\frac{1}{4}XG_{5,X}.
\end{aligned}
\end{equation}

\section{Integrability conditions for equations of motion}
\label{App:integrability}

As we discussed in section~\ref{Sec:char-BiHondeski} and also in \cite{Ohashi:2015fma}, 
we need to impose integrability conditions 
to the equations of motion of the bi-Horndeski theory summarized in appendix~\ref{App:BiHorndeskiEoM},
in order to guarantee the existence of Lagrangian that gives these equations of motion.
The integrability conditions are summarized as
\begin{equation}
 \frac{\delta E^{a b}(x)}{\delta g_{c d}(y)}-\frac{\delta E^{c d}(y)}{\delta g_{a b}(x)}=0,
\qquad
  \frac{\delta E^{a b}(x)}{\delta \phi_{I}(y)}- \frac{\delta E_I(y)}{\delta g_{a b}(x)}=0,
\qquad
 \frac{\delta E_I(x)}{\delta \phi_{J}(y)}-\frac{\delta E_J(y)}{\delta \phi_{I}(x)}=0,
\end{equation}
where the equations of motion of the bi-Horndeski theory have the structure
\begin{equation}
E^{ab} =
E^{ab}\bigl(
g_{cd} \, ,
g_{cd,e} \, ,
g_{cd,ef} \, ,
\phi_{I} \, ,
\phi_{I,c} \, ,
\phi_{I,cd} 
\bigr),
\qquad
E_I =
E_I\bigl(
g_{cd} \, ,
g_{cd,e} \, ,
g_{cd,ef} \, ,
\phi_{J} \, ,
\phi_{J,c} \, ,
\phi_{J,cd} 
\bigr).
\end{equation}
Applying the method of \cite{Deffayet:2013tca} to these equations,
we can derive the integrability conditions as
\begin{align}
&E^{a b ;c d ,e f}-E^{c d ; a b ,e f} =0
\label{invarianceid}
\\
&E^{a b ; c d , e}+E^{ c d ; a b , e}-2\partial_{f}E^{ c d ;a b,e f}=0\\
& E^{a b ; c d}-E^{c d ; a b}+\partial_{e} \left( E^{c d ; a b ,e} -\partial_{f} E^{c d ; a b ,e f}\right) =0\\ \notag \\
&E_I{}^{;a b}_{J}-E_J{}^{;a b}_{I}=0\\
&E_I{}^{;a}_{J}+E_J{}^{;a}_{I}-2\partial_{b}E_J{}^{;a b}_{I}=0\\
&E_{I,J}-E_{J,I}+\partial_{a} \left( E_J{}^{;a}_{I}-\partial_{b}E_J{}^{;a b}_{I} \right) =0\\ 
\notag \\
&E^{a b}{}_I^{;c d}-E_I{}^{;ab ,c d }=0 \\
&E^{a b}{}_I^{;c}+E_I{}^{;a b ,c}-2\partial_{d}E_I{}^{;a b ,c d} =0\\
&E^{a b}{}_{,I}-E_I{}^{;a b}
+\partial_{c} \left( 
E_I{}^{;a b,c} -\partial_{d} E_I{}^{;a b,c d}
\right)=0,
\end{align}
where
\begin{equation}
\begin{aligned}
&E^{a b ;c d ,e f} \equiv \frac{\partial E^{a b }}{\partial g_{c d , e f} } ,
\quad
E^{a b ;c d ,e } \equiv \frac{\partial E^{a b }}{\partial g_{c d , e } } ,
\quad
E^{a b ;c d } \equiv \frac{\partial E^{a b }}{\partial g_{c d } },
\quad
E_I{}^{;a b}_{J}\equiv \frac{\partial E_I}{\partial \phi_{J,a b  } } ,
\quad
E_I{}^{;a}_{J}\equiv \frac{\partial E_I}{\partial \phi_{J,a} } ,
\quad
E_I{}_{,J}\equiv \frac{\partial E_I}{\partial \phi_{J  } } ,
\\&
E^{a b}{}_{,I} \equiv \frac{\partial E^{a b }}{\partial \phi_I } ,
\quad
E_I{}^{;a b} \equiv \frac{\partial E_I}{\partial g_{a b} },
\quad
E^{a b}{}_{I}^{;c} \equiv \frac{\partial E^{a b }}{\partial \phi_{I,c} } ,
\quad
E_I{}^{;a b,c} \equiv \frac{\partial E_I}{\partial g_{a b ,c} },
\quad
E^{a b}{}_{I}^{;c d} \equiv \frac{\partial E^{a b }}{\partial \phi_{I,c d} } ,
\quad
E_I{}^{;a b,c d } \equiv \frac{\partial E_I}{\partial g_{a b ,c d} } .
\end{aligned}
\end{equation}
The condition~(\ref{invarianceid}) is a part of the so-called invariance identity following from covariance of the theory, and has already been imposed on the equations of motion of the bi-Horndeski theory in \cite{Ohashi:2015fma}. 
The other equations above have not been imposed yet, and would give constraints to the arbitrary functions of the theory.

\section{Principal symbol of shift-symmetric Horndeski theory}
\label{App:ssHorndeski}

In this appendix,
we summarize the explicit formula of the principal symbol~(\ref{ssHorndeskiP}), which is obtained by taking derivatives of the field equations~(\ref{geq}), (\ref{Jeq}) of the shift-symmetric Horndeski theory.
Since we have the symmetry~(\ref{IPsymmetry}), below we show only
$\xi_s\xi_t r^{ab}\frac{\partial{\cal G}_{ab}^n}{\partial \phi_{,st}}$, 
which is equal to
$r_{qr}\xi_s \xi_t \frac{\partial \nabla^a {\cal J}_a^n}{\partial g_{qr,st}}$.
To derive the principal symbol, we use
\begin{equation}
 R_{abcd} = -2 g_{[a|[c,d]|b]}+\cdots,
  \qquad
  \frac{\partial R_{a_1 a_2}^{b_1 b_2}}{\partial g_{qr,st}}
  r_{qr}\xi_s \xi_t
  =
  -2 r_{[a_1}^{[b_1}\xi_{a_2]}\xi^{b_2]}.
\end{equation}

\subsection{Principal symbol}
\label{App:ssHorndeski_P}

The metric part of the principal symbol is given by
\begin{align}
\xi_s \xi_t r^{ab}r_{qr}\frac{\partial {\cal G}^2_{ab}}{\partial g_{qr,st}} 
&=
\xi_s \xi_t r^{ab}r_{qr}\frac{\partial {\cal G}^3_{ab}}{\partial g_{qr,st}} 
= 0
\\
\xi_s \xi_t r^{ab}r_{qr}\frac{\partial {\cal G}^4_{ab}}{\partial g_{qr,st}} 
&=
 \frac12 \left( G_4- 2X G_{4X}\right)
 \delta_{a a_1 a_2}^{b b_1 b_2} \xi^a \xi_b r^{a_1}_{b_1}r^{a_2}_{b_2}
-\frac12  G_{4X} \delta_{a a_1 a_2 a_3}^{b b_1 b_2 b_3} \xi^a \xi_b r^{a_1}_{b_1}r^{a_2}_{b_2}\phi^{|a_3} \phi_{|b_3}
\\
\xi_s \xi_t r^{ab}r_{qr}
\frac{\partial {\cal G}^5_{ab}}{\partial g_{qr,st}} 
&=
\frac12 X
G_{5X}
 \delta_{a a_1 a_2 a_3}^{b b_1 b_2 b_3}\xi^a \xi_b r^{a_1}_{b_1}r^{a_2}_{b_2}\phi^{|a_3}_{|b_3}.
\end{align}
The mixed part between the metric and scalar parts is given by
\begin{align}
\xi_\mu \xi_\nu r^{ab}
 \frac{\partial {\cal G}^2_{ab}}{\partial \phi_{,st}} 
&= 0
\\
 \xi_s \xi_t r^{ab}
 \frac{\partial {\cal G}^3_{ab}}{\partial \phi_{,st}} 
&=
-\frac12 G_{3X}\left(
 \delta_{a a_1 a_2}^{b b_1 b_2} \xi^a \xi_b r_{a_1}^{b_1}
 \phi^{|a_2}\phi_{|b_2}
+2X
\delta_{b b_1}^{a a_1} \xi^a \xi_b r^{a_2}_{b_2}
\right)
\\
 \xi_s \xi_t
r^{ab}
 \frac{\partial {\cal G}^4_{ab}}{\partial \phi_{,st}}
 &=
 \left(G_{4X} + 2X G_{4XX}\right)
\delta_{a a_1 a_2}^{b b_1 b_2}  \xi^a \xi_b r^{a_1}_{b_1} \phi^{|a_2}_{|b_2}
 + G_{4XX}
 \delta_{a a_1 a_2 a_3}^{b b_1 b_2 b_3} \xi^a \xi_b r^{a_1}_{b_1} \phi^{|a_2}_{|b_2} \phi^{|a_3} \phi_{|b_3}
 \\
 \xi_s \xi_t
r^{ab} 
 \frac{\partial {\cal G}^5_{ab}}{\partial \phi_{,st}}
&=
-\frac12 \left(G_{5X} + X G_{5XX}\right)
\delta_{a a_1 a_2 a_3}^{b b_1 b_2 b_3} \xi^a \xi_b r^{a_1}_{b_1} \phi^{|a_2}_{|b_2} \phi^{|a_3}_{|b_3} 
- \frac14 X G_{5X} 
\delta_{a a_1 a_2 a_3}^{b b_1 b_2 b_3} \xi^a \xi_b r^{a_1}_{b_1} R^{a_2 a_3}_{b_2 b_3}.
\end{align}
Finally the scalar part is given by (where $\xi^2\equiv \xi_a \xi^a$ and $\xi\cdot\phi \equiv \xi_a \phi^{|a}$)
\begin{align}
\xi_s \xi_t \frac{\partial \nabla^a {\cal J}_a^2}{\partial \phi_{,st}}
&=
- K_X \xi^2 + K_{XX}(\xi\cdot \phi)^2
\label{ddivJdphipp}
\\
\xi_s \xi_t \frac{\partial \nabla^a {\cal J}_a^3}{\partial \phi_{,st}}
&=
2\left(G_{3X} + X G_{3XX}\right)\delta_{a a_1}^{b b_1}\xi^a \xi_b \phi^{|a_1}_{|b_1}
+ G_{3XX} \delta_{a a_1 a_2}^{b b_1 b_2}\xi^a \xi_b \phi^{|a_1}_{|b_1}\phi^{|a_2}\phi_{|b_2}
\\
\xi_s \xi_t \frac{\partial \nabla^a {\cal J}_a^4}{\partial \phi_{,st}}
 &=
 - \left(3 G_{4XX} + 2X G_{4XXX} \right)
 \delta_{a a_1 a_2}^{b b_1 b_2} \xi^a \xi_b \phi^{|a_1}_{|b_1}\phi^{|a_2}_{|b_2}
  - G_{4XXX}
 \delta_{a a_1 a_2 a_3}^{b b_1 b_2 b_3} \xi^a \xi_b \phi^{|a_1}_{|b_1}\phi^{|a_2}_{|b_2} \phi^{|a_3} \phi_{|b_3}
 \notag \\ &\quad
  - \frac12 \left( G_{4X} +2 X G_{4XX} \right) \delta_{a a_1 a_2}^{b b_1 b_2} \xi^a \xi_b R^{a_1 a_2}_{b_1 b_2}
 -\frac12 G_{4XX} \delta_{a a_1 a_2 a_3}^{b b_1 b_2 b_3} \xi^a \xi_b R^{a_1 a_2}_{b_1 b_2} \phi^{|a_3} \phi_{|b_3}
\\
\xi_s \xi_t \frac{\partial \nabla^a {\cal J}_a^5}{\partial \phi_{,st}}
 &=
  \frac{2G_{5XX} +X G_{5XXX}}{3}
\delta_{a a_1 a_2 a_3}^{b b_1 b_2 b_3} \xi^a \xi_b\phi^{|a_1}_{|b_1} \phi^{|a_2}_{|b_2} \phi^{|a_3}_{|b_3} 
+
\frac{G_{5X} + X G_{5XX}}{2}
\delta_{a a_1 a_2 a_3}^{b b_1 b_2 b_3} \xi^a \xi_b\phi^{|a_1}_{|b_1} R^{a_2 a_3}_{b_2 b_3}.
\end{align}
Expressions with non-contracted indices is obtained just by removing one of $r^a_b$ from the expressions above. 

\subsection{Trace-reversed principal symbol}
\label{App:ssHorndeski_trrev}

For the analysis in sections~\ref{Sec:planewave} and \ref{Sec:shock}, we need also the trace-reversed expressions of the principal symbol defined by
\begin{equation}
\tilde P = P - \frac12 \left({\mathrm tr} P\right) g.
\end{equation}
We summarize its expression below. Since the scalar part of it is the same as the above expressions, we show only the metric part.
The pure metric part is given by
\begin{align}
\xi_s \xi_t r_{qr}\frac{\partial \tilde{\cal G}^2{}_a^b}{\partial g_{qr,st}} 
&=
\xi_s \xi_t r_{qr}\frac{\partial \tilde {\cal G}^3{}_a^b}{\partial g_{qr,st}} 
= 0
\label{dtildeGdGpp}
\\
\xi_s \xi_t r_{qr}\frac{\partial \tilde{\cal G}^4{}_a^b}{\partial g_{qr,st}} 
&=
\frac{G_4 - 2 X G_{4X}}{2} 
\left(
\delta_{a a_1 a_2}^{b b_1 b_2}
-\delta_a^b \delta_{a_1 a_2}^{b_1 b_2}
\right)
\xi^{a_1} \xi_{b_1} r^{a_2}_{b_2}
- 
\frac{G_{4X}}{2}
\left(
\delta_{a a_1 a_2 a_3}^{b b_1 b_2 b_3}
-\frac12 \delta_a^b \delta_{a_1 a_2 a_3}^{b_1 b_2 b_3}
\right)
\xi^{a_1} \xi_{b_1} r^{a_2}_{b_2}\phi^{|a_3} \phi_{|b_3}
\\
\xi_s \xi_t r_{qr}
\frac{\partial \tilde{\cal G}^5{}_a^b}{\partial g_{qr,st}} 
&=
\frac12 X
G_{5X}
\left(
\delta_{a a_1 a_2 a_3}^{b b_1 b_2 b_3}
-\frac12 \delta_a^b \delta_{a_1 a_2 a_3}^{b_1 b_2 b_3}
\right)
\xi^{a_1} \xi_{b_1} r^{a_2}_{b_2}\phi^{|a_3}_{|b_3},
\end{align}
and the mixed part between the metric and scalar parts is given by
\begin{align}
\xi_s \xi_t 
\frac{\partial \tilde {\cal G}^2{}_a^b}{\partial \phi_{,st}} 
&= 0
\\
\xi_s \xi_t \frac{\partial \tilde {\cal G}^3{}_a^b}{\partial \phi_{,st}} 
&=
-\frac12 G_{3X}\left\{
\left(
\delta_{a a_1 a_2}^{b b_1 b_2}
-\delta_a^b \delta_{a_1 a_2}^{b_1 b_2}
\right)
\xi^{a_1} \xi_{b_1} \phi^{|a_2}\phi_{|b_2}
+X\left(
2\delta_{a a_1}^{b b_1} \xi^{a_1} \xi_{b_1} 
- 3 \xi^2 \delta_a^b
\right)
\right\}
\\
\xi_s \xi_t \frac{\partial \tilde {\cal G}^4{}_a^b}{\partial \phi_{,st}} 
&=
\left(
G_{4X} + 2X G_{4XX}
\right)
\left(
\delta_{a a_1 a_2}^{b b_1 b_2}
-\delta_a^b \delta_{a_1 a_2}^{b_1 b_2}
\right)
\xi^{a_1} \xi_{b_1} \phi^{|a_2}_{|b_2}
+ G_{4XX}
\left(
\delta_{a a_1 a_2 a_3}^{b b_1 b_2 b_3}
-\frac12 \delta_a^b \delta_{a_1 a_2 a_3}^{b_1 b_2 b_3}
\right)
\xi^{a_1} \xi_{b_1} \phi^{|a_2}_{|b_2}
\phi^{|a_3} \phi_{|b_3} 
\\
\xi_s \xi_t \frac{\partial \tilde {\cal G}^5{}_a^b}{\partial \phi_{,st}} 
&=
\left(
\delta_{a a_1 a_2 a_3}^{b b_1 b_2 b_3}
-\frac12 \delta_a^b \delta_{a_1 a_2 a_3}^{b_1 b_2 b_3}
\right)
\left\{
-\frac12 \left(G_{5X} + X G_{5XX}\right)
\xi^{a_1} \xi_{b_1}  \phi^{|a_2}_{|b_2} \phi^{|a_3}_{|b_3} 
- \frac14 X G_{5X} 
\xi^{a_1} \xi_{b_1} R^{a_2 a_3}_{b_2 b_3}
\right\}.
\end{align}

\section{$\cal N$ of the shift-symmetric Horndeski theory}
\label{App:N}

We summarize the explicit formula of $\cal N$ in the shift-symmetric Horndeski theory discussed in section~\ref{Sec:N}.
This quantity is derived by taking derivatives of the principal symbol obtained in section~\ref{Sec:EoM} and appendix~\ref{App:ssHorndeski}.
In this appendix, we use the coordinate $x^0$ where $\xi_a = \left(dx^0\right)_a$ is a normal of a $x^0$-constant surface.
We also utilize the following formula for derivatives by
$\xi_c \frac{\partial}{\partial g_{ab,c}} = \frac{\partial}{\partial g_{ab,0}}$ and $\xi_a \frac{\partial}{\partial \phi_{,a}} = \frac{\partial}{\partial \phi_{,0}}$:
\begin{equation}
\begin{aligned}
&r_{ab}\frac{\partial R^{a_1 a_2}_{b_1 b_2}}{\partial g_{ab,0}}
=
2 \Gamma^0 {}^{[a_1}_{[b_1} r{}^{a_2]}_{b_2]},
\qquad
\frac{\partial \phi^2}{\partial \phi_{,0}} = 2\phi^{|0},
\qquad
\frac{\partial X}{\partial \phi_{,0}} = -\phi^{|0},
\qquad
\frac{\partial\phi_a}{\partial \phi_{,0}} = \delta^0_a = \xi_a,
\\
&\frac{\partial\phi_{ab}}{\partial \phi_{,0}}
= 
- \Gamma^0{}^a_b,
\qquad
 r_{cd}
\frac{\partial \phi^a_b}{\partial g_{cd,0}}
=
\frac12\left(
\phi^{|0}r^a_b - \xi^a r_{bc}\phi^{|c} - \xi_b r^a_c\phi^{|c}
\right)
\simeq \frac12 \phi^{|0} r^a_b.
\end{aligned}
\end{equation}
The last equality ($\simeq$) holds only when $\frac{\partial \phi^a_b}{\partial g_{cd,0}}$ is contracted with the generalized Kronecker delta multiplied by $\xi^a\xi_b$, by which the terms involving $\xi$ vanish identically.

We summarize the terms appearing in $\cal N$~(\ref{Ndef2}) in general spacetime below.
First, the pure metric terms proportional to $r_{ab}{}^3$ are given by
\begin{align}
r^{ab} r_{cd}r_{ef}\frac{\partial}{\partial g_{ef,0}}
\frac{\partial {\cal G}^{2,3,4}_{ab}}{\partial g_{cd,00}} 
&=
0
\\
 r^{ab} r_{cd}r_{ef}\frac{\partial}{\partial g_{ef,0}}
\frac{\partial {\cal G}^5_{ab}}{\partial g_{cd,00}}
&=
 \frac{\phi^{|0}}{4}
 X G_{5X}
 \delta_{a a_1 a_2 a_3}^{b b_1 b_2 b_3} 
\xi^a \xi_b r^{a_1}_{b_1} r^{a_2}_{b_2} r^{a_3}_{b_3}.
\label{d2G5dgpdgpp}
\end{align}
Next, the mixed terms proportional to $r_{ab}{}^2 r_\phi$ are given by
\begin{align}
r^{ab}r_{ef}\frac{\partial}{\partial g_{ef,0}}
\frac{\partial {\cal G}^{2,3}_{ab}}{\partial \phi_{,00}} 
&=
0
\\
r^{ab}r_{ef}\frac{\partial}{\partial g_{ef,0}}
\frac{\partial {\cal G}^4_{ab}}{\partial \phi_{,00}} 
&=
\frac{\phi^{|0}}{2}\left(G_{4X} + 2XG_{4XX}\right)
\delta_{a a_1 a_2}^{b b_1 b_2} \xi^a \xi_b r^{a_1}_{b_1} r^{a_2}_{b_2} 
+ \frac{\phi^{|0}}{2}G_{4XX}
\delta_{a a_1 a_2 a_3}^{b b_1 b_2 b_3} \xi^a \xi_b r^{a_1}_{b_1} r^{a_2}_{b_2} \phi^{|a_3}\phi_{|b_3}
 \\
r^{ab}r_{ef}\frac{\partial}{\partial g_{ef,0}}
\frac{\partial {\cal G}^5_{ab}}{\partial \phi_{,00}} 
&=
-\frac{\phi^{|0}}{2}\left(G_{5X} + X G_{5XX}\right)
\delta_{a a_1 a_2 a_3}^{b b_1 b_2 b_3} \xi^a \xi_b r^{a_1}_{b_1} r^{a_2}_{b_2} \phi^{|a_3}_{|b_3}
- \frac{X G_{5X}}{2}  
\delta_{a a_1 a_2 a_3}^{b b_1 b_2 b_3} \xi^a \xi_b r^{a_1}_{b_1} r^{a_2}_{b_2} \Gamma^0{}^{a_3}_{b_3},
\end{align}
where the overall factor $r_\phi$ is omitted.
The other mixed terms proportional to $r_{ab}r_\phi{}^2$ are given by
\begin{align}
r_{ef}\frac{\partial}{\partial g_{ef,0}} \frac{\partial \nabla^a {\cal J}_a^2}{\partial \phi_{,00}}
 &= 0
\\
r_{ef}\frac{\partial}{\partial g_{ef,0}} \frac{\partial \nabla^a {\cal J}_a^3}{\partial \phi_{,00}}
&=
\phi^{|0} \left(G_{3X} +X G_{3XX}\right) \delta_{a a_1}^{b b_1} \xi^a \xi_b r^{a_1}_{b_1} 
+ \frac{\phi^{|0}}{2}G_{3XX} \delta_{a a_1 a_2}^{b b_1 b_2} \xi^a \xi_b r^{a_1}_{b_1}\phi^{|a_2}\phi_{|b_2}
\\
r_{ef}\frac{\partial}{\partial g_{ef,0}} \frac{\partial \nabla^a {\cal J}_a^4}{\partial \phi_{,00}}
 &=
 -\phi^{|0}
\xi^a \xi_b \phi^{|a_1}_{|b_1} r^{a_2}_{b_2}
\left\{
\left(
 3 G_{4XX} + 2 X G_{4XXX}
 \right)\delta_{a a_1 a_2}^{b b_1 b_2} 
+ G_{4XXX} \delta_{a a_1 a_2 a_3}^{b b_1 b_2 b_3} \phi^{|a_3}\phi_{|b_3}
\right\}
\notag \\ & \quad
 -\left( G_{4X} + 2X G_{4XX} \right)
 \delta_{a a_1 a_2}^{b b_1 b_2} \xi^a\xi_b r^{a_1}_{b_1} \Gamma^0{}^{a_2}_{b_2}
- G_{4XX} 
 \delta_{a a_1 a_2 a_3}^{b b_1 b_2 b_3}
 \xi^a\xi_b r^{a_1}_{b_1} \Gamma^0{}^{a_2}_{b_2}\phi^{|a_3}\phi_{|b_3}
\\
r_{ef}\frac{\partial}{\partial g_{ef,0}} \frac{\partial \nabla^a {\cal J}_a^5}{\partial \phi_{,00}}
&=
\frac14 \left(G_{5X} + X G_{5XX}\right)
\delta_{a a_1 a_2 a_3}^{b b_1 b_2 b_3}\xi^a\xi_b \left(
\phi^{|0} r^{a_1}_{b_1}R^{a_2 a_3}_{b_2 b_3} + 4 \phi^{|a_1}_{|b_1}\Gamma^0{}^{a_2}_{b_2} r^{a_3}_{b_3}
\right)
\notag \\
&\quad
+ \frac{\phi^{|0}}{2}\left(2G_{5XX} + X G_{5XXX}\right) 
\delta_{a a_1 a_2 a_3}^{b b_1 b_2 b_3}\xi^a\xi_b \phi^{|a_1}_{|b_1}\phi^{|a_2}_{|b_2} r^{a_3}_{b_3}.
\end{align}
Last, the pure scalar terms proportional to $r_\phi{}^3$ are given by
\begin{align}
\frac{\partial}{\partial\phi_{,0}}  \frac{\partial \nabla^a {\cal J}^2_a}{\partial \phi_{,00}}
&=
3 \phi^{|0} K_{XX} \xi^2 - (\phi^{|0})^3 K_{XXX}
\label{d2J2dphippdphip}
\\
\frac{\partial}{\partial\phi_{,0}}  \frac{\partial \nabla^a {\cal J}^3_a}{\partial \phi_{,00}}
&=
-\phi^{|0} \left\{
2\left(2G_{3XX}+ XG_{3XXX}\right) \delta_{a a_1}^{b b_1}\xi^a\xi_b \phi^{|a_1}_{|b_1}
+ G_{3XXX} \delta_{a a_1 a_2}^{b b_1 b_2} \xi^a\xi_b \phi^{|a_1}_{|b_1} \phi^{|a_2}\phi_{|b_2}
\right\}
\notag \\ & \quad
- 2\left(G_{3X} + X G_{3XX}\right) \delta_{a a_1}^{b b_1} \xi^a\xi_b \Gamma^0{}^{a_1}_{b_1}
- G_{3XX} \delta_{a a_1 a_2}^{b b_1 b_2}\xi^a\xi_b \Gamma^0{}^{a_1}_{b_1} \phi^{|a_2}\phi_{|b_2}
\\
 \frac{\partial}{\partial\phi_{,0}}  \frac{\partial \nabla^a {\cal J}^4_a}{\partial \phi_{,00}}
 &=
 \frac{\phi^{|0}}{2}
\delta_{a a_1 a_2}^{b b_1 b_2} \xi^a\xi_b 
 \left\{
2 \left(5G_{4XXX} + 2X G_{4XXXX}\right)\phi^{|a_1}_{|b_1}\phi^{|a_2}_{|b_2}
 +  \left(3 G_{4XX} + 2 X G_{4XXX}\right)R^{a_1 a_2}_{b_1 b_2}
\right\}
 \notag \\&\quad
 +\frac{\phi^{|0}}{2}
\delta_{a a_1 a_2 a_3}^{b b_1 b_2 b_3} \xi^a\xi_b\phi^{|a_1}\phi_{|b_1}
 \left(
2 G_{4XXXX}  \phi^{|a_2}_{|b_2}\phi^{|a_3}_{|b_3}
 +  G_{4XXX}   R^{a_2 a_3}_{b_2 b_3}
 \right)
 \notag \\&\quad
+ 2\left(3 G_{4XX} + 2X G_{4XXX}\right)
\delta_{a a_1 a_2}^{b b_1 b_2} \xi^a\xi_b \phi^{|a_1}_{|b_1}\Gamma^0{}^{a_2}_{b_2}
 + 2 G_{4XXX}
 \delta_{a a_1 a_2 a_3}^{b b_1 b_2 b_3}
 \xi^a\xi_b \Gamma^0{}^{a_1}_{b_1} \phi^{|a_2}_{|b_2} \phi^{|a_3}\phi_{|b_3}
 \\
 \frac{\partial}{\partial\phi_{,0}}  \frac{\partial \nabla^a {\cal J}^5_a}{\partial \phi_{,00}}
 &=
 -\delta_{a a_1 a_2 a_3}^{b b_1 b_2 b_3} \xi^a\xi_b
 \biggl\{
 \left(2G_{5XX} + X G_{5XXX}\right) \phi^{|a_1}_{|b_1} \phi^{|a_2}_{|b_2} \Gamma^0{}^{a_3}_{b_3}
 +\frac12 \left(G_{5X} + X G_{5XX}\right) \Gamma^0{}^{a_1}_{b_1} R^{a_2 a_3}_{b_2 b_3}
 \notag \\ & \qquad 
 +\frac{\phi^{|0}}{3}\left( 3 G_{5XXX} + X G_{5XXXX} \right) \phi^{|a_1}_{|b_1} \phi^{|a_2}_{|b_2} \phi^{|a_3}_{|b_3}
 + \frac{\phi^{|0}}{2}\left( 2 G_{5XX} + X G_{5XXX} \right) \phi^{|a_1}_{|b_1}R^{a_2 a_3}_{b_2 b_3}
\biggr\}.
\end{align}

\section{$\cal N$ on the plane wave solution}
\label{App:Npp}

In this appendix, we show the explicit form of $\cal N$ on the plane wave solution, which can be obtained by plugging the background solution given in 
section~\ref{Sec:BG}
into expressions in appendix~\ref{App:N}.
We also need to use the expressions of the eigenvectors (\ref{EVgrav}), (\ref{EVscalar}) of the tensor and scalar modes to find ${\cal N}$ for each mode.

For the tensor mode  $r = (r^\text{(T)}_{ab},0)$, the only nontrivial term is (\ref{d2G5dgpdgpp}), but this term is zero for the plane wave solution since $X=0$. Hence ${\cal N}=0$ for the tensor mode on the plane wave solution.

Let us move on to the scalar mode~(\ref{EVscalar}), whose eigenvector is given by
\begin{equation}
r = 
\left( r_{ab}, r_\phi  \right)
= 
\left(
2 \tilde r_{\ell n} \ell_{(a}n_{b)},
r_\phi
\right),
\quad
\tilde r_{\ell n} = 
\frac{2 K_X}{G_4} \, 
\frac{ -\frac12 G_{3X} \phi'^2 + G_{4X} \phi''}
{K_{XX}\phi'^2 - 2 G_{3X}\phi'' - G_{4X} \Delta F} 
\, r_\phi.
\label{evPP}
\end{equation}
When this eigenvector is contracted with the generalized Kronecker delta multiplied by $\xi_a \xi^b$, it simplifies to (by dropping the $\xi$ terms)
\begin{equation}
r_a^b = r_{\ell n}\left(\ell_a n^b + \ell^b n_a\right)
\simeq
\omega r_{\ell n} \ell_a \ell^b.
\end{equation}

Below, we summarize the terms comprising $\cal N$ given in appendix~\ref{App:N} using the eigenvector given by~(\ref{evPP}).
It is useful to use the following formula for evaluation:
\begin{gather}
\ell_a = \left(du\right)_a,
\quad
\ell^a = \left(\partial_v\right)^a
= \left(\ell^u,\ell^0\right) = \left(0,1\right),
\quad
\xi_a = \left(dx^0\right)_a ,
\quad
\xi^a = (\xi^u,\xi^0) = (1,-\omega),
\\
\ell\cdot \xi = g^{0u} = 1,
\quad
\xi^2 = g^{00} = -\omega,
\quad
n_a = \frac\omega2 \ell_a + \xi_a,
\quad
n\cdot \xi = -\frac\omega2,
\quad
\phi^{|0} = \xi_a \phi^{|a} = \phi' \xi\cdot\ell = \phi'.
\end{gather}

First, we find that the pure metric terms 
and also
the mixed terms proportional to $r_{ab}{}^2 r_\phi$
identically vanish:
\begin{equation}
r^{ab} r_{cd} r_{ef}\frac{\partial}{\partial g_{ef,0}}
\frac{\partial {\cal G}^{2,3,4,5}_{ab}}{\partial g_{cd,00}} 
=
r^{ab}r_{ef}\frac{\partial}{\partial g_{ef,0}}
\frac{\partial {\cal G}^{2,3,4,5}_{ab}}{\partial \phi_{,00}} 
=0.
\end{equation}
The other mixed terms proportional to $r_{ab}r_\phi{}^2$ are given by
\begin{equation}
r_{ef}\frac{\partial}{\partial g_{ef,0}} \frac{\partial \nabla^a {\cal J}_a^{2,5}}{\partial \phi_{,00}}
 = 0,
\quad
r_{ef}\frac{\partial}{\partial g_{ef,0}} \frac{\partial \nabla^a {\cal J}_a^3}{\partial \phi_{,00}}
=
\phi' G_{3X} \delta_{a a_1}^{b b_1} \xi^a\xi_b r^{a_1}_{b_1} 
,
\quad
r_{ef}\frac{\partial}{\partial g_{ef,0}} \frac{\partial \nabla^a {\cal J}_a^4}{\partial \phi_{,00}}
=
- G_{4X}  \delta_{a \alpha_1 a_2}^{b \beta_1 b_2} \xi^a\xi_b \Gamma^{0}{}^{\alpha_1}_{\beta_1} r^{a_2}_{b_2},
 \end{equation}
where $x^\alpha =x,y$ as defined in section~\ref{Sec:Npp_geometry}.
Last, the pure scalar terms are given by
\begin{align}
\frac{\partial}{\partial\phi_{,0}}  \frac{\partial \nabla^\alpha {\cal J}^2_\alpha}{\partial \phi_{,00}}
&=
-3\omega \phi' K_{XX}  - \phi'^3 K_{XXX}
\\
\frac{\partial}{\partial\phi_{,0}}  \frac{\partial \nabla^\alpha {\cal J}^3_\alpha}{\partial \phi_{,00}}
&=
- 4\phi'\phi'' G_{3XX} \delta_{a a_1}^{b b_1}\xi^a\xi_b \ell^{a_1}\ell_{b_1}
- 2 G_{3X} \delta_{a \alpha_1}^{b \beta_1}\xi^a\xi_b \Gamma^0{}^{\alpha_1}_{\beta_1}
- \phi'^2 G_{3XX} \delta_{a \alpha_1 a_2}^{b \beta_1 b_2}\xi^a\xi_b \Gamma^0{}^{\alpha_1}_{\beta_1} \ell^{a_2}\ell_{b_2}
\\
\frac{\partial}{\partial\phi_{,0}}  \frac{\partial \nabla^\alpha {\cal J}^4_\alpha}{\partial \phi_{,00}}
&=
-3 \phi' G_{4XX} \delta_{a \alpha_1 a_2}^{b \beta_1 b_2}\xi^a\xi_b F^{,\alpha_1}_{,\beta_1} \ell^{a_2} \ell_{b_2}
+ 6 G_{4XX} \phi'' \delta_{a a_1 \alpha_2}^{b b_1 \beta_2} \xi^a\xi_b \ell^{a_1} \ell_{b_1} \Gamma^0{}^{\alpha_2}_{\beta_2}
 \\
 \frac{\partial}{\partial\phi_{,0}}  \frac{\partial \nabla^\alpha {\cal J}^5_\alpha}{\partial \phi_{,00}}
&=
  G_{5X} \delta_{a \alpha_1 \alpha_2 a_3}^{b \beta_1 \beta_2 b_3}
\xi^a\xi_b  \Gamma^0{}^{\alpha_1}_{\beta_1} F^{,\alpha_2}_{,\beta_2} \ell^{a_3}\ell_{b_3}.
\end{align}

Summing up the above terms, we find $\cal N$ for the scalar mode on the plane wave solution to be given by
\begin{align}
{\cal N} &=
3\omega r_\phi{}^2 \tilde r_{\ell n}\left(
\phi' G_{3X} \delta_{a a_1}^{b b_1}\xi^a\xi_b \ell^{a_1}\ell_{b_1}
- G_{4X}\delta_{a \alpha_1 a_2}^{b \beta_1 b_2}\xi^a\xi_b \Gamma^0{}^{\alpha_1}_{\beta_1} \ell^{a_2}\ell_{b_2}
\right)
\notag \\ & \quad
+ r_\phi{}^3 \Bigl\{
-3\omega K_{XX} \phi'
- K_{XXX}\phi'^3
- 4\phi' \phi'' G_{3XX}\delta_{a a_1}^{b b_1}\xi^a\xi_b \ell^{a_1}\ell_{b_1}
\notag \\ & \qquad \qquad
- 2 G_{3X} \delta_{a \alpha_1}^{b \beta_1}\xi^a\xi_b \Gamma^0{}^{\alpha_1}_{\beta_1}
+\left(
 - \phi'^2G_{3X}
 + 6 \phi'' G_{4XX}
\right)
 \delta_{a \alpha_1 a_2}^{b \beta_1 b_2}\xi^a\xi_b \Gamma^0{}^{\alpha_1}_{\beta_1}\ell^{a_2}\ell_{b_2}
\notag \\ & \qquad \qquad
-3 \phi' G_{4XX} \delta_{a \alpha_1 a_2}^{b \beta_1 b_2}\xi^a\xi_b F^{,\alpha_1}_{,\beta_1}\ell^{a_2}\ell_{b_2}
+ G_{5X} \delta_{a \alpha_1 a_2 \alpha_3}^{b \beta_1 b_2 \beta_3}\xi^a\xi_b \Gamma^0{}^{\alpha_1}_{\beta_1}\ell^{a_2}\ell_{b_2} F^{,\alpha_3}_{,\beta_3}
\Bigr\}.
\label{Nppgen}
\end{align}
If we further assume that  $F$ in the metric is given by Eq.~(\ref{Fexp}) and also $\phi''=0$,
we can introduce the metric~(\ref{gppmod}) adapted to geodesics, with which we can evaluate Eq.~(\ref{Nppgen}) more explicitly to arrive at Eq.~(\ref{Nscalar_pp}) by setting $r_\phi=1$.

\bibliography{Causality_v2.bbl}

\end{document}